\documentclass[twocolumn,nofootinbib,superscriptaddress,preprintnumbers,floatfix,eqsecnum]{revtex4}

\usepackage[small]{subfigure}
\usepackage[bookmarks=false,hyperfootnotes=false]{hyperref}
\usepackage{graphicx}
\usepackage{amsmath}
\usepackage{xspace}

\newcommand{\eq}[1]{Eq.~\eqref{eq:#1}}
\newcommand{\eqs}[2]{Eqs.~\eqref{eq:#1} and \eqref{eq:#2}}
\renewcommand{\sec}[1]{Sec.~\ref{sec:#1}}
\newcommand{\secs}[2]{Secs.~\ref{sec:#1} and \ref{sec:#2}}
\newcommand{\subsec}[1]{Sec.~\ref{subsec:#1}}
\newcommand{\subsecs}[2]{Secs.~\ref{subsec:#1} and \ref{subsec:#2}}
\newcommand{\fig}[1]{Fig.~\ref{fig:#1}}
\newcommand{\figs}[2]{Figs.~\ref{fig:#1} and \ref{fig:#2}}
\newcommand{\app}[1]{Appendix~\ref{app:#1}}

\newcommand{\abs}[1]{\lvert#1\rvert}
\newcommand{\ord}[1]{\mathcal{O}(#1)}

\newcommand{\mae}[3]{\langle#1\lvert#2\rvert#3\rangle}
\newcommand{\Mae}[3]{\bigl\langle#1\bigl\lvert#2\bigr\rvert#3\bigr\rangle}

\newcommand{\df}{\mathrm{d}}
\newcommand{\img}{\mathrm{i}}
\newcommand{\tr}{\mathrm{tr}}
\newcommand{\id}{\mathbf{1}}

\newcommand{\e}{\epsilon}
\newcommand{\la}{\lambda}
\newcommand{\w}{\omega}
\newcommand{\Tau}{\mathcal{T}}

\newcommand{\bn}{\bar{n}}
\newcommand{\bq}{{\bar{q}}}

\newcommand{\vC}{\vec{C}}
\newcommand{\vO}{\vec{O}}

\newcommand{\hq}{\hat{q}}
\newcommand{\hs}{\hat{s}}
\newcommand{\hP}{\hat{P}}
\newcommand{\hga}{\widehat{\gamma}}

\newcommand{\hS}{\widehat{S}}
\newcommand{\hX}{\widehat{X}}
\newcommand{\hY}{\widehat{Y}}

\newcommand{\cB}{\mathcal{B}}
\newcommand{\cL}{\mathcal{L}}
\newcommand{\cM}{\mathcal{M}}

\newcommand{\Dslash}{D\!\!\!\!\slash}

\newcommand{\bnslash}{\bar{n}\!\!\!\slash}

\newcommand{\GeV}{\,\mathrm{GeV}}

\newcommand{\nn}{\nonumber}

\newcommand{\mcdot}{\!\cdot\!}

\newcommand{\lqcd}{\Lambda_\mathrm{QCD}}

\newcommand{\lp}{\tilde{p}}       
\newcommand{\Pop}{\mathcal{P}}    

\newcommand{\lbD}{\mathcal{D}}    
\newcommand{\Dc}{D}               
\newcommand{\Dcslash}{\Dslash}

\newcommand{\nt}{{n_t}}
\newcommand{\bnt}{\bn_t}
\newcommand{\tN}{\mathcal{T}}     

\newcommand{\X}{{X}}              
\newcommand{\collF}{\mathcal{C}}  

\newcommand{\T}{\mathbf{T}}       

\newcommand{\usoft}{{us}}
\newcommand{\csoft}{{cs}}
\newcommand{\coll}{{c}}
\newcommand{\cusp}{\mathrm{cusp}}
\newcommand{\cut}{\mathrm{cut}}

\newcommand{\zero}{{(0)}}
\newcommand{\one}{{(1)}}

\newcommand{\SCETp}{SCET$_+$\xspace}
\newcommand{\SCETI}{SCET\xspace}

\allowdisplaybreaks[4]

\begin{document}


\preprint{\vbox{\hbox{MIT--CTP 4273}}}

\title{Factorization and Resummation for Dijet Invariant Mass Spectra}

\author{Christian W.~Bauer}
\affiliation{Ernest Orlando Lawrence Berkeley National Laboratory,
University of California, Berkeley, CA 94720\vspace{0.5ex}}

\author{Frank J.~Tackmann}
\affiliation{Center for Theoretical Physics, Massachusetts Institute of
Technology, Cambridge, MA 02139\vspace{0.5ex}}

\author{Jonathan R.~Walsh}
\affiliation{Ernest Orlando Lawrence Berkeley National Laboratory,
University of California, Berkeley, CA 94720\vspace{0.5ex}}

\author{Saba Zuberi}
\affiliation{Ernest Orlando Lawrence Berkeley National Laboratory,
University of California, Berkeley, CA 94720\vspace{0.5ex}}

\date{June 29, 2011}

\begin{abstract}

Multijet cross sections at the LHC and Tevatron are sensitive to several distinct kinematic energy scales. When measuring the dijet invariant mass $m_{jj}$ between two signal jets produced in association with other jets or weak bosons, $m_{jj}$ will typically be much smaller than the total partonic center-of-mass energy $Q$, but larger than the individual jet masses $m$, such that there can be a hierarchy of scales $m \ll m_{jj} \ll Q$. This situation arises in many new-physics analyses at the LHC, where the invariant mass between jets is used to gain access to the masses of new-physics particles in a decay chain. At present, the logarithms arising from such a hierarchy of kinematic scales can only be summed at the leading-logarithmic level provided by parton-shower programs. We construct an effective field theory, \SCETp, which is an extension of soft-collinear effective theory that applies to this situation of hierarchical jets. It allows for a rigorous separation of different scales in a multiscale soft function and for a systematic resummation of logarithms of both $m_{jj}/Q$ and $m/Q$. As an explicit example, we consider the invariant mass spectrum of the two closest jets in $e^+e^-\!\to 3$ jets. We also give the generalization to $pp\to N$ jets plus leptons relevant for the LHC.
\end{abstract}

\maketitle

\section{Introduction}
\label{sec:intro}

A detailed understanding of events with several hadronic jets in the final state is of central importance at the Large Hadron Collider (LHC) and the Tevatron. This is because the Standard Model and most of its extensions produce energetic partons in the underlying short-distance processes, which appear as collimated jets of energetic hadrons in the detectors.

Multijet processes are inherently multiscale problems. For generic high-$p_T$ jets, there are at least three relevant energy scales present: the $p_T$ or total energy of a jet, which is of order the partonic center-of-mass energy $Q$ of the collision, the invariant mass $m$ of a jet, which is typically much smaller than $Q$, and $\lqcd$, the scale of nonperturbative physics in the strong interaction. To separately treat the physics at these different energy scales one has to rely on factorization theorems. With a factorization theorem in hand, the long-distance physics sensitive to $\lqcd$ can be determined from experimental data, while short-distance contributions can be obtained through perturbative calculations.

For most multijet events, the jets will not be equally separated and they will not have equal energy. Instead there will be some hierarchy in the invariant masses between jets or the jet energies due to the soft and collinear enhancement of emissions in QCD. Depending on the measurement, the cuts imposed to enhance the sensitivity to physics beyond the Standard Model introduce sensitivity to these additional kinematic scales. For example, requiring a large $p_T$ for the leading signal jet(s) and a smaller $p_T$ for additional jets introduces sensitivity to a new kinematic scale, namely the $p_T$ of the subleading jet. Another example is the measurement of the dijet invariant mass of jets, which is used to identify new particles decaying to jets, a special case being the identification of boosted heavy particles by measuring the invariant mass of two subjets within a larger jet. As a final example, consider the CDF measurement of $W+2$ jets~\cite{Aaltonen:2011mk}, which shows an excess in the invariant mass of the two jets around $m_{jj} \sim 150\GeV$. The excess is dominated by back-to-back jets, where the subleading jet is rather soft with $p_T\sim 40\GeV$, whereas the total invariant mass of the $Wjj$ system is of order $Q\sim 300\GeV$.  The challenges in this type of measurement are clear: the recent D0 measurement~\cite{Abazov:2011af} finds no evidence for an excess. A precise theoretical understanding of multijet processes with multiple scales is clearly valuable.

Such configurations with multiple different kinematic scales present are not necessarily well described in fixed-order perturbation theory. The sensitivity to different scales can give rise to Sudakov double logarithms of the ratio of those scales, such as $\ln^2(m_{jj}/Q)$ or $\ln^2(p_T/m_{jj})$, at each order in perturbation theory. If the scales are widely separated, the logarithmic terms become the dominant contribution and the convergence of fixed-order perturbation theory deteriorates. A well-behaved perturbative expansion that allows for reliable predictions is obtained by performing a resummation of the logarithmic terms to all orders in perturbation theory. At present, the resummation of such kinematic Sudakov logarithms in exclusive multijet processes can only be carried out at the leading-logarithmic level using parton-shower Monte Carlos.

\begin{figure*}[ht!]
\subfigure[\hspace{1ex}All jets equally separated.]{%
\parbox{0.55\columnwidth}{\includegraphics[scale=0.4]{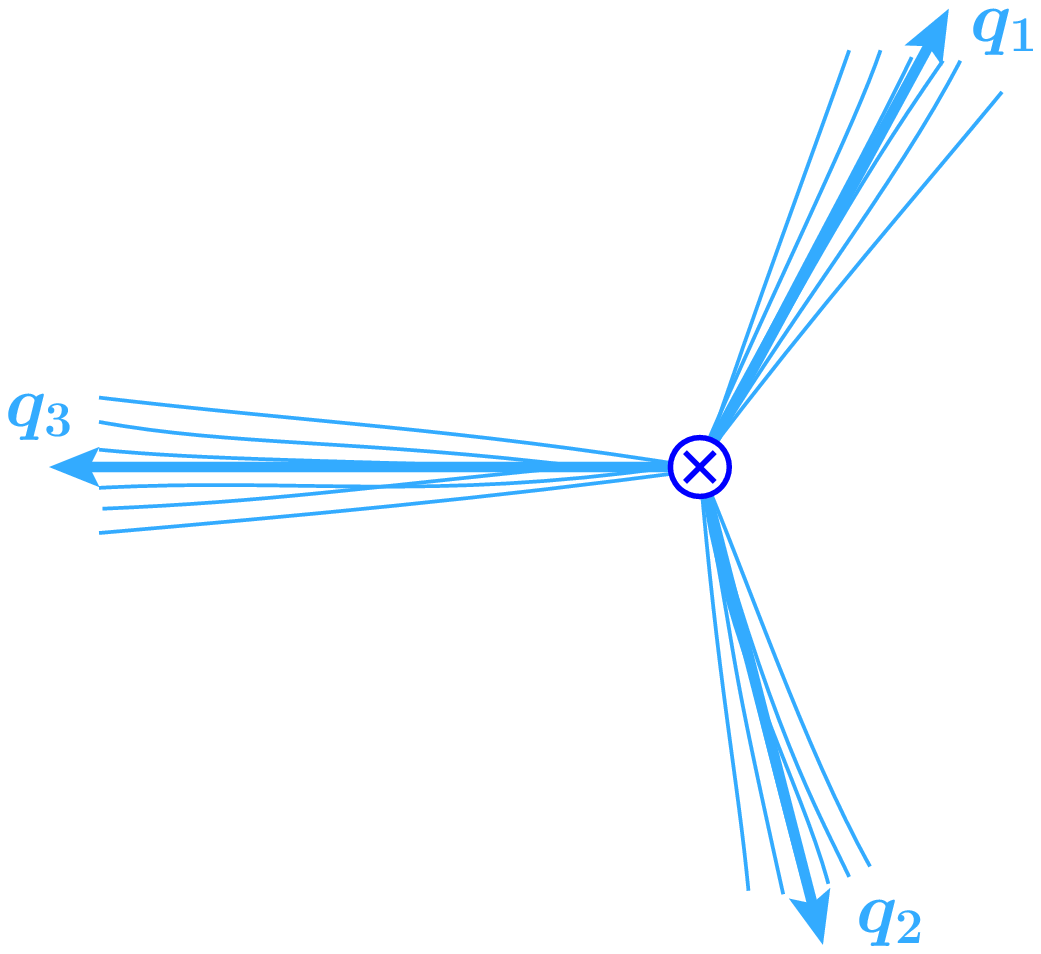}}%
{\LARGE$\Rightarrow$}%
\parbox{0.3\columnwidth}{\includegraphics[scale=0.5]{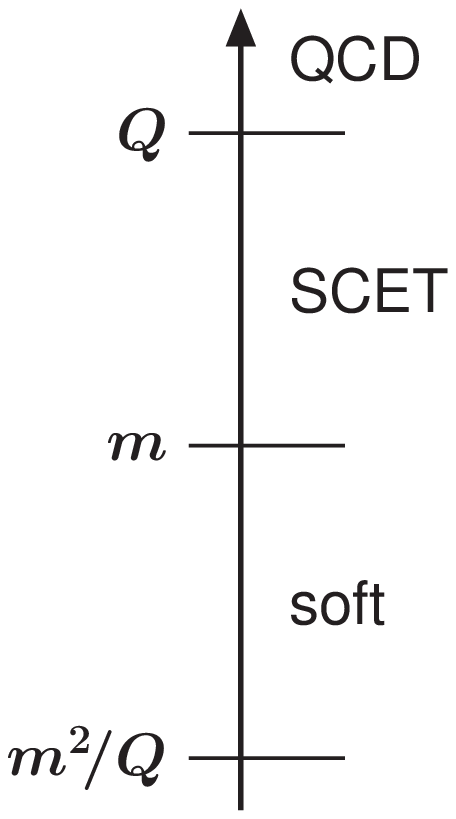}}%
\label{fig:SCET_3jets}}%
\hfill%
\subfigure[\hspace{1ex}Two jets close to each other.]{%
\parbox{0.63\columnwidth}{\includegraphics[scale=0.4]{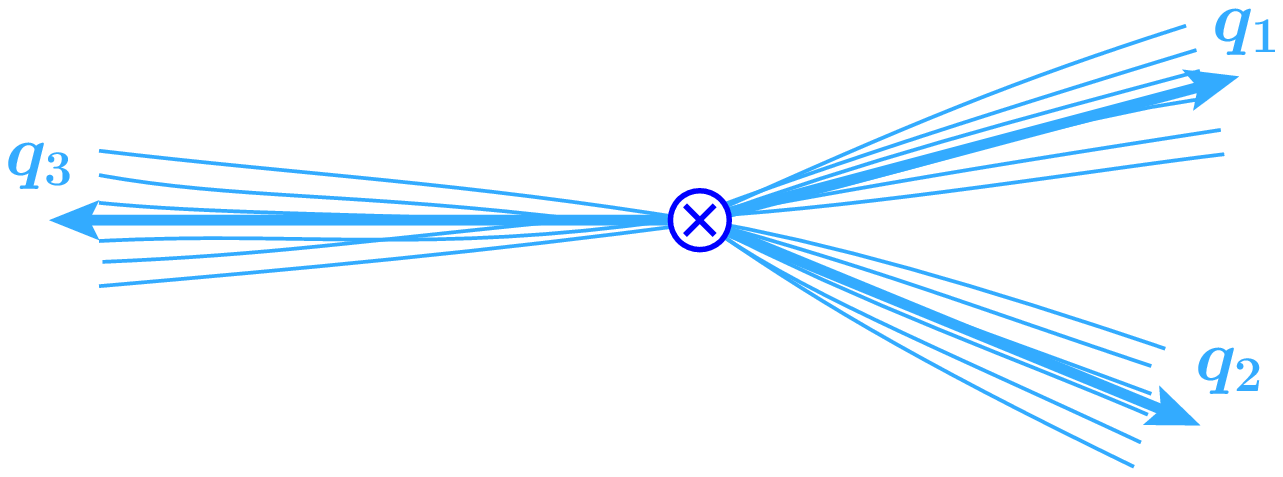}}%
{\LARGE$\Rightarrow$}%
\parbox{0.3\columnwidth}{\includegraphics[scale=0.5]{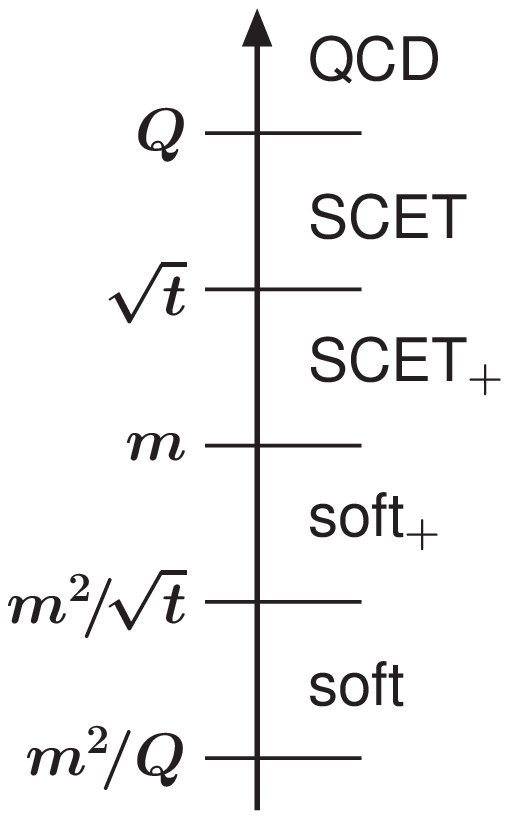}}%
\label{fig:SCET+_3jets}}%
\vspace{-0.5ex}
\caption{Different kinematic situations and relevant scales for the case of three jets with invariant mass $m$. On the left, the invariant masses between any two jets are comparable, $s_{ij} = 2q_i\cdot q_j \sim Q^2$, and the only relevant scales are $Q$, $m$, and $m^2/Q$. On the right, the dijet invariant mass between jets $1$ and $2$, $t = 2q_1\cdot q_2$, is parametrically smaller than that between any other pair of jets, so there are two more relevant scales, $\sqrt{t}$ and $m^2/\sqrt{t}$.}
\label{fig:3jets}
\end{figure*}

In this paper, we consider exclusive jet production in the kinematical situation where all jets have comparable energy $\sim Q$ and two of the jets become close%
\footnote{We fondly refer to this kinematic region as ``ninja'', since it corresponds to $n_i\cdot n_j \ll 1$, where $n_i$ and $n_j$ are the light-cone directions for nearby jets. We thank Iain Stewart for publicizing this codename so that this kinematic region has now become known as the ninja region.}, as illustrated in \fig{SCET+_3jets} for the case of three jets. We are interested in the dijet invariant mass $m_{jj}$ between the two close jets, which is much smaller than the other dijet invariant masses of order $Q$, but much larger than the invariant mass $m$ of the individual jets, i.e., there is a hierarchy of scales $m \ll m_{jj} \ll Q$. In this case, the cross section contains two types of logarithms, those related to the mass of the jets, $\ln^2(m/Q)$, as well as kinematic logarithms $\ln^2(m_{jj}/Q)$. For $m_{jj}\sim Q$, all jets are well separated, as in \fig{SCET_3jets}, and the jet-mass logarithms $\ln^2(m/Q)$ in the exclusive jet cross section can be resummed~\cite{Stewart:2009yx, Ellis:2009wj, Ellis:2010rwa, Stewart:2010tn, Jouttenus:2011wh} using soft-collinear effective theory (SCET)~\cite{Bauer:2000ew, Bauer:2000yr,Bauer:2001ct, Bauer:2001yt}.

In this paper, we construct a new effective theory, \SCETp, which is valid in the limit $m \ll m_{jj} \ll Q$. The added complication in this case arises from the fact that one needs to separate the soft radiation within a given jet from the radiation between the two close jets, giving rise to two different scales. In regular SCET, both of these processes are described by the same soft function, which therefore contains multiple scales. Soft functions with multiple scales have been observed in SCET before, and it has been suggested that this requires one to ``refactorize'' the soft function into more fundamental pieces depending on only a single scale. This was first pointed out in Ref.~\cite{Ellis:2010rwa}. Here we explicitly construct for the first time an effective theory that accomplishes a refactorization of the soft sector and separates different scales in a soft function. Using \SCETp, we derive the factorization of multijet processes in the limit $m \ll m_{jj} \ll Q$, where each function in the factorization theorem depends only on a single scale. The renormalization group evolution in \SCETp then allows us to sum all large logarithms arising from this scale hierarchy, including those in the soft sector.

It is worthwhile to note that the multijet events we consider in this paper are part of a broader class of kinematic configurations that give rise to multiple disparate scales. The case we address here of small dijet invariant masses belongs to the class of configurations for which the kinematics of the final-state jets introduces additional kinematic scales.  In our case this gives rise to large logarithms of ratios of dijet masses $\ln (m_{jj}/Q)$. Other configurations which give rise to large kinematic logarithms, such as those with a hierarchy of jet $p_T$s, may require a different effective-theory treatment, which we leave to future work. These kinematic logarithms are in contrast to so-called ``nonglobal'' observables~\cite{Dasgupta:2001sh},  which introduce additional scales by imposing parametrically different cuts in different phase space regions. This corresponds for example to a hierarchy between individual jet masses $m_i\ll m_j$, giving rise to logarithms of the form $\ln(m_i/m_j)$. The structure of such logarithms has been recently explored using \SCETI in Refs.~\cite{Kelley:2011ng, Hornig:2011iu}.

In the next section, we explain the physical picture of the effective-theory setup. In \sec{SCET+}, we discuss the construction of \SCETp, which requires a new mode with collinear-soft scaling to properly describe the soft radiation between the two close jets. As an explicit example of the application of \SCETp, we consider the simplest case of $e^+ e^-\!\to 3$ jets, for which in \sec{3jetfact} we derive the factorized cross section in the limit $m\ll m_{jj} \ll Q$, and in \sec{3jetNLO} we obtain all ingredients at next-to-leading order (NLO). In \sec{3jetNLO}, we also discuss the consistency of the factorized result in \SCETp, and show how the usual $3$-jet hard and soft functions in SCET are separately factorized into two pieces each. Readers not interested in the technical details of this example can skip over \secs{3jetfact}{3jetNLO}. In \sec{njets}, we generalize our results to the case of $pp\to N$ jets plus leptons. In \sec{Applications}, we present numerical results for the dijet invariant mass spectrum for $e^+e^-\!\to 3$ jets with all logarithms of $m/Q$ and $m_{jj}/Q$ resummed at next-to-leading logarithmic (NLL) order. We conclude in \sec{Conclusions}.

\clearpage
\section{Overview of the Effective Field Theory Setup}
\label{sec:PhysPicture}

Effective field theories provide a natural way to systematically resum large logarithms of ratios of scales appearing in perturbation theory. This is achieved by integrating out the relevant degrees of freedom at each scale. The renormalization group evolution within the effective theory is then used to sum the logarithms between the different scales. SCET provides the appropriate effective-theory framework to resum the logarithms arising from collinear and soft radiation in QCD.

For later convenience, for each jet we define a massless reference momentum
\begin{equation} \label{eq:qi_def}
q_i^\mu = E_i (1, \vec{n}_i) = \abs{\vec{q}_{iT}} (\cosh\eta_i, \vec{n}_{i\perp}, \sinh\eta_i)
\,,\end{equation}
where $E_i$ and $\vec{n}_i$ are the energy and direction of the $i$th jet, and $\vec{q}_{iT}$, $\eta_i$, and $\vec{n}_{i\perp} = \vec{q}_{iT}/\abs{\vec{q}_{iT}}$ are its transverse momentum, pseudorapidity, and transverse direction. The four-vector $(1,\vec{n}_i)$ is called $n_i$.

Given two light-cone vectors $n_i$ and $n_j$ with $n_i^2 = n_j^2 = 0$, we decompose any four-vector into light-cone components
\begin{equation}
p^\mu = n_j\cdot p\,\frac{n_i^\mu}{n_i\cdot n_j} + n_i\cdot p\,\frac{n_j^\mu}{n_i\cdot n_j} + p^\mu_{\perp_{ij}}
\,.\end{equation}
The subscripts on $\perp_{ij}$ specify with respect to which light-cone directions the perpendicular components are defined. To simplify the notation we will mostly drop them, unless there are potential ambiguities. To discuss the scaling of a four-vector with respect to the $n_i$ direction, we use
\begin{equation}
p^\mu = (n_i\cdot p, \bn_i\cdot p, p_{\perp_{i}}) \equiv (p^+, p^-, p_\perp)
\,,\end{equation}
where $\bn_i$ stands for any other light-cone vector in a direction that is considered parametrically different from $n_i$, i.e., parametrically $n_i\cdot\bn_i\sim 1$.  In terms of these light-cone coordinates we have $p^2 = p^+ p^- + p_{\perp}^2$.

\subsection{Equally Separated Jets}
\label{subsec:SCETpicture}

We first review the situation for equally separated jets, as depicted for three jets in \fig{SCET_3jets}, where we show the relevant energy scales. Formally, this case is defined by considering the pairwise invariant masses between any two jets to be parametrically of the same typical size $Q$. In addition, the invariant masses of all jets are parametrically of the same typical size $m$. Hence, we have the scaling
\begin{equation} \label{eq:scetscaling}
s_{ij} = 2 q_i\cdot q_j \sim Q^2
\,,\qquad
m_i^2  = P_i^2 \sim m^2
\,,\end{equation}
where $P_i$ denotes the total four-momentum of the jets. For $m$ much smaller than $Q$, we have the hierarchy of scales
\begin{equation} \label{eq:scaling}
\mu_S \simeq \frac{m^2}{Q} \ll \mu_J \simeq m \ll \mu_H \simeq Q
\,.\end{equation}

The cross section contains logarithms scaling like $\ln(m^2/Q^2)$. To resum these in the effective theory, one first matches QCD onto SCET at the hard scale $Q$; see \fig{SCET_3jets}. In this step, one integrates out all degrees of freedom with virtualities $\gtrsim Q^2$. The relevant modes in SCET (technically SCET$_{\rm I}$) below that scale are collinear and ultrasoft (usoft) modes. They have momentum scaling
\begin{equation}
p_\coll\sim Q(\lambda^2, 1, \lambda)
\,,\quad
p_\usoft \sim Q(\lambda^2, \lambda^2, \lambda^2)
\,,\quad
\lambda = \frac{m}{Q}
\,.\end{equation}
(The scaling for the collinear momentum $p_\coll$ here is understood to be defined with respect to its corresponding jet direction.) To see this, first note that to describe the collinear emissions within each jet we must have one set of collinear modes in each jet direction. Since the typical invariant mass of the jets is $m^2$, the collinear modes must have virtuality $\sim Q^2\lambda^2 = m^2$. Direct interactions between two collinear modes in different directions are not allowed, since they would produce modes with virtuality $\sim Q^2$, which have already been integrated out, i.e.,
\begin{equation}
Q(\lambda^2, 1, \lambda) + Q(1, \lambda^2, \lambda) \sim Q(1, 1, \lambda)
\,.\end{equation}
Hence, interactions between collinear modes can only happen via usoft modes, which can couple to any collinear mode without changing its virtuality:
\begin{equation}
Q(\lambda^2, 1, \lambda) + Q(\lambda^2, \lambda^2, \lambda^2) \sim Q(\lambda^2, 1, \lambda)
\,.\end{equation}
In the next step, one integrates out the collinear modes at the scale $m$, leaving only the soft theory with usoft modes of virtuality $\sim Q^2\lambda^4 = m^4/Q^2$.

In each step one performs a power expansion in the ratio of the lower scale divided by the higher scale. In the first step $m/Q = \lambda$, and in the second step $(m^2/Q)/m = \lambda$, so the expansion parameter is the same in both cases. By Lorentz invariance, in the end the expansion is actually in the ratio of the virtualities, i.e., in $\lambda^2$. There will be no power corrections of $\ord{\lambda}$.

Using this effective theory, one can derive a factorization theorem for the cross section for $pp \to N$ jets, which has the schematic form~\cite{Bauer:2002nz, Bauer:2008jx, Stewart:2009yx, Ellis:2010rwa, Stewart:2010tn}
\begin{equation} \label{eq:sigmaN}
\sigma_N
= \vec{C}_N(\mu)\, \vec{C}^{\dagger}_N(\mu) \!\times\!
\Bigl[ B_a(\mu)\, B_b(\mu) \prod_{i = 1}^N\! J_i(\mu) \Bigr] \otimes \widehat{S}_N(\mu)
\,.\end{equation}
The hard matching coefficient $\vec{C}_N$ describes the short-distance partonic $2\to N$ process. It arises from integrating out the hard modes at the scale $Q$ when matching full QCD to SCET. The beam functions $B_{a,b}$ and jet functions $J_i$ describe the collinear initial-state and final-state radiation, respectively, forming the jets around the ingoing and outgoing primary hard partons. They arise from integrating out the collinear modes at the intermediate jet scale $m$. The remaining matrix element in the soft theory yields the soft function $\widehat{S}_N$.  In general, $\vec{C}_N$ is a vector in color space and $\widehat{S}_N$ is a matrix, while the beam and jet functions are diagonal in color.

Each function in \eq{sigmaN} explicitly depends on the renormalization scale $\mu$. This dependence cancels in the product and convolutions of all functions on the right-hand side, since the cross section is $\mu$ independent. Since the different functions each only contain physics at a single energy scale, they can only contain logarithms of $\mu$ divided by that physical scale.
In general, the logarithms appearing in the hard, jet, and soft function are of the form
\begin{align} \label{eq:logs}
\vec{C}_N: \quad & \ln\frac{\mu^2}{s_{ij}}
\,,\nn\\
B_{a,b}, J_i: \quad &  \ln\frac{\mu^2}{m_i^2}
\,,\nn\\
\hS_N: \quad & \ln\frac{\mu^2s_{ij}}{m_i^4}
\,.\end{align}
(For the $N$-jet soft function this can be seen explicitly from the results in Ref.~\cite{Jouttenus:2011wh}.)
Hence, since all $s_{ij}\sim Q^2$ and all $m_i \sim m$ as in \eq{scetscaling}, there are no large logarithms when evaluating each function at its own natural scale,
\begin{equation}
\mu_H = Q
\,, \qquad
\mu_J = \mu_B = m
\,, \qquad \mu_S = \frac{m^2}{Q}
\,.\end{equation}
Using the renormalization group evolution in the effective theory, each function can then be evolved from its own natural scale to the common arbitrary scale $\mu$, which sums the logarithms in \eq{logs}. Combining all functions evolved to $\mu$ as in \eq{sigmaN} then sums all logarithms of the form $\ln(m^2/Q^2)$ in the cross section.

\subsection{Two Jets Close To Each Other}
\label{subsec:SCET+picture}

In the situation depicted in \fig{SCET+_3jets}, the invariant mass of two of the jets becomes parametrically smaller than all the other pairwise invariant masses between jets. In the following, we take the two jets that are close to each other to be jets 1 and 2 and use $t \equiv s_{12}$ to denote their invariant mass, and $s_{ij}$ to denote all other dijet invariant masses. We then have
\begin{equation} \label{eq:tscaling}
m^2 \ll t = s_{12} \ll s_{ij} \sim Q^2
\,.\end{equation}

In principle, the factorization theorem in \eq{sigmaN} can still be applied in this case, since the invariant masses of all jets are still much smaller than any of the dijet invariant masses. However, the hard matching coefficient $C_N$ now depends on two parametrically different hard scales, $\sqrt{t}$ and $Q$, and from \eq{logs} it contains corresponding logarithms $\ln(\mu^2/Q^2)$ and $\ln(\mu^2/t)$. This means there is no single hard scale $\mu_H$ that we can choose that would minimize all logarithms in the hard matching. In particular, choosing $\mu_H = Q$ as before, there are now unresummed large logarithms $\ln(t/Q^2)$ in the hard matching coefficient.

Similarly, the soft function now depends on two parametrically different soft scales, $m^2/\sqrt{t}$ and $m^2/Q$, containing logarithms $\ln(\mu^2 t/m^4)$ as well as $\ln(\mu^2 Q^2/m_i^4)$. Hence, there is not a single soft scale $\mu_S$ we can choose to minimize all logarithms in the soft function. Choosing $\mu_S = m^2/Q$ as before, there are still unresummed large logarithms $\ln(t/Q^2)$ in the soft function. In the soft function these naturally arise as $\ln n_i\mcdot n_j$.

To be able to resum the logarithms of $\ln(t/Q^2)$ we have to perform additional matching steps at each of the new intermediate scales $\sqrt{t}$ and $m^2/\sqrt{t}$, as shown in \fig{SCET+_3jets}. At the scale $Q$ we match QCD onto SCET as before, integrating out hard modes of virtuality $\gtrsim Q^2$. This effective theory has collinear modes with virtuality $t$ and corresponding soft modes,
\begin{equation}
p_\coll \sim Q(\lambda_t^2, 1, \lambda_t)
\,,\quad
p_\usoft \sim Q(\lambda_t^2, \lambda_t^2, \lambda_t^2)
\,,\quad
\lambda_t = \frac{\sqrt{t}}{Q}
\,.\end{equation}
There is one set of collinear modes for each of the jets, except for the two close jets 1 and 2. The latter are described at the hard scale $Q$ by a single set of collinear modes with virtuality $t$ in a common direction $n_t$. Since the total invariant mass between the two jets is $t$, such $n_t$-collinear modes can freely exchange momentum between the two close jets without changing their virtuality. This matching corresponds to performing an expansion in $\lambda_t$.

In the next step, at the scale $\sqrt{t}$, we match SCET onto a new effective theory \SCETp, integrating out all modes of virtuality $t$. Below this scale we now have separate collinear modes with virtuality $m^2$ for each jet, including jets 1 and 2,
\begin{equation}
p_\coll = Q(\lambda^2, 1, \lambda)
\,,\qquad
\lambda = \frac{m}{Q}
\,.\end{equation}
Note that for the well-separated jets, this matching at $\sqrt{t}$ will have no effect, since we do not perform a measurement in those directions that is sensitive to this scale. This means the virtuality of their collinear modes is simply lowered from $t$ to $m^2$. On the other hand, for jets 1 and 2 we match a single $n_t$ collinear sector in SCET onto two independent $n_1$ collinear and $n_2$ collinear sectors in \SCETp.

As before, the collinear modes cannot directly interact with each other. Interactions between the jets are possible via ultrasoft modes
\begin{equation}
p_{us} = Q(\lambda^2, \lambda^2, \lambda^2)
\,,\end{equation}
which have virtuality $Q^2\lambda^4 = (m^2/Q)^2$. In addition we can still have collinear modes in the $n_t$ direction which are soft enough to interact between jets 1 and 2 without changing the virtuality of the $n_1$ and $n_2$ collinear modes. These collinear-soft (csoft) modes have momentum scaling in the $\nt$ direction as
\begin{equation} \label{eq:csoft}
p_\csoft \sim Q(\lambda^2, \eta^2, \eta\lambda)
\,,\qquad
\eta = \frac{\lambda}{\lambda_t} = \frac{m}{\sqrt{t}}
\,.\end{equation}
We will see explicitly in \subsec{Consistency3jets} that these csoft modes are required to correctly reproduce the IR structure of QCD in this limit. Note also that the csoft modes are not allowed to couple to other collinear modes, since they would give them a virtuality $\sim Q^2\eta^2 = m^2/\lambda_t^2 \gg m^2$.

To derive the csoft scaling in \eq{csoft}, first note that in order for a soft gluon to separately interact with jet 1 or jet 2 it must be able to resolve the difference between the two directions $n_1$ and $n_2$. Hence, its angle with respect to the directions $n_1$ or $n_2$ has to be of order of the separation between these two directions, so it must have the scaling
\begin{equation}
p_\csoft \sim  Q_\csoft (\lambda_t^2, 1, \lambda_t)
\,.\end{equation}
At the same time, for the contribution of $p_\csoft$ to the invariant mass of either jet 1 or jet 2 (or equivalently the virtuality of a $n_1$ or $n_2$ collinear mode) to be $\sim  m^2$, its plus component must have the same size as a collinear plus component, i.e.,
\begin{equation}
Q_\csoft \lambda_t^2 \sim Q \lambda^2
\quad\Rightarrow\quad
Q_\csoft \sim Q \eta^2
\,,\end{equation}
which implies the scaling in \eq{csoft}. Hence, the $\nt$ csoft modes can be thought of as the soft remnant of the original $\nt$ collinear modes in the SCET above the scale $\sqrt{t}$.  Note that the csoft degrees of freedom can resolve the two directions $n_1$ and $n_2$ dynamically, such that only a single csoft mode is required that couples to both of these collinear directions. We will continue to use the direction $\nt$ to label this mode in the rest of the paper.

The virtuality of the csoft modes is
\begin{equation}
p_\csoft^2 \sim Q^2 \lambda^2 \eta^2 = \frac{m^4}{t}
\,,\end{equation}
which is the intermediate soft scale we already encountered. Hence, the main feature of \SCETp is that its soft sector contains two separate degrees of freedom, a csoft mode with virtuality $m^4/t$, and a usoft mode with virtuality $m^4/Q^2$. In the next step, we integrate out the collinear modes with virtuality $m^2$ at the scale $m$, leaving only the soft sector of \SCETp [denoted by soft$_+$ in \fig{SCET+_3jets}] consisting of csoft and usoft modes. Finally, at the csoft scale $m^2/\sqrt{t}$ we integrate out the csoft modes, which leaves only the usoft modes.

In \SCETp, the cross section for $pp\to N$ jets takes the schematic form
\begin{align}
\sigma_{N}
&= \abs{C_+(\mu)}^2\, \vec{C}_{N-1}(\mu)\,\vec{C}^{\dagger}_{N-1}(\mu)\,
\\ \nn &\quad
\times \Bigl[ B_a(\mu) B_b(\mu) \prod_{i = 1}^N J_i(\mu) \Bigr]
\otimes S_+(\mu) \otimes \hS_{N-1}(\mu)
\,.\end{align}
Here, each piece arises as the matching coefficient from one of the matching steps described above, and only depends on a single scale, allowing us to resum all large logarithms.
In effect, the additional matching at $\sqrt{t}$ factorizes the original hard matching coefficient $\vec{C}_N$ in \eq{sigmaN} into two pieces, $\vec{C}_{N-1}$ and $C_+$, each only depending on a single scale $Q$ and $\sqrt{t}$, which enables us to resum the logarithms $\ln(t/Q^2)$ present in the original hard matching. Similarly, the additional matching step at the scale $m^2/\sqrt{t}$ effectively factorizes the original multiscale soft function $\hS_N$ in \eq{sigmaN} into two separate pieces, $S_+$ and $\hS_{N-1}$, each only depending on a single soft scale, $m^2/\sqrt{t}$ and $m^2/Q$, respectively. This then enables us to sum all logarithms $\ln(t/Q^2)$ that were present in the original soft function.

Note that for the hard matching coefficient, the kinematic situation of $t\ll s_{ij}$ has been addressed before~\cite{Bauer:2006qp, Bauer:2006mk, Baumgart:2010qf} using a two-step matching procedure similar to our matching step at the scale $\sqrt{t}$. It was shown through an explicit one-loop calculation that this matching separates the two scales $Q^2$ and $t$ in the hard function from one another. In \app{RPIMatchingHard} we show that this holds to all orders in perturbation theory using reparametrization invariance  (RPI)~\cite{Manohar:2002fd} of the effective theory. In these previous analyses, however, the soft sector of the theory below $\sqrt{t}$ was not fully considered (or assumed to be that of standard SCET).
In this paper we give a complete description of the effective theory below the scale $\sqrt{t}$ including the appropriate soft sector, allowing us to accomplish a similar scale separation and resummation of the logarithms of $t/Q^2$ in the soft sector of the theory.  We stress that the unresummed logarithms in the original $\vec{C}_N$ and $\hS_N$ are equally large, so it is essential to be able to resum the logarithms in the soft sector.

\section{Construction of \SCETp}
\label{sec:SCET+}

\begin{table}[t!]
\begin{tabular}{c|cc}
\hline\hline
Mode & Scaling $(+,-,\perp)$ & Virtuality $p^2$ \\\hline
collinear & $Q\, (\lambda^2, 1, \lambda)$ & $(\lambda Q)^2 = m^2$ \\
csoft & $Q\,(\lambda^2, \eta^2, \eta\lambda)$ &  $(\eta\lambda Q)^2 = m^4/t$\\
usoft & $Q\,(\lambda^2, \lambda^2, \lambda^2)$ & $(\lambda^2 Q)^2 = m^4/Q^2$ \\\hline\hline
\end{tabular}
\caption{Modes in \SCETp.}
\label{table:modes}
\end{table}

In this section we construct \SCETp, an effective theory containing the usual collinear and usoft modes as well as the new intermediate csoft mode. The modes and their scaling are summarized in Table~\ref{table:modes}. We will show that the interactions between the three different types of modes can all be decoupled at the level of the Lagrangian by appropriate field redefinitions, similar to the BPS field redefinition to decouple collinear and usoft modes in regular SCET.

\subsection{Review of Standard SCET}
\label{subsec:SCET}

In SCET, the momentum of collinear particles in the $n$ direction are separated into a large label momentum $\lp$ and a small residual momentum $k$,
\begin{equation}
p^\mu = \lp^\mu + k^\mu
\,,\qquad
\lp^\mu = \bn\cdot\lp\, \frac{n^\mu}{2} + \lp_{\perp}^\mu
\,.\end{equation}
The momentum components scale as $\bn\cdot\lp\sim Q$, $\lp_\perp \sim Q\lambda$, and $k\sim Q\lambda^2$. The corresponding quark and gluon fields, $\xi_{n,\lp}(x)$ and $A_{n,\lp}(x)$, are multipole expanded with expansion parameter $\lambda$. They have fixed label momentum, and particles with different label momenta are described by different fields. Derivatives acting on the fields pick out the residual momentum dependence, $\img\partial^\mu \sim k^{\mu}$, while the large label momentum is obtained using the label momentum operator~\cite{Bauer:2001ct}
\begin{equation}
\Pop_n^\mu\, \xi_{n,\lp} = \lp^\mu\, \xi_{n,\lp}
\,.\end{equation}
When acting on several collinear fields, $\Pop_n^\mu$ returns the sum of the label momenta of all
$n$-collinear fields.

The interactions between collinear fields can only change the label momentum but not the collinear direction $n$, so it is convenient to define fields with only the direction $n$ fixed,
\begin{equation} \label{eq:collfields}
\xi_n(x) = \sum_{\lp\neq 0} \xi_{n,\lp}(x)
\,,\qquad
A_n(x) = \sum_{\lp\neq 0} A_{n,\lp}(x)
\,.\end{equation}
The sum over $\lp$ here excludes the zero-bin $\lp = 0$. This avoids double-counting the usoft modes, which are described by separate usoft quark and gluon fields. When calculating matrix elements, we implement this by summing over all $\lp$ and then subtracting the zero-bin contribution, which is obtained by taking the limit $\lp\to 0$~\cite{Manohar:2006nz}.

The Lagrangian for a collinear quark in the $n$ direction in SCET at leading order in $\lambda$ is well known and given by~\cite{Bauer:2000yr}
\begin{align} \label{eq:Ln}
\cL_n &= \bar{\xi}_n \Bigl[\img n\mcdot \Dc_n\! + g\, n\mcdot A_\usoft\! + \img\Dcslash_{n\perp} W_n \frac{1}{\bn\mcdot \Pop_n} W_n^\dagger \img\Dcslash_{n\perp}\! \Bigr]\frac{\bnslash}{2}\xi_n
,\end{align}
where the collinear covariant derivatives are
\begin{align}
\img n\cdot \Dc_n &= \img n\cdot\partial + g\,n\cdot A_n
\,,\nn\\
\img \Dc_{n\perp}^\mu &= \Pop^\mu_{n\perp} + g\, A^\mu_{n\perp}
\,.\end{align}
The Wilson line $W_n$ in \eq{Ln} is constructed out of $n$-collinear gluons. In momentum space, one has
\begin{equation} \label{eq:Wn}
W_n(x) = \biggl[\sum_\text{perms} \exp\Bigl(\frac{-g}{\bn\mcdot\Pop_n}\,\bn\mcdot A_n(x)\Bigr)\biggr]
\,,\end{equation}
where the label operator only acts inside the square brackets. $W_n$ sums up arbitrary emissions of $n$-collinear gluons from an $n$-collinear quark or gluon, which are $\ord{1}$ in the power counting.

The Lagrangian for usoft quarks and gluons is identical to the full QCD Lagrangian written in terms of usoft quark and gluon fields. It cannot contain any interactions with collinear modes, since the usoft fields do not have sufficient momentum to pair-produce collinear modes.

Because of the multipole expansion, at leading order in $\la$ the only coupling to usoft gluons in the collinear Lagrangian, \eq{Ln}, is through $n\cdot A_\usoft$. This coupling is removed by the BPS field redefinition~\cite{Bauer:2001yt},
\begin{align} \label{eq:BPS}
\xi^\zero_n(x) &= Y^\dagger_n(x)\,\xi_n(x)
\,,\nn\\
A^\zero_n(x) &= Y^\dagger_n(x)\,A_n(x)\, Y_n(x)
\,,\end{align}
where $Y_n$ is a usoft Wilson line in the direction $n$,
\begin{equation}
Y_n^\dagger(x) = \mathrm{P} \exp\biggl[\img g \int_0^{\infty} \!\df s \, n\mcdot A_\usoft(x + s\, n) \biggr]
\,,\end{equation}
and $\mathrm{P}$ denotes path ordering along the integration path. Since $W_n(x)$ is localized with respect to the residual position $x$, we have
\begin{align} \label{eq:Wnzero}
W_n^\zero(x) &= Y^\dagger_n(x)\, W_n(x)\, Y_n(x)
\nn\\
&= \biggl[\sum_\text{perms} \exp\Bigl(\frac{-g}{\bn\mcdot\Pop_n}\,\bn\mcdot A_n^\zero(x)\Bigr)\biggr]
\,.\end{align}
Therefore, using \eq{BPS} in \eq{Ln} together with
\begin{equation}
Y^\dagger_n(x) \bigl[\img n\cdot\partial + g\,n\cdot A_\usoft(x) \bigr] Y_n(x)
= \img n\cdot\partial
\,,\end{equation}
eliminates the dependence of $\cL_n$ on $n\mcdot A_\usoft$,
\begin{align} \label{eq:Lnzero}
\cL_n^\zero &= \bar{\xi}^\zero_n \Bigl[\img n\mcdot \Dc^\zero_n + \img\Dcslash^\zero_{n\perp} W_n^\zero \frac{1}{\bn\mcdot \Pop_n} W_n^{\zero\dagger} \img\Dcslash^\zero_{n\perp} \Bigr]\frac{\bnslash}{2}\, \xi^\zero_n
,\end{align}
where now
\begin{align} \label{eq:Dnzero}
\img n\cdot \Dc^\zero_n &= \img n\cdot\partial + g\,n\cdot A_n^\zero
\,,\nn\\
\img \Dc_{n\perp}^{\zero\mu} &= \Pop^\mu_{n\perp} + g\, A^{\zero\mu}_{n\perp}
\,.\end{align}
Hence, after the field redefinition there are no more interactions between usoft and collinear fields at leading order in the power counting, and the redefined fields no longer transform under usoft gauge transformations.

With more than one collinear sector, there are separate collinear Lagrangians for each sector, which decouple from each other and the usoft Lagrangian, $\cL_\usoft$. The total Lagrangian is then given by the sum
\begin{equation}
\cL_\mathrm{SCET} = \sum_i \cL_{n_i}^\zero + \cL_\usoft + \dotsb
\,,\end{equation}
where the ellipses denote the terms that are of higher order in the power counting.

\subsection{\SCETp}
\label{subsec:SCET+}

To construct \SCETp, we follow the same logic as in \subsec{SCET+picture}. To be concrete, we start from  SCET with two collinear sectors along $n_3$ and $\nt$ that have been decoupled from the usoft sector,
\begin{equation}
\cL_\mathrm{SCET} = \cL_{n_3}^\zero + \cL_{\nt}^\zero + \cL_\usoft + \dotsb
\,,\end{equation}
and where the scaling of the momenta is set by $\lambda_t = \sqrt{t}/Q$. (Additional collinear sectors for other well-separated jets are treated identically to $n_3$.) Since the three sectors are all decoupled in SCET, we can discuss them independently. When matching onto \SCETp, nothing happens for the $n_3$-collinear and usoft modes, whose momentum scaling is simply lowered from $\lambda_t$ to $\lambda = m/Q$.

On the other hand, as explained in \subsec{SCET+picture}, when we lower the scaling from $\lambda_t$ to $\lambda$, the $n_t$-collinear modes are separated into $n_1$-collinear and $n_2$-collinear modes, which cannot interact with each other any longer, plus a csoft mode in the $n_t$ direction,
\begin{equation}
\xi_\nt^\coll \to \xi_{n_1}^\coll + \xi_{n_2}^\coll + \xi_\nt^\csoft
\,,\quad
A_\nt^\coll \to A_{n_1}^\coll + A_{n_2}^\coll + A_\nt^\csoft
\,.\end{equation}
In \SCETp the labels $\nt$, $n_1$, $n_2$ uniquely specify whether we are dealing with csoft or collinear modes, so we will mostly drop the explicit labels ``$\csoft$'' and ``$\coll$''.

\subsubsection{Collinear-Soft Sector}

The csoft modes are just a softer version of the $\nt$ collinear modes. Hence, they still have a label direction $\nt$ and are multipole expanded, i.e.,\ their momentum is written in terms of a csoft label momentum and residual momentum,
\begin{equation}
p^\mu_\csoft = \lp_\csoft^\mu + k^\mu
\,,\qquad
\lp_\csoft^\mu = \bnt\cdot\lp_\csoft\, \frac{n_t^\mu}{2} + \lp_{\csoft\perp}^\mu
\,,\end{equation}
with momentum scaling
\begin{align}
\bnt\cdot \lp_\csoft &\sim Q \eta^2
\,,\qquad
\lp_{\csoft\perp} \sim (Q \eta^2) \lambda_t = Q\eta\lambda
\,,\nn\\
k^\mu &\sim (Q\eta^2) \lambda_t^2 = Q\lambda^2
\,,\end{align}
and an associated csoft label momentum operator,
\begin{equation}
\Pop_\nt^\mu\, \xi_{\nt,\lp} = \lp_\csoft^\mu\, \xi_{\nt,\lp}
\,.\end{equation}

The Lagrangians for csoft quarks and gluons are simply scaled down versions of the original $\nt$ collinear quark and gluon Lagrangians, $\cL_\nt^\zero$. For example, for the csoft quarks
\begin{align}
\cL_\nt^{\csoft\zero}
&= \bar{\xi}^\zero_\nt \Bigl[\img \nt \cdot \Dc^\zero_\nt
\nn\\ & \quad
+ \img\Dcslash^\zero_{\nt\perp} V_\nt^\zero\, \frac{1}{\bnt\mcdot \Pop_\nt}\, V_\nt^{\zero\dagger} \img\Dcslash^\zero_{\nt\perp} \Bigr]\frac{\bnslash_t}{2}\, \xi^\zero_\nt
\,,\end{align}
where the csoft covariant derivatives are defined exactly as in \eq{Dnzero} but in terms of csoft gluons.
Note that the csoft modes are decoupled from the usoft modes, as indicated by the superscript $\zero$, since they are obtained from the decoupled $\nt$-collinear modes of regular SCET. The $V_\nt^\zero$ Wilson lines are the csoft version of the $W_n^\zero$ in \eq{Wnzero},
\begin{equation} \label{eq:Vnt}
V_\nt^\zero(x) = \biggl[\sum_\text{perms} \exp\Bigl(\frac{-g}{\bn\mcdot\Pop_\nt}\,\bnt\mcdot A_\nt^\zero(x)\Bigr)\biggr]
\,.\end{equation}
Just like the $W_n$ in SCET, they sum up arbitrary emissions of $\nt$-csoft gluons from an $\nt$-csoft quark or gluon, which are $\ord{1}$ in the power counting, and are required to ensure csoft gauge invariance.

Similarly, the csoft gluon Lagrangian follows directly from the collinear gluon Lagrangian in SCET, and we just state the result here for completeness,
\begin{align}
\cL_\csoft^\zero &= \frac{1}{2 g^2} {\rm Tr} \Bigl\{ \Bigl[ \img \lbD_\nt^{\zero\mu}\!
+ g A^{\zero\mu}_\nt ,\, \img \lbD_\nt^{\zero\nu}\! + g A^{\zero\nu}_\nt \Bigr] \Bigr\}^2
\nn \\ & \quad
+ 2\, {\rm Tr} \Bigl\{ \bar{c}^\zero_\nt \Bigl[ \img \lbD_{\nt\mu}^\zero ,
\bigl[ \img \lbD_\nt^{\zero\mu}\! + g A^{\zero\mu}_\nt ,\, c^\zero_\nt \bigr] \Bigr] \Bigr\}
\nn \\ & \quad
+ \frac{1}{\alpha} {\rm Tr} \Bigl\{ \Bigl[\img \lbD_{\nt \, \mu}^\zero ,\, A^{\zero\mu}_\nt
\Bigr] \Bigr\}^2
\,.\end{align}
where $\lbD_\nt^{\zero \mu} = \bnt \cdot \Pop_\nt n_t^\mu/2 + \Pop^\mu_\perp + \img \nt \cdot \partial^\mu$, $c_{\nt}$ are ghost fields and $\alpha$ is a gauge fixing parameter.

As we will see below, csoft gluons can still couple to $n_1$ and $n_2$ collinear modes but only through their $n_1\mcdot A_\nt$ and $n_2\mcdot A_\nt$ components, respectively. From that point of view, they appear more like usoft gluons, hence the name csoft. However, their $n_1$ component is given by
\begin{align} \label{eq:n1Ant}
n_1 \mcdot A_\nt
&= \frac{n_1 \mcdot \nt}{2}\, \bnt\mcdot A_\nt + \frac{n_1 \mcdot \bnt}{2}\, \nt\mcdot A_\nt
+ (n_1)_{\perp_t} \mcdot A_{\nt\perp}
\nn\\
&\sim \quad \lambda_t^2 \times Q\eta^2 \quad\!+\,\,\quad 1 \times Q\lambda^2
\quad\! + \quad \lambda_t \times Q\eta\lambda
\nn\\
&\sim \qquad Q\lambda^2 \qquad\, + \qquad\! Q\lambda^2 \qquad +\qquad\! Q\lambda^2
\,,\end{align}
and similarly for $n_2$. The essential feature of the csoft scaling is that all three terms here contribute equally and must be kept. This is precisely the reason why despite being collinear modes the csoft modes are still able to couple to both $n_1$ and $n_2$ collinear modes. However, since the csoft modes are already multipole expanded with respect to the $\nt$ direction, $n_1\mcdot A_\nt$ is not an independent component with its own power counting, but is just a short-hand notation for the combination in \eq{n1Ant}. The same applies to $n_2\mcdot A_\nt$.

It is important to properly implement the usoft zero-bin subtraction for the csoft modes, which is necessary to avoid double counting the usoft region. The zero-bin limit of the csoft modes is defined by taking each light-cone component with respect to $\nt$ to have usoft scaling $\sim Q\lambda^2$. Hence, in the zero-bin limit, only the second term in \eq{n1Ant} contributes,
\begin{equation}\label{eq:csoftZeroBin}
n_1 \cdot A_\nt \,\xrightarrow{\text{zero-bin}}\, \nt\cdot A_\nt
\,,\end{equation}
while the other terms are suppressed by $\lambda_t$.

Note that csoft fields only couple to collinear fields whose direction are in the same csoft equivalence class as $n_t$, as discussed above. For all other collinear fields, the interaction with a csoft field would increase the virtuality of the field such that these interactions are integrated out of the theory.

\subsubsection{Collinear Sectors}

We now turn to the $n_1$ and $n_2$ collinear modes. To be specific we will use $n_1$; the discussion is identical for $n_2$. In the SCET above the scale $\sqrt{t}$, $n_1$ and $\nt$ belong to the same equivalence class.%
\footnote{This can be understood formally using RPI~\cite{Manohar:2002fd}, which is a symmetry of the effective theory that restores Lorentz invariance of the full theory that was broken by choosing a fixed direction $n_i^\mu$ for each collinear degree of freedom. One can show that $n_1$, $n_2$ and $n_t$ can all be obtained from one another by an RPI transformation, see Ref.~\cite{Marcantonini:2008qn} for a detailed discussion.}
This means the leading-order Lagrangian for $n_1$ collinear quarks directly follows from expanding \eq{Lnzero} in $\eta$,
\begin{align} \label{eq:Ln1}
\cL_{n_1}^\zero
&= \bar{\xi}^\zero_{n_1} \Bigl[\img n_1\mcdot \Dc^\zero_{n_1} + g\, n_1\mcdot A_\nt^\zero
\nn\\ & \quad
+ \img\Dcslash^\zero_{n_1\perp} W_{n_1}^\zero \frac{1}{\bn_1\mcdot \Pop_{n_1}} W_{n_1}^{\zero\dagger} \img\Dcslash^\zero_{n_1\perp} \Bigr]\frac{\bnslash_1}{2}\, \xi^\zero_{n_1}
\,,\end{align}
where the collinear covariant derivatives, $\Dc^\zero_{n_1}$, and Wilson line, $W_{n_1}^\zero$, are as defined in \eqs{Dnzero}{Wnzero} with $n = n_1$. As anticipated, the csoft modes couple to the $n_1$ collinear modes via $n_1\mcdot A_\nt^\zero \sim Q\lambda^2$. As in \eq{n1Ant}, all components of $A_\nt$ contribute equally to this coupling. However, below the scale $\sqrt{t}$, the $n_1$ collinear modes know nothing about the $\nt$ direction, so from their point of view the csoft modes behave just like ordinary soft modes with eikonal coupling in the $n_1$ direction. In particular, just as in standard SCET, we can remove the coupling between csoft and collinear modes from the Lagrangian by performing a field redefinition,
\begin{align} \label{eq:csoftBPS}
\xi^{(0,0)}_{n_1}(x) &= \X_{n_1}^{\zero\dagger}(x)\,\xi_{n_1}^\zero(x)
\,,\\\nn
A^{(0,0)}_{n_1}(x) &= \X_{n_1}^{\zero\dagger}(x)\,A_{n_1}^\zero(x) \X^\zero_{n_1}(x)
\,,\end{align}
where the superscript $(0,0)$ indicates that the collinear fields are decoupled from both usoft and csoft interactions. Here, $X_{n_1}^\zero$ is now a Wilson line in the $n_1$ direction built out of (usoft-decoupled) csoft gluons,
\begin{align}
X_{n_1}^{\zero\dagger}(x) &= \mathrm{P} \exp\biggl[\img g \int_0^{\infty}\!\df s \, n_1\mcdot A^\zero_\nt(x + s\, n_1) \biggr]
\,.\end{align}

After the csoft field redefinition for $n_1$ and $n_2$, there are no more interactions between any of the sectors. The above discussion is not affected by additional collinear sectors like $n_3$. The Lagrangian of \SCETp thus completely factorizes into independent collinear, csoft, and usoft sectors,
\begin{equation}
\cL_\text{\SCETp}
= \sum_{i=1,2} \cL_{n_i}^{(0,0)} + \cL_\nt^\zero + \sum_{i\geq 3} \cL_{n_i}^\zero + \cL_\usoft + \dotsb
\,.\end{equation}

\subsection{Operators in SCET and \SCETp}
\label{subsec:operators}

In this section we discuss how operators in \SCETp are constructed from gauge-invariant building blocks. As an explicit example, we use $e^+ e^- \to$ 3 jets with jets $1$ and $2$ getting close as in \fig{SCET+_3jets} since we will use it in \sec{3jetfact}. For simplicity, we assume here that jets 1, 2, and 3 are created by an outgoing quark, gluon, and antiquark, respectively, such that $n_1\equiv n_q$, $n_2\equiv n_g$, $n_3 \equiv n_\bq$. The operators with the quark and antiquark interchanged simply follow from Hermitian conjugation. Note that the case where the quark and antiquark jets get close to each other is power suppressed, so there is no corresponding operator in \SCETp at leading order in the power counting.

The allowed operators one can construct in SCET are constrained by local gauge invariance. It is well known that
using the collinear Wilson line $W_n(x)$ one can construct gauge-invariant collinear quark and gluon fields
\begin{align} \label{eq:chiB}
\chi_n(x) &= W_n^\dagger(x)\, \xi_n(x)
\,,\nn\\
\cB_{n\perp}^\mu(x)
&= \frac{1}{g}\Bigl[W_n^\dagger(x)\,\img \Dc_{n\perp}^\mu W_n(x)\Bigr]
\,,\end{align}
which are local with respect to soft interactions. Hence, we can use them to construct local collinear gauge-invariant operators in SCET.

For example, for widely separated jets as in \fig{SCET_3jets}, we match the matrix element for $e^+e^-\!\to 3$ jets in full QCD onto the operator
\begin{equation}
O_3 = \bar\chi_{n_1}\, \cB_{n_2}\, \chi_{n_3}
\,,\end{equation}
where for simplicity we neglect the Dirac structure. When matching QCD onto SCET in the situation with two close jets as in \fig{SCET+_3jets}, we first match onto the SCET operator for $e^+e^-\!\to 2$ jets,
\begin{equation} \label{eq:O2}
O_2 = \bar\chi_\nt \chi_{n_3}
\,,\end{equation}
describing a quark and antiquark jet in the $\nt$ and $n_3$ directions. Under local usoft gauge transformations, the fields in different collinear sectors all transform in the same way, so $O_2$ and $O_3$ are also explicitly gauge invariant under usoft gauge transformations.

After the field redefinition in \eq{BPS}, we obtain corresponding redefined fields $\chi_n^\zero(x)$ and $\cB_{n\perp}^{\zero\mu}(x)$ which are gauge invariant under both collinear and usoft gauge transformations. All usoft interactions are now described by usoft Wilson lines explicitly appearing in the operators, e.g.,
\begin{align} \label{eq:O2zero}
O_3 &= \bar\chi_{n_1}^\zero\, Y^\dagger_{n_1} Y_{n_2}\, \cB^\zero_{n_2\perp}\, Y_{n_2}^\dagger Y_{n_3}\,\chi^\zero_{n_3}
\,,\nn\\
O_2 &= \bar\chi^\zero_\nt\,  Y^\dagger_\nt Y_{n_3}\, \chi^\zero_{n_3}
\end{align}

In \SCETp we can use the same definitions as in \eq{chiB} to define collinear fields that are gauge invariant under collinear gauge transformations. The $n_{1,2}$ collinear fields in addition transform under csoft gauge transformations, $U_\nt(x)$,
\begin{align}
\chi_{n_1}(x) &\to U_\nt(x)\, \chi_{n_1}(x)
\,,\nn\\
\cB_{n_2\perp}(x) &\to U_\nt(x)\, \cB_{n_2\perp}(x)\, U^\dagger_\nt(x)
\,,\nn\\
V_\nt(x) &\to U_\nt(x)\, V_\nt(x)
\,.\end{align}
As discussed in~\subsec{SCET+}, only the $n_1$ and $n_2$ collinear fields couple to csoft gluons, thus the $n_3$ collinear fields do not transform under csoft gauge transformations. Therefore, to form gauge-invariant operators in \SCETp we have to include factors of the csoft $V_\nt$ Wilson lines. For example, after the usoft field redefinition in SCET, an $\nt$ collinear quark field in SCET is matched onto \SCETp as
\begin{equation} \label{eq:chint_matching}
\chi^\zero_\nt(x) \to V_\nt^{\zero\dagger}(x)\, \cB^\zero_{n_2\perp}(x)\, \chi^\zero_{n_1}(x)
\,.\end{equation}
Since $V^{\zero\dagger}_\nt$ does not transform under $n_{1,2}$ collinear gauge transformations, the right-hand side is invariant under both $n_{1,2}$ collinear and $n_\nt$ csoft gauge transformations. We can think of $V^{\zero\dagger}_\nt$ here effectively as arising from the csoft limit of the $W_\nt^{\zero\dagger}$ Wilson line inside $\chi^\zero_\nt$.

Hence, when matching SCET onto \SCETp for the situation in \fig{SCET+_3jets}, the SCET operator $O_2$ in \eq{O2zero} is matched onto the \SCETp operator
\begin{equation}
O_3^+ = \bigl[\bar\chi^\zero_{n_1}\, \cB^\zero_{n_2\perp} V_\nt^\zero\bigr] \bigl[Y^\dagger_{n_t}\, Y_{n_3} \bigr] \bigl[\chi^\zero_{n_3}\bigr]
\,,\end{equation}
where the different factors in square brackets do not interact with each other. Finally, we perform the csoft field redefinition in \eq{csoftBPS} to decouple the csoft fields from the $n_{1,2}$ collinear fields, which yields
\begin{align} \label{eq:O3plus}
O_3^+
&= \bigl[\bar\chi^{(0,0)}_{n_1}\bigr]_i \bigl[\cB^{(0,0)A}_{n_2\perp}\bigr]
\bigl[ \X_{n_1}^{\dagger\zero} \X_{n_2}^\zero T^A \X_{n_2}^{\zero\dagger} V_\nt^\zero\bigr]_{ij}
\nn\\ & \quad
\times  \bigl[Y^\dagger_{n_t} Y_{n_3}\bigr]_{jk} \bigl[\chi^\zero_{n_3} \bigr]_k
\,.\end{align}
Here, each factor in square brackets now belongs to a different sector in \SCETp, and we have shown how the adjoint and fundamental color indices are contracted. Since the different sectors are now completely factorized, we will drop the superscripts $(0)$ and $(0,0)$ in the following sections.

\subsection{Alternative Construction of \SCETp}
\label{subsec:altSCET+}

When constructing \SCETp in \subsec{SCET+} we started from the usoft-decoupled version of SCET, for which the csoft modes arise from the usoft-decoupled $\nt$ collinear sector in SCET. By simply lowering the scaling in the usoft sector from $\lambda_t$ to $\lambda$, we have implicitly used the fact that to be consistent and maintain the usoft decoupling one has to simultaneously lower the scaling of the usoft subtractions for the $\nt$-collinear sector. This is the reason why the csoft modes arise as the csoft limit of the $\nt$ collinear modes. The advantage of this approach is that the matching onto \SCETp really only happens within one collinear sector of SCET.

Alternatively, we can also be completely agnostic about the theory above the scale $\sqrt{t}$, and simply write down the Lagrangians for all the modes in \SCETp and use appropriate field redefinitions to decouple them. In the end, the matching calculation will ensure that \SCETp reproduces the correct UV physics, while having the right degrees of freedom ensures that the IR physics of the theory above is reproduced. The latter can be checked explicitly by testing whether the IR divergences in the theory above are reproduced in \SCETp.

This procedure should of course give the same final result. Since it is instructive to see how it does, we will briefly go through it here. The discussion for the $n_2$ collinear sector is again identical to that for $n_1$, so we will ignore it. We start by writing down the Lagrangians for the collinear and csoft modes,
\begin{align}
\cL_{n_1}
&= \bar \xi^c_{n_1}
\Bigl[ \img n_1 \mcdot \Dc^c_{n_1} + g\, n_1\mcdot A^{cs}_\nt + g\, n_1\mcdot A_\usoft + \dotsb \Bigr] \xi^c_{n_1}
\,,\nn\\
\cL_{n_3}
&= \bar \xi^c_{n_3}
\Bigl[ \img n_3 \mcdot \Dc^c_{n_3} + g\, n_3\mcdot A_\usoft + \dotsb \Bigr] \xi^c_{n_3}
\,,\nn\\
\cL_\nt
&= \bar \xi^{cs}_\nt
\Bigl[ \img \nt\mcdot \Dc^{cs}_\nt + g\, \nt\mcdot A_\usoft + \dotsb \Bigr] \xi^{cs}_\nt
\,,\end{align}
where the power counting in the multipole expansion restricts the possible interactions. Note that we have added collinear, $c$, and csoft, $cs$, labels here for clarity. For simplicity, we only write down the quark Lagrangians and drop the perpendicular pieces, indicated by the ellipses, which are not relevant for this discussion. Note that the csoft gluons, $A^{cs}_\nt$, only couple to the $n_1$ (and $n_2$) collinear modes, while the usoft gluons, $A_\usoft$, couple to all collinear sectors as well as the csoft sector.

We first perform the usual usoft field redefinition in \eq{BPS} on all three sectors,
\begin{align}
\xi_{n_i}^\zero(x) &= Y_{n_i}^\dagger(x)\,\xi_{n_i}(x)
\,,\nn\\
A_{n_i}^\zero(x) &= Y_{n_i}^\dagger(x)\, A_{n_i}(x)\, Y_{n_i}(x)
\,.\end{align}
As far as the coupling to usoft gluons is concerned, the csoft sector is just another collinear sector, so we get
\begin{align} \label{eq:Lnizero}
\cL_{n_1}^\zero
&= \bar \xi_{n_1}^{c\, \zero}
\Bigl[ \img n_1 \mcdot \Dc_{n_1}^{c \, \zero} + Y_{n_1}^\dagger Y_\nt g\, n_1\mcdot A_\nt^{cs \, \zero} Y_\nt^\dagger Y_{n_1}
\nn\\ & \qquad\qquad
+ \dotsb \Bigr] \xi^ {c \,\zero}_{n_1},
\nn\\
\cL_{n_3}^\zero
&= \bar \xi_{n_3}^{c\, \zero} \Bigl[ \img n_3 \mcdot \Dc_{n_3}^{c \, \zero} + \dotsb \Bigr] \xi_{n_3}^{c\, \zero}
\,,\nn\\
\cL_\nt^\zero
&= \bar \xi_\nt^{cs\, \zero}
\Bigl[ \img \nt\mcdot \Dc_\nt^{cs\, \zero} + \dotsb \Bigr] \xi_\nt^{cs\, \zero}
\,,\end{align}
where both the csoft and the $n_3$ collinear sectors are now decoupled from the usoft.

To decouple the $n_1$ collinear sectors, we have to eliminate the products of usoft Wilson lines in $\cL_{n_1}^\zero$. Using
\begin{align} \label{eq:n1INntBasis}
n_1 \cdot A_\usoft
&= \frac{n_1 \mcdot \nt}{2}\, \bnt\mcdot A_\usoft + \frac{n_1 \mcdot \bnt}{2}\, \nt\mcdot A_\usoft + (n_1)_{\perp_t} \mcdot A_\usoft
\nn\\
&=  \nt\mcdot A_\usoft\, \bigl[1 + \ord{\lambda_t}\bigr]
\,,\end{align}
it follows that
\begin{align}
Y_{n_1} (x)
& = \mathrm{P} \exp \biggl[\img g \!\int_{-\infty}^0\!\df s \, \nt \mcdot A_\usoft(x+s \, \nt) \biggr] + \ord{\lambda_t}
\nn\\
&= Y_\nt(x)+ \ord{\lambda_t}
\,,\end{align}
and therefore
\begin{equation} \label{eq:Yexp}
Y^\dagger_\nt Y_{n_1}  = 1 + \ord{\lambda_t}
\,.\end{equation}
Using \eq{Yexp} in \eq{Lnizero}, the $n_1$ collinear sector also decouples from the csoft one. We have now arrived at the same point as in \eq{Ln1}. The remaining coupling of the (usoft-decoupled) csoft gluons to the $n_1$ collinear sector via $n_1\cdot A_\nt^\zero$ is eliminated using the additional csoft field redefinition in \eq{csoftBPS}.

It is essential to perform the field redefinitions in this order. If we first perform a csoft field redefinition on $n_1$, we would get a term
\begin{equation}
\X_{n_1}^\dagger g\, n_1\mcdot A_\usoft \X_{n_1}
\end{equation}
in $\cL_{n_1}$, which cannot be eliminated anymore by a usoft field redefinition. The fact that we have to perform the usoft field redefinition first and that it requires us to expand in $\lambda_t$, shows that this step is really linked to the SCET above the scale $\sqrt{t}$, which has $\lambda_t$ as its expansion parameter.

\section{\boldmath Factorization for $e^+e^-\!\to 3$ Jets}
\label{sec:3jetfact}

\begin{figure}[t]
\centering
\includegraphics[scale=0.4]{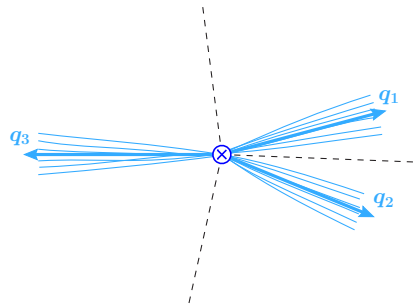}
\vspace{-0.5ex}
\caption{The kinematic configuration of $e^+e^-\! \to 3$ jets we consider. The boundaries of the jet regions determined by 3-jettiness are illustrated by the dashed lines. Note that the location of the boundaries depends only on the jet reference momenta $q_i$.}
\label{fig:3jetconfig}
\end{figure}

In the previous section, we constructed a new effective theory, \SCETp, which extends SCET with an additional mode that has csoft scaling.  As a concrete example, in this section we apply \SCETp to $3$-jet production in $e^+e^-$ collisions. We are interested in the kinematic configuration shown in \fig{3jetconfig}.  We use $N$-jettiness~\cite{Stewart:2010tn} with $N = 3$ to define the exclusive 3-jet final state, where the individual $3$-jettiness contributions $\tN_i$ of each jet determine the mass of the jets. We show how the factorization in \SCETp works and how the logarithms of the scales $Q$, $t$, and $\tN_i$ are simultaneously resummed.

We note that the applicability of \SCETp is not limited to the class of $N$-jettiness observables. However, $N$-jettiness provides a convenient observable well suited for factorization because it is linear in momentum, does not depend on additional parameters (such as a jet radius $R$), and covers all of phase space (i.e., there is no out-of-jet region).

\subsection{Definition of Observable and Power Counting}

\subsubsection{Observables}

In terms of the lightlike jet reference momenta $q_i^\mu$ in \eq{qi_def}, $N$-jettiness is defined as~\cite{Stewart:2010tn}
\begin{equation}\label{eq:TauN_def}
\Tau_N = \sum_k \min_i \{ 2 \hq_i\cdot p_k \}
\,,\qquad
\hq_i^\mu = \frac{q_i^\mu}{Q_i}
\,,\end{equation}
where for convenience we defined the dimensionless reference vectors $\hq_i$. $\Tau_N$ divides the phase space into $N$ jet (or beam) regions, where a particle with momentum $p_k$ is in jet region $i$ if $\hq_i \cdot p_k < \hq_j \cdot p_k$ for every $j\neq i$, i.e., the particle is closest to $\hq_i$. The boundaries between the jet regions are illustrated by the dashed lines in \fig{3jetconfig}. The $Q_i$ in \eq{TauN_def} are hard scales, such as the jet energies, $p_T$, or the total invariant mass. Different choices of $Q_i$ give different reference vectors $\hq_i$, which lead to different choices of the distance measure used in dividing up the phase space into jet regions. The distance between two different jets is measured by the dimensionless quantity
\begin{equation}
\hs_{ij} = 2 \hq_i \cdot \hq_j = \frac{s_{ij}}{Q_i Q_j}
\,.\end{equation}
For $Q_i = E_i$, we have a geometric measure with $\hq_i \equiv n_i$, and $\hs_{ij} = 2n_i\cdot n_j$ measures the angle between jets $i$ and $j$. For $Q_i = Q$, we have an invariant-mass measure and $\hs_{ij} = s_{ij}/Q^2$ is equivalent to the invariant mass between jets $i$ and $j$. By using $\hq_i$ and $\hs_{ij}$ we will keep our notation measure independent. (We will specify specific conditions on the used measure when necessary below.)

We can write $\Tau_N$ as
\begin{equation} \label{eq:TauNsum}
\Tau_N = \sum_i \tN_i
\,,\end{equation}
where $\tN_i$ is the contribution to $\tN_N$ from the $i$th jet region, which is given by
\begin{equation} \label{eq:TauNi_def}
\tN_i = 2\hq_i \cdot \sum_k p_k\, \Theta_i(p_k)
\,.\end{equation}
Here, the function
\begin{equation}\label{eq:thetaFull}
\Theta_i(p) = \prod_{j\neq i} \theta\bigl(\hq_j \cdot p - \hq_i \cdot p \bigr)
\end{equation}
imposes the phase space constraints for a particle with momentum $p$ to lie in jet region $i$. Note that this constraint only depends on the jet reference momenta (in addition to $p$ itself).

In the following, we will consider the cross section differential in each of the $\tN_i$, as well as the minimum dijet invariant mass, $t$, and the jet energy fraction, $z$, defined as
\begin{equation}
z = \frac{E_1}{E_1 + E_2}
\,,\end{equation}
where the observed jets are numbered such that $t \equiv s_{12}$ and $E_1<E_2$. Since experimentally we cannot determine the type of hard parton initiating a jet, we will sum over all relevant partonic channels in the end. For simplicity, we also integrate over the three angles which together with $t$ and $z$ describe the full $3$-body phase space of the three jets. (Two angles determine the overall orientation of the final state with respect to the beam axis. The third angle can be taken as the azimuthal angle of the two close jets.)

\subsubsection{Power Counting in SCET}

We consider the regime where all jets have similar energies, such that $Q_i \sim Q$ and $z \sim 1 - z$, and take the distance between jets $1$ and $2$ to be parametrically smaller than each of their distance to jet $3$, such that
\begin{align}
\hs_t &\equiv \hs_{12} \ll \hs_{13} \sim \hs_{23} \sim 1
\,,\nn\\
t &\equiv s_{12} \ll s_{13} \sim s_{23} \sim Q^2
\,,\end{align}
corresponding to \eq{tscaling}.

To define the power expansion in our two-step matching procedure we now have to specify some power-counting properties of the distance measure. In the following, we assume that we have chosen a measure such that the large components in $\hq_1$ and $\hq_2$ are equal up to power corrections, such that $\hs_{13} = \hs_{23} + \ord{\lambda_t}$. This is always the case for a geometric measure, where $\hs_{ij}$ are effectively angles.  For the invariant-mass measure, this is satisfied if the energies of jets $1$ and $2$ are equal up to power corrections. For measures where $\hs_{13}$ and $\hs_{23}$ differ by an amount of $\ord{1}$, the factorization still goes through but will have a somewhat different structure from what we will find below, and we leave the discussion of this case to future work.

The power expansion of the SCET above the scale $\sqrt{t}$ in terms of $\lambda_t$ is defined by choosing a common reference vector $\hq_t$ for jets $1$ and $2$ in the direction of $n_t$, such that
\begin{align} \label{eq:hqtscaling}
\hq_t &= \hq_1[1 + \ord{\lambda_t}] = \hq_2[1 + \ord{\lambda_t}]
\,,\nn\\
\hs_Q \equiv 2\hq_t\cdot \hq_3 &= \hs_{13} [1 + \ord{\lambda_t}] =\hs_{23} [1 + \ord{\lambda_t}]
\,,\nn\\
\hs_{1t} \equiv 2\hq_t\cdot \hq_1 &= \hs_t [1 + \ord{\lambda_t}]
\,,\nn\\
\hs_{2t} \equiv 2\hq_t\cdot \hq_2 &= \hs_t [1 + \ord{\lambda_t}]
\,.\end{align}
The choice of $\hq_t$ is constrained by label momentum conservation in SCET,
\begin{equation}
(Q, \vec{0}) = (Q_1 + Q_2)\, \hq_t^\mu + Q_3\, \hq_3^\mu
\,,\end{equation}
which upon squaring yields
\begin{equation}
\hs_Q = \frac{Q^2}{(Q_1 + Q_2) Q_3}
\,.\end{equation}
The dijet invariant masses $s_{13}$ and $s_{23}$ in the SCET above $\sqrt{t}$ are thus given by
\begin{equation} \label{eq:si3}
s_{13} = Q_1 Q_3\, \hs_Q
\,,\quad
s_{23} = Q_2 Q_3\, \hs_Q
\,,\end{equation}
which we can also write as
\begin{equation} \label{eq:si3x}
s_{13} = x Q^2
\,,\quad
s_{23} = (1-x) Q^2
\,,\quad
x = \frac{Q_1}{Q_1 + Q_2}
\,.\end{equation}
In particular, for the geometric measure with $Q_i = E_i$, we have $x = z$. Note that by counting all $Q_i \sim Q$ we in particular count $Q_{1,2}/Q_3 \sim 1$, which is necessary to have a consistent power expansion, such that
\begin{equation}
\frac{\hs_t}{\hs_Q} \sim \frac{s_{12}}{s_{13}} \sim \frac{s_{12}}{s_{23}} \sim \lambda_t^2
\end{equation}
are all counted in the same way.

\subsubsection{Power Counting in \SCETp}

To setup the power expansion in \SCETp in $\lambda$ (or equivalently $\eta$), we first note that the invariant mass of the $i$th jet is given by~\cite{Jouttenus:2011wh}
\begin{equation}
m_i^2 = P_i^2 = Q_i \tN_i\, \bigl[1 + \ord{\tN_i/Q_i}\bigr]
\,,\end{equation}
where $P_i = \sum_k p_k \Theta_i(p_k)$ and so the invariant mass is determined by $\tN_i$. Hence, the condition $m^2 \ll s_{ij}$ in \eq{tscaling}, which requires the jet size to be small compared to the jet separation, corresponds to $\tN_i\, Q_i \ll s_{ij}$. The power-counting parameters $\lambda^2 = m^2/Q^2$ and $\eta^2 = m^2/t$ are then determined by
\begin{equation}
\frac{\tN_i}{\hs_Q} \sim Q \lambda^2
\,,\qquad
\frac{\tN_i}{\hs_t} \sim Q \eta^2
\,,\qquad
\frac{\tN_i}{\sqrt{\hs_t}} \sim Q \eta\lambda
\,.\end{equation}
Note that to keep the power expansion in $\lambda_t \sim \lambda/\eta$ consistent, we still have to use the same vector $\hq_t$ (or $n_t$) as in SCET to define the csoft modes in \SCETp. This also applies when expanding the usoft measurement in $\lambda_t$ [see \eq{3texp} below].

All quantities related to the hard jet kinematics that enter in the final factorized cross section are uniquely determined in terms of the observables $t$ and $z$ by the label momentum conservation of the collinear fields in \SCETp,
\begin{equation} \label{eq:labelcons}
(Q, \vec{0}) = Q_1\, \hq_1^\mu + Q_2\, \hq_2^\mu + Q_3\, \hq_3^\mu
\,.\end{equation}
Recall that the large components of the collinear fields in \SCETp are determined up to $\ord{\lambda}$, which means we have to keep terms of $\ord{\lambda_t}$ in \eq{labelcons}. For example, for the geometric measure where we choose $Q_i = E_i$ so $\hq_i = n_i$, \eq{labelcons} leads to
\begin{align} \label{eq:Qi}
Q_1 &= z \frac{Q}{2} \Bigl(1 + \frac{t}{Q^2} \Bigr)
\,,\qquad
Q_2 = (1 - z) \frac{Q}{2} \Bigl(1 + \frac{t}{Q^2} \Bigr)
\,,\nn\\
Q_3 &= \frac{Q}{2} \Bigl(1 - \frac{t}{Q^2} \Bigr)
\,.\end{align}

\subsection{Factorization}
\label{subsec:fact}

The SCET factorization theorem for the cross section fully differential in the $\tN_i$ for equally separated jets was derived in Refs.~\cite{Stewart:2010tn, Jouttenus:2011wh}. The derivation in \SCETp follows the same logic, but we now have to take into account the presence of the new csoft modes.

We first separate $\tN_i$ for each $i$ into its contributions from the collinear, csoft, and usoft sectors,
\begin{equation} \label{eq:tNifact}
\tN_i = \tN_i^\coll + \tN_i^\csoft + \tN_i^\usoft
\,,\end{equation}
where the individual contribution from different sectors are defined by restricting the sum over particles $k$ in \eq{TauNi_def} to a given sector. We will now determine the resulting measurement and phase space constraints for each sector. They are most easily obtained by expanding the full-theory measurement in \eq{TauNi_def} using the appropriate momentum scaling of each mode.

For a collinear mode in sector $j$ with momentum $p^\coll_j$, the distance to jet $i$, $\hq_i\cdot p^\coll_j$, is by definition minimized for $i = j$, so
\begin{equation} \label{eq:Thetacoll}
\Theta_i^\coll (p^\coll_j) = \delta_{ij}
\,,\end{equation}
and therefore
\begin{equation} \label{eq:TauNcoll}
\tN_i^\coll = 2\hq_i\cdot \sum_{k\in \mathrm{i-coll}} p_k = 2\hq_i \cdot P_i^\coll = \frac{s_i}{Q_i}
\,,\end{equation}
where $P_i^\coll$ is the total momentum in the $i$-collinear sector, and (up to power corrections) $s_i$ is the
total invariant mass in the $i$-collinear sector. Note that there are no phase space constraints from the jet boundaries in the collinear sectors, which leads to inclusive jet functions, $J(s_i)$, in the factorization theorem.

\begin{figure*}[th!]
\subfigure[\hspace{1ex}$\tN_i$]{\parbox[c]{0.215\textwidth}{\includegraphics[angle=90,width=0.215\textwidth]{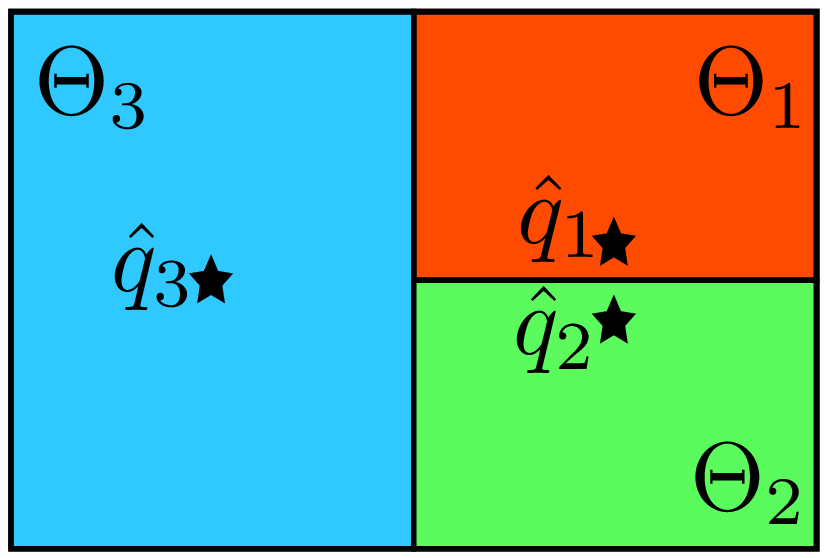}}}
\parbox[c]{4ex}{\boldmath\centering\LARGE$=$}
\subfigure[\hspace{1ex}$\tN_i^\usoft$]{\parbox[c]{0.215\textwidth}{\includegraphics[angle=90,width=0.215\textwidth]{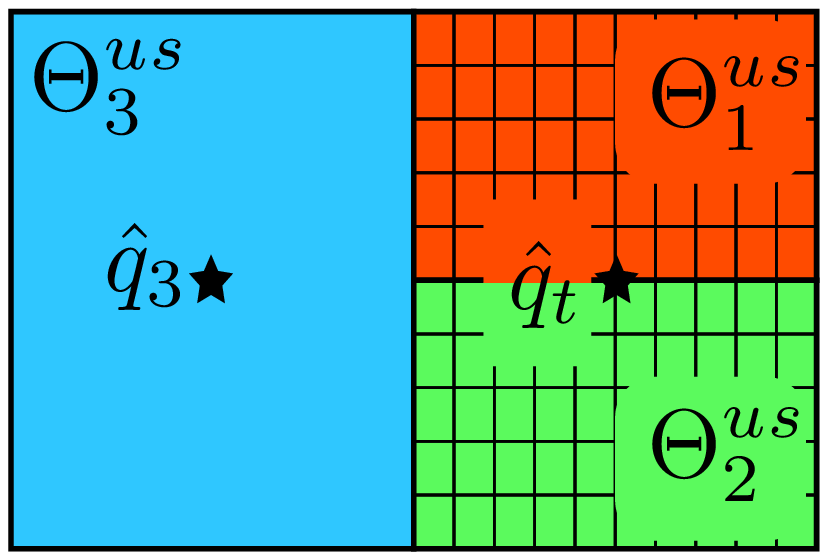}}}
\parbox[c]{4ex}{\boldmath\centering\LARGE$+$}
\subfigure[\hspace{1ex}$\tN_i^\csoft$]{\parbox[c]{0.215\textwidth}{\includegraphics[angle=90,width=0.215\textwidth]{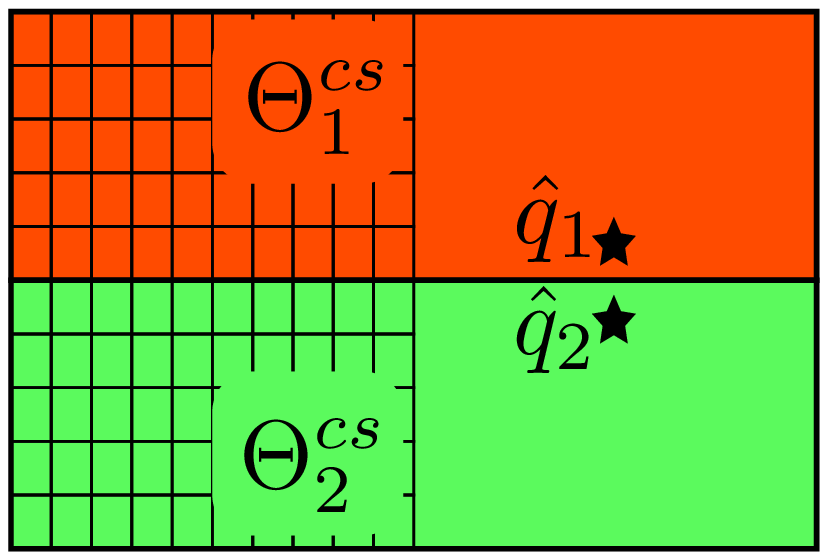}}}
\parbox[c]{4ex}{\boldmath\centering\LARGE$-$}
\subfigure[\hspace{1ex}$\tN_i^{\csoft\,0}$]{\parbox[c]{0.215\textwidth}{\includegraphics[angle=90,width=0.215\textwidth]{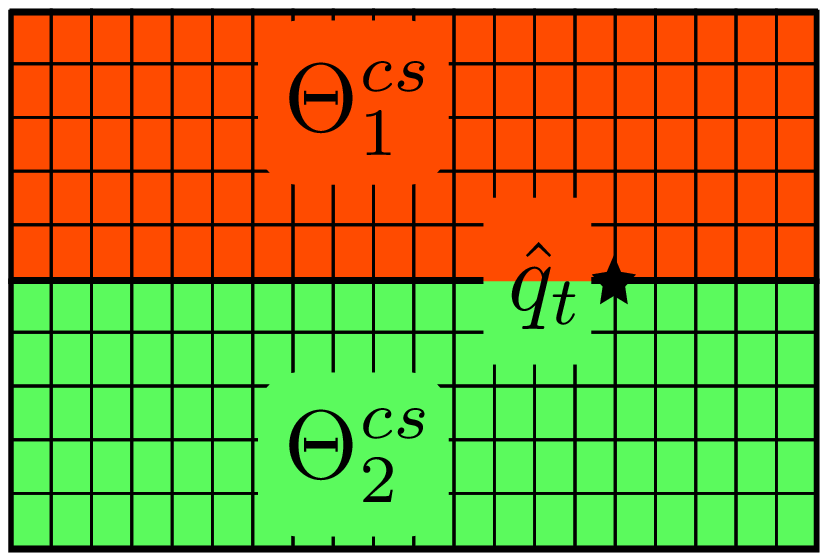}}}
\vspace{-0.5ex}
\caption{Graphical representation of the different measurement functions in the soft sectors in the $\theta$-$\phi$ plane for the geometric measure. The regions with different colors represent the phase space regions identified by the $\Theta_i(p)$, while the stars represent the directions of the dimensionless reference vectors $\hq_i$ used to calculate the observable. The full $3$-jettiness measurement is shown on the left. The hatching on the right indicates a region where a different reference vector than on the left is used to compute the $3$-jettiness observable. The contributions from the different hatched regions cancel on the right.}
\label{fig:measurements}
\end{figure*}

The division of the full measurement for the soft degrees of freedom between the csoft and usoft sectors is more complicated and is illustrated in \fig{measurements} in the $\theta$-$\phi$ plane.  In \fig{measurements}(a) we show the full measurement as determined by $\hq_{1,2,3}$ and $\Theta_{1,2,3}$. The three figures on the right show the various soft contributions which we will discuss next.

To determine $\tN_1^\usoft$, we write the $N$-jettiness measure for jet 1 for a usoft mode with momentum $p_\usoft$ in terms of the reference vectors $\hq_3$ and $\hq_t$,
\begin{align} \label{eq:3texp}
\hq_1\cdot p_\usoft &= \frac{\hs_{13}}{\hs_Q}\, \hq_t\cdot p_\usoft + \frac{\hs_{1t}}{\hs_Q}\, \hq_3\cdot p_\usoft + (\hq_1)_{\perp_{3t}} \cdot p_\usoft
\nn\\
&= \hq_t\cdot p_\usoft \bigl[1 + \ord{\lambda_t}\bigr]
\,,\end{align}
where we used the power counting in \eq{hqtscaling} and the fact that all components of $p_\usoft$ have a common scaling. The same is true for $\tN_2$ and $\hq_2$, which means we can replace $\hq_{1,2}$ by $\hq_t$ in the usoft contributions $\tN_i^\usoft$.

To determine the boundary between the jet regions 1 and 2 we have to compare $\hq_1\cdot p_\usoft$ with $\hq_2\cdot p_\usoft$. For this comparison the subleading terms in \eq{3texp} become relevant, so we have to be more careful. At the leading nontrivial order in the power counting this comparison only depends on the relative orientation of $\hq_1$ and $\hq_2$ in the transverse direction. To have a simple way of writing the constraint, we can choose $\hq_t$ such that $\hq_1$ and $\hq_2$ are back-to-back in the $\perp_t$ transverse plane and define the angle $\phi_{t}(p_\usoft)$ as the angle between $\hq_1$ and $p_\usoft$ in that plane. Then, $p_\usoft$ is in region 1 for $\cos\phi_{t}(p_\usoft) > 0$ and in region 2 for $\cos\phi_{t}(p_\usoft) < 0$. To summarize, we have
\begin{align} \label{eq:TauNusoft}
\tN_1^\usoft &= 2\hq_t\cdot \sum_{k\in\mathrm{usoft}} p_k\, \Theta_1^\usoft(p_k)
\,,\nn\\
\tN_2^\usoft &= 2\hq_t\cdot \sum_{k\in\mathrm{usoft}} p_k\, \Theta_2^\usoft(p_k)
\,,\nn\\
\tN_3^\usoft &= 2\hq_3\cdot \sum_{k\in\mathrm{usoft}} p_k\, \Theta_3^\usoft(p_k)
\,,\end{align}
where the boundaries are given by
\begin{align} \label{eq:Thetasoft}
\Theta_1^\usoft(p) &= \theta(\hq_3\cdot p - \hq_t\cdot p)\,\theta[\cos\phi_{t}(p)]
\,, \nn \\
\Theta_2^\usoft(p) &= \theta(\hq_3\cdot p - \hq_t\cdot p)\,\theta[-\cos\phi_{t}(p)]
\,, \nn \\
\Theta_3^\usoft(p) &=  \theta(\hq_t\cdot p - \hq_3\cdot p)
\,.\end{align}
These are illustrated in \fig{measurements}(b). The hatching in regions $1$ and $2$ denotes the fact that $\tN_{1,2}^\usoft$ are defined in terms of the common $\hq_t$ rather than their own $\hq_{1}$ or $\hq_{2}$.

Note that the standard $2$-jet soft function depends on only two variables, whereas ours depends on three. However, the only information about $\hq_{1,2}$ that is retained in the usoft measurement is their collective direction, given by $\hq_t$, and their relative orientation, given by $\phi_{t}$. In particular, the usoft measurement contains no information about the angle between $\hq_1$ and $\hq_2$, or equivalently $\hs_t$, at leading order in the power counting. This is in direct correspondence with the fact that the usoft modes only couple to the $n_1$ and $n_2$ collinear sectors through a common Wilson line in the $n_t$ direction. Physically, the usoft modes are not energetic enough to resolve the difference between the $n_1$ and $n_2$ directions. As a result, the usoft function will only be sensitive to the scale $\tN_i/\sqrt{\hs_{Q}} \sim Q\lambda^2$ but not $\tN_i/\sqrt{\hs_t} \sim Q\eta\lambda$, which is consistent with our expectations from the physical picture as discussed in \subsec{SCET+picture}.

The $n_t$ csoft modes are by definition collinear with jets 1 and 2, so as with collinear modes their scaling implies that they are always closest to either $\hq_1$ or $\hq_2$. Hence, only the boundary between jets 1 and 2 remains, so the csoft phase space constraints are
\begin{align}
\Theta_1^\csoft (p) &= \theta(\hq_2\cdot p - \hq_1\cdot p)
\,, \nn \\*
\Theta_2^\csoft (p) &= \theta(\hq_1\cdot p - \hq_2\cdot p)
\,, \nn \\*
\Theta^\csoft _3 (p) &= 0
\,,\end{align}
and the csoft contributions to $\tN_i$ are given by
\begin{align} \label{eq:TauNcsoft}
\tN_1^\csoft &=  2\hq_1 \cdot \sum_{k\in\mathrm{csoft}} p_k \,\Theta^\csoft_1(p_k)
\,,\nn\\
\tN_2^\csoft &=  2\hq_2 \cdot \sum_{k\in\mathrm{csoft}} p_k \,\Theta^\csoft_2(p_k)
\,,\nn\\
\tN_3^\csoft &= 0
\,.\end{align}
The csoft measurement is illustrated in \fig{measurements}(c). We now have only two different measurements, $\tN_1$ and $\tN_2$. In regions 1 and 2 they are computed with their proper reference vectors $\hq_{1}$ and $\hq_{2}$, reproducing the correct measurement in \fig{measurements}(a) for jets 1 and 2. At the same time, a different measurement is made in region 3, as indicated by the hatching. However, in region 3 the csoft modes are far away from $n_t$, and so can only have usoft scaling there. Hence, the zero-bin subtraction of the csoft modes, which removes the double-counting with the usoft modes, will remove this region of phase space.

Taking the usoft limit of \eq{TauNcsoft} using \eqs{3texp}{Thetasoft}, we obtain the csoft zero-bin contribution
\begin{align} \label{eq:TauNcsoft0}
\tN_1^{\csoft\,0} &=  2\hq_t \cdot \sum_{k\in\mathrm{csoft}\to \mathrm{usoft}} p_k \,\Theta^{\csoft\,0}_1(p_k)
\,,\nn\\
\tN_2^{\csoft\,0} &=  2\hq_t \cdot \sum_{k\in\mathrm{csoft}\to \mathrm{usoft}} p_k \,\Theta^{\csoft\,0}_2(p_k)
\,,\nn\\
\tN_3^{\csoft\,0} &= 0
\,,\end{align}
where the sum runs over all momenta in the csoft sector that actually have usoft scaling, and
\begin{align}
\Theta_1^{\csoft\,0} (p) &= \theta[\cos\phi_{t}(p)]
\,, \nn \\
\Theta_2^{\csoft\,0} (p) &= \theta[-\cos\phi_{t}(p)]
\,.\end{align}
The pictorial representation of this measurement is shown in \fig{measurements}(d). As for the naive csoft, there are only two different measurements, but as indicated by the hatching in all regions the measurement is now performed with a different reference vector than the one used in the full $3$-jettiness measurement. The complete csoft contribution is given by subtracting the zero-bin contributions in \eq{TauNcsoft0} from \eq{TauNcsoft}.

From \fig{measurements} one can see how the total soft measurement in the full theory is reproduced by the combination of the usoft and csoft measurements. The zero-bin csoft measurement cancels both the csoft measurement in region 3 made with a different reference vector than $\hq_3$ and the usoft measurements in regions 1 and 2 made with different reference vectors than $\hq_1$ and $\hq_2$. The remaining csoft contribution in regions 1 and 2 and usoft contribution in region 3 make up the correct measurement. To see this, consider the contribution of a generic soft gluon with momentum $p$ to $\tN_1$. Summing up all its contributions, we find
\begin{align}
&(\tN_1^\csoft - \tN_1^{\csoft\,0} + \tN_1^\usoft)(p)
\nn\\ & \quad
=  2\hq_1 \cdot p \,\,\theta(\hq_2\cdot p - \hq_1\cdot p)
- 2\hq_t \cdot p \, \theta[\cos\phi_{t}(p)]
\nn\\ & \qquad
+ 2\hq_t\cdot p\, \theta(\hq_3\cdot p - \hq_t\cdot p)\,\theta[\cos\phi_{t}(p)]
\nn\\ & \quad
=  2\hq_1 \cdot p \bigl[\Theta_1(p)
+ \theta(\hq_2\cdot p - \hq_1\cdot p)\, \theta(\hq_1\cdot p - \hq_3\cdot p) \bigr]
\nn\\ & \qquad
- 2\hq_t \cdot p \, \theta[\cos\phi_{t}(p)] \theta(\hq_t\cdot p-\hq_3\cdot p)
\nn\\ & \quad
= \tN_1(p) \bigl[1 + \ord{\lambda_t}\bigr] \,,
\end{align}
where $\Theta_1(p)$ is given in \eq{thetaFull}. A similar equation is obtained for  $\tN_2$. For $\tN_3$ we find
\begin{align}
&(\tN_3^\csoft - \tN_3^{\csoft\,0} + \tN_3^\usoft)(p)
\nn\\ & \quad
= 2\hq_3 \cdot p\, \Theta_3(p)
+ 2\hq_3\cdot p \bigr[\theta(\hq_t\cdot p - \hq_3\cdot p) - \Theta_3(p)\bigl]
\nn\\ & \quad
= \tN_3(p) \bigl[1 + \ord{\lambda_t}\bigr]
\,,\end{align}
We will see this cancellation again explicitly in our one-loop calculation below.

To formulate the measurement of $\tN_i$ at the operator level, we define momentum operators which pick out the total momentum of all particles in each region according to Eqs.~\eqref{eq:TauNi_def}, \eqref{eq:TauNcoll}, \eqref{eq:TauNusoft}, and \eqref{eq:TauNcsoft}:
\begin{align} \label{eq:hPidef}
\hP_i &\equiv \sum_k p_k \Theta_i(p_k)
\,,\qquad
\hP_i^\coll \equiv \sum_k p_k \Theta_i^\coll(p_k)
\,,\nn\\
\hP_i^\csoft &\equiv \sum_k p_k \Theta_i^\csoft(p_k)
\,,\qquad
\hP_i^\usoft \equiv \sum_k p_k \Theta_i^\usoft(p_k)
\,.\end{align}

The differential cross section in $\tN_1$, $\tN_2$, $\tN_3$ in \SCETp is obtained from the forward scattering matrix element of the operator $O_3^+$ in \eq{O3plus},
\begin{equation} \label{eq:sigmaO3+}
\mae{0}{O_3^{+\dagger}  \, \cM_3(\tN_1,\tN_2,\tN_3) \, O_3^+}{0}
\,,\end{equation}
with the $3$-jettiness measurement function
\begin{align} \label{eq:3jettinessMfull}
\mathcal{M}_{3}(\tN_1,\tN_2,\tN_3) =  \prod_i \delta\bigl(\tN_i - 2\hq_i\cdot \hP_i\bigr)
\,.\end{align}
Using $\tN_i = \tN_i^\coll + \tN_i^\csoft + \tN_i^\usoft$ from \eq{tNifact} together with Eqs.~\eqref{eq:TauNcoll}, \eqref{eq:TauNusoft}, \eqref{eq:TauNcsoft}, and the momentum operators in \eq{hPidef}, we can factorize the measurement function,
\begin{widetext}
\begin{align}
\label{eq:3jettinessMfactor}
\mathcal{M}_{3}(\tN_1,\tN_2,\tN_3) &=
\biggl[\prod_{i=1}^{3} \int\! \frac{\df s_i}{Q_i} \, \cM^\coll (s_i)\biggr]
\int \!\df \tN_1^\csoft \df \tN_2^\csoft \, \cM^\csoft(\tN_1^\csoft,\tN_2^\csoft)
\int \!\df \tN_1^\usoft \df \tN_2^\usoft \df \tN_3^\usoft \, \mathcal{M}_{3}^\usoft(\tN_1^\usoft,\tN_2^\usoft,\tN_3^\usoft)
\nn\\ & \quad \times
\prod_{i=1,2} \delta\Bigl(\tN_i - \frac{s_i}{Q_i} - \tN_i^\csoft - \tN_i^\usoft\Bigr) \, \delta\Bigl(\tN_3 - \frac{s_3}{Q_3} - \tN_3^\usoft \Bigr)
\,,\end{align}
where the collinear, csoft, and usoft measurement functions are
\begin{align} \label{eq:softmeas}
\mathcal{M}^c (s_i)
&= \delta(s_i - Q_i \, 2 \hq_i\cdot P_i^{c} )
\,,\qquad
\mathcal{M}^\csoft(\tN_1^\csoft,\tN_2^\csoft)
= \prod_{i=1,2} \delta\bigl(\tN_i^\csoft - 2\hq_i\cdot \hP_i^\csoft\bigr)
\,,\nn\\
\mathcal{M}_{3}^\usoft(\tN_1^\usoft,\tN_2^\usoft,\tN_3^\usoft)
&= \delta\bigl(\tN_3^\usoft - 2\hq_3\cdot \hP_3^\usoft\bigr) \prod_{i=1,2} \delta\bigl(\tN_i^\usoft - 2\hq_t\cdot \hP_i^\usoft\bigr)
\,.\end{align}
This factorization of the measurement function together with the factorization of the operator $O_3^+$ discussed in \subsec{operators} allows us to factorize \eq{sigmaO3+} into separate collinear, csoft, and usoft matrix elements. This is the cornerstone in obtaining the factorization theorem for the differential cross section. The derivation of the final factorization formula now only requires one to properly deal with the phase space sums over label and residual momentum and to provide an operator definition of all components in the factorization theorem. The required steps in \SCETp are straightforward and the same as in SCET, see Refs.~\cite{Bauer:2002nz, Fleming:2007qr, Bauer:2008dt, Bauer:2008jx, Stewart:2009yx}. The final factorized cross section, differential in the $\tN_i$, $t$, and $z$ is given by
\begin{align} \label{eq:factSCETp}
\frac{\df \sigma}{\df\tN_1\,\df\tN_2\,\df\tN_3\, \df t\, \df z}
&= \frac{\sigma_0}{Q^2}
\sum_{\kappa}
H_2(Q^2, \mu)\, H_+^\kappa(t,z, \mu) \prod_i \, \int \df s_i \,  J_{\kappa_i}(s_i, \mu)
\nn\\ &\quad \times
\int \df k_1 \df k_2 \, S_+^\kappa(k_1, k_2, \mu)\,
S_2\Bigl(\tN_1 - \frac{s_1}{Q_1} - k_1, \tN_2 - \frac{s_2}{Q_2} - k_2, \tN_3 - \frac{s_3}{Q_3}, \mu\Bigr)
\,.\end{align}
Here, $\sigma_0 = (4\pi \alpha_{\rm em}^2 /3Q^2) N_C \sum_q Q_q^2$ is the tree-level cross section for $e^+e^-\! \to \textrm{hadrons}$.

Since jets initiated by different types of partons are not distinguished experimentally, we sum over the relevant partonic channels to produce the observed jets, which are labeled such that the minimum dijet invariant mass $t$ is $s_{12}$ and $E_1<E_2$. The sum over partonic channels is denoted by the sum over $\kappa \equiv \{ \kappa_1,\kappa_2,\kappa_3\}$, which runs over the four partonic channels $\kappa = \{q,g,\bq\}$, $\{g,q,\bq \}$, $\{\bq,g,q \}$ and $\{g,\bq ,q \}$. For the first two channels, jets $1$ and $2$ effectively arise from a $q\to qg$ splitting, and for the last two from a $\bq \to \bq g$ splitting. For each splitting there are two channels, depending on whether the gluon or (anti)quark has the larger energy fraction. (The contribution where the quark and antiquark form the two jets with the smallest invariant mass does not enter in the sum because it is power suppressed.)

The hard function $H_2$ is the squared Wilson coefficient of $O_2$ from matching QCD onto SCET, and in our case is independent of $\kappa$. The hard function $H_+^\kappa$ is the squared Wilson coefficient of $O_3 ^+$ from matching SCET onto \SCETp. The $J_{\kappa_i}(s, \mu)$ are the standard inclusive jet functions in SCET and the soft functions $S_2$ and $S_+^\kappa$ denote the matrix elements of the usoft and csoft fields, respectively,
\begin{align} \label{eq:softdefs}
S_2(\tN^\usoft_1, \tN^\usoft_2, \tN^\usoft_3, \mu)
& =\frac{1}{N_C} \Mae {0}{\bar{T} \bigl[Y_{n_3}^{\dag} \, Y_{\nt}\bigr]_{ji}\, \cM_{3}^\usoft(\tN^\usoft_1, \tN^\usoft_2, \tN^\usoft_3)\, T \bigl[Y_{\nt}^{\dagger} Y_{n_3}\bigr]_{ij} }{0}
\,,\nn\\
S_+^{\{q,g,\bq\}}(\tN^\csoft_1,\tN^\csoft_2,\mu)
&= \frac{1}{N_C\,C_F} \Mae{0}{\bar{T} \bigl[V_{\nt}^{\dagger} \X_{n_g} T^{A} \X_{n_g}^\dagger \X_{n_q}]_{ji}\, \cM^\csoft(\tN^\csoft_1,\tN^\csoft_2)\, T\bigl[ \X_{n_q}^{\dagger} \X_{n_g} T^A \X_{n_g}^{\dagger} V_\nt\bigr]_{ij} }{0}
\,.\end{align}
\end{widetext}
The soft functions implicitly depend on the reference vectors $\hq_i$ through the combinations $\hs_Q$ and $\hs_t$, respectively, which is suppressed in our notation. The definition for $S_2$ is given for $\nt$ and $n_3$ corresponding to a quark and antiquark, respectively, but $S_2$ itself is independent of $\kappa$, i.e., it is the same for $q\leftrightarrow \bq$, which only switches $Y\leftrightarrow Y^\dagger$. The definition of $S_+$ is given for $\kappa = \{q,g,\bq\}$ for which $n_1 = n_q$, $n_2 = g$ and $n_t$ corresponds to a quark. The definitions for the other channels follow from the obvious interchanges of the appropriate Wilson lines.

In the next section we discuss all the ingredients in \eq{factSCETp} in detail, and obtain their explicit one-loop expressions. We also discuss the relation of the hard and soft functions in \eq{factSCETp} to the $3$-jet hard and soft functions in SCET, and derive the structure of the anomalous dimensions of $H_+$ and $S_+$ to all orders in perturbation theory. Readers not interested in these details can skip to \sec{njets} where we give the generalization of \eq{factSCETp} to $pp\to N$ jets or to \sec{Applications} where we present explicit numerical results for the dijet invariant-mass spectrum resulting from \eq{factSCETp} at NLL$'$.

\section{\boldmath Perturbative Results for $e^+e^-\!\to 3$ Jets}
\label{sec:3jetNLO}

To exhibit the color structure and be able to easily generalize our results in \sec{njets}, we will use the standard color-charge notation, where $\T_i^A$ denotes the color charge of the $i$th external parton when coupling to a gluon with color $A$. In general $\T_i\cdot \T_j \equiv \sum_A \T_i^A \T_j^A$ are matrices in the color space of the external partons. In particular
\begin{equation}
\T_i^2 = \id\, C_i
\quad\text{where}\quad
C_q = C_{\bar q} = C_F
\,,\quad
C_g = C_A
\,.\end{equation}
In the following, we will have three external partons, $q\bar qg$, for which the color space is still one-dimensional and the color matrices reduce to numbers,
\begin{align} \label{eq:3color}
\id &= 1
\,,\qquad
\T^2_q = \T^2_{\bar{q}} = C_F
\,,\qquad
\T^2_g = C_A
\,,\nn \\
\T_q \cdot \T_{\bar{q}} &=  \frac{C_A}{2} - C_F
\,,\qquad
\T_{q,\bq} \cdot \T_g = - \frac{C_A}{2}
\,.\end{align}

\subsection{Hard Functions}
\label{subsec:HardFuncs}

As discussed in \subsecs{SCET+picture}{operators}, for the jet configuration in \fig{3jetconfig} we are interested in, the matching onto the operator $O_3^+$ proceeds in two steps. This allows the dependence on the two parametrically different scales, $t = s_{12} \ll s_{13}\sim s_{23}\sim Q^2$ to be separated.

In the first step we match at the hard scale $Q$ from QCD onto SCET as shown in \fig{SCET+_3jets}. In our case we match the QCD current, $\bar{\psi} \gamma^\mu \psi$, onto the SCET two-jet operator, $O_2 = \bar{\chi}_{\nt} \gamma^\mu_\perp \chi_{n_\bq}$, by computing and comparing the $q\bar q$ matrix elements in both theories,
\begin{equation} \label{eq:H2match}
\cM_\mathrm{QCD}(0\to q\bar q)
= C_2(Q^2, \mu)\, \mae{q \bq}{O_2(\mu)}{0}
\,.\end{equation}
Here, $O_2(\mu)$ denotes the $\overline{\mathrm{MS}}$ renormalized operator. This matching is well-known (see, e.g., Ref.~\cite{Stewart:2009yx} for a detailed discussion) and was first performed at one loop in Refs.~\cite{Manohar:2003vb,Bauer:2003di}. The resulting matching coefficient is
\begin{align} \label{eq:C2oneloop}
C_2(Q^2, \mu) &= 1 + \frac{\alpha_s(\mu)\,C_F}{4\pi} \biggl[-\ln^2 \Bigl(\frac{-Q^2-\img 0}{\mu^2}\Bigr)
\nn\\ &\quad
+ 3 \ln \Bigl(\frac{-Q^2-\img 0}{\mu^2}\Bigr) - 8 + \frac{\pi^2}{6} \biggr]
\,,\end{align}
and satisfies the RGE
\begin{equation} \label{eq:C2RGE}
\mu \frac{\df}{\df\mu} C_2(Q^2, \mu) = \gamma_{C_2}(Q^2, \mu)\, C_2(Q^2, \mu)
\,.\end{equation}
The one-loop anomalous dimension is given by
\begin{align} \label{eq:gammaC2}
\gamma_{C_2}(Q^2, \mu)
&= \frac{\alpha_s(\mu) C_F}{4\pi}\biggl[ 4 \ln\frac{- Q^2 - \img 0}{\mu^2} - 6 \biggr]
\,.\end{align}
The hard function $H_2(Q^2, \mu)$ in \eq{factSCETp} and its anomalous dimension are given by
\begin{align} \label{eq:H2gamma2}
H_2(Q^2, \mu) &= \abs{C_2(Q^2, \mu)}^2
\,,\nn\\
\gamma_{H_2}(Q^2, \mu) &= 2\mathrm{Re}[\gamma_{C_2}(Q^2, \mu)]
\,.\end{align}

We then run down to the scale $\sqrt{t}$, and match from $O_2$ in SCET to the $O_3^+$ operator in \SCETp, as shown in \fig{SCET+_3jets}. In principle, this matching is computed in an analogous way by calculating the relevant $3$-parton matrix elements in both theories (suppressing any spin indices),
\begin{align}\label{eq:C3+match}
\mae{q\bq g}{O_2(\mu)}{0}
= C_+^\kappa(t, x,\mu)\, \mae{q\bq g}{O_3^+(\mu)}{0}
\,.\end{align}
The full one-loop calculation for the matrix element of $O_2$ is quite involved. However, we can extract the one-loop result for the hard function $H_+^\kappa$, given by
\begin{equation}
H_+^\kappa(t, x, \mu) = \abs{C_+^\kappa(t, x, \mu)}^2
\,,\end{equation}
using the known one-loop result for $e^+e^-\!\to 3$ jets from Ref.~\cite{Ellis:1980wv}. Since the operator matching \eq{H2match} is independent of the final state, it follows that%
\footnote{Since we are only interested in the cross section integrated over angles, we can consider the spin-averaged matrix element, which removes the dependence on the azimuthal angle in the $q\to qg$ splitting. Including this dependence requires to explicitly take into account the spin structure of $O_3^+$.}
\begin{align}\label{eq:qcd1loop}
&\abs{\cM_\mathrm{QCD}(0\to q\bar qg)}^2
\Big|_{s_{12}\ll s_{13} \sim s_{23}}
\nn\\ & \quad
=  H_2(Q^2,\mu)\, \abs{\mae{q\bq g}{O_2(\mu)}{0}}^2
\nn\\ & \quad
= H_2(Q^2,\mu)\, H_+^\kappa(t, x, \mu)\, \abs{\mae{q\bq g}{O_3^+(\mu)}{0}}^2
\,.\end{align}
In pure dimensional regularization, the virtual one-loop corrections to the bare $O_3^+$ matrix element are scaleless and vanish. Hence, the renormalized matrix element of $O_3^+(\mu)$ on the right-hand side is given by the tree-level result plus the counter-term contribution, which effectively supplies the proper $1/\epsilon$ divergences to cancel the IR divergences in the left-hand side matrix element. The remaining finite terms then determine the one-loop corrections to $H_+^\kappa(\mu)$.

For $\kappa = \{q, g, \bq\}$, we take $t = s_{qg}$, $s_{13} = s_{q\bq} $, $s_{23} = s_{\bq g}$. Expanding the one-loop virtual corrections for $\abs{\cM_\mathrm{QCD}(0\to q\bar qg)}^2$ from Ref.~\cite{Ellis:1980wv} in the limit $t \ll Q^2$ with $s_{q \bq}=x Q^2$ and $s_{\bq g}= (1-x)Q^2$ [see \eq{si3x}], and combining them with the one-loop corrections to $H_2(Q^2, \mu)$, we find
\pagebreak
\begin{widetext}
\begin{align} \label{eq:H3Plus}
H_+^{\{q,g,\bq\}} (t, x, \mu)
&=  Q^2\,\frac{\alpha_s(\mu) C_F}{2\pi}\, \frac{1}{t} \frac{1+x^2}{1-x}
\biggl\{ 1+\frac{\alpha_s(\mu)}{2 \pi} \biggl[
\Bigl(\frac{C_A}{2} - C_F \Bigr) \Bigl(2 \ln\frac{t}{\mu^2} \ln x + \ln^2 x + 2 {\rm Li}_2(1-x) \Bigr)
\nn\\ & \quad
- \frac{C_A}{2} \biggl( \ln^2\frac{t}{\mu^2} -\frac{7 \pi^2}{6} + 2\ln\frac{t}{\mu^2}\ln(1-x) + \ln^2(1-x) + 2 {\rm Li}_2(x)\biggr)
+ (C_A - C_F) \frac{1-x}{1+x^2}  \biggr] \biggr\}
\,.\end{align}
\end{widetext}
The overall factor of $Q^2$ here is included to make $H_+$ dimensionless. Note that at tree level $H_+$ takes the form of the common $q\to qg$ splitting function. As discussed below \eq{QCDC+}, beyond tree level it is related to universal splitting amplitudes (which are not the same as splitting functions). The results for the other partonic channels are given by
\begin{align}
H_+^{\{\bq,g,q\}} (t, x, \mu) &= H_+^{\{q,g,\bq\}} (t, x, \mu)
\,,\\
H_+^{\{g,q,\bq\}} (t, x, \mu) &= H_+^{\{g, \bq,q\}} (t, x, \mu) = H_+^{\{q,g,\bq\}} (t, 1 - x, \mu)
\nn
\,.\end{align}
They follow from the fact that $\abs{\cM_\mathrm{QCD}(0\to q\bar qg)}^2$ is symmetric under the interchange $q \leftrightarrow \bq$.

The hard function $H_+$ satisfies the RGE
\begin{equation}\label{eq:RGEgammaH3Plus}
\mu \frac{\df}{\df\mu} H_+^\kappa(t, x, \mu)
= \gamma_{H_+}^\kappa(t, x, \mu)\, H_+^\kappa(t, x, \mu)
\,.\end{equation}
At one loop we find for $\kappa = \{q,g,\bq \}$ using \eq{H3Plus}
\begin{align}
\gamma_{H_+}^{\{q,g,\bq\}}(t, x, \mu)
&=\frac{\alpha_s(\mu)}{2\pi} \biggl[2C_A \ln\frac{t}{\mu^2} +
4 \Bigl( C_F -\frac{C_A}{2} \Bigr) \ln x
\nn\\ & \quad
+ 2 C_A \ln(1-x)- \beta_0\biggr]
\,,\end{align}
where $\beta_0 = (11C_A - 4 T_F n_f)/3$. For general $\kappa$, the anomalous dimension can be written as
\begin{align} \label{eq:gammaH3Plus}
\gamma_{H_+}^\kappa(t, x, \mu)
&=-\frac{\alpha_s(\mu)}{2\pi} \biggl[4\,\T_1\cdot \T_2 \ln\frac{t}{\mu^2}
+4\, \T_1 \mcdot \T_3 \ln x
\nn\\ & \quad
+ 4\, \T_2 \mcdot \T_3  \ln (1-x) + \beta_0\biggr]
\,.\end{align}
Its all-order structure is derived from consistency in \subsec{Consistency3jets}.

\subsection{Jet Functions}
\label{subsec:JetFuncs}

The jet functions are given by the matrix elements of collinear fields, and are the standard inclusive jet functions as in many other SCET applications. We give the one-loop renormalized jet function in $\overline{\textrm{MS}}$ for completeness~\cite{Lunghi:2002ju,Bauer:2003pi,Fleming:2003gt,Becher:2009th}
\begin{align}
\label{eq:jetfuncs}
J_q(s, \mu)
&= \delta(s) + \frac{\alpha_s(\mu)C_F}{4\pi} \biggl[ \frac{4}{\mu^2}\cL_1\Bigl(\frac{s}{\mu^2} \Bigr)
- \frac{3}{\mu^2} \cL_0\Bigl(\frac{s}{\mu^2} \Bigr)
\nn \\ & \quad
+ (7 - \pi^2) \delta(s) \biggr]
\,, \nn \\
J_g(s, \mu) &= \delta(s) + \frac{\alpha_s(\mu)}{4\pi} \biggl\{C_A \frac{4}{\mu^2}\cL_1 \Bigl(\frac{s}{\mu^2} \Bigr)
- \beta_0\frac{1}{\mu^2} \cL_0\Bigl(\frac{s}{\mu^2} \Bigr)
\nn \\* & \quad
+ \biggl[\Bigl(\frac{4}{3} - \pi^2\Bigr)C_A + \frac{5}{3}\beta_0 \biggr] \delta(s) \biggr\}
\,,\end{align}
where $\beta_0 = (11C_A - 4 T_F n_f)/3$ and $\cL_n(x)$ denotes the standard plus distribution,
\begin{equation}
\cL_n(x) = \biggl[\frac{\theta(x)\ln^n\!x}{x}\biggr]_+
\,.\end{equation}
The jet functions satisfy the RGE
\begin{align} \label{eq:J_RGE}
\mu\,\frac{\df}{\df\mu} J_i(s, \mu)
&= \int\!\df s'\, \gamma^i_J(s - s', \mu)\, J_i(s', \mu)
\,,\end{align}
with the anomalous dimensions
\begin{align} \label{eq:gammaJ}
\gamma_J^i(s, \mu)
&= -2\, C_i\, \Gamma_\cusp[\alpha_s(\mu)] \,\frac{1}{\mu^2}\cL_0\Bigl(\frac{s}{\mu^2}\Bigr)
\nn\\ &\quad
+ \gamma_J^i[\alpha_s(\mu)]\,\delta(s)
\,,\end{align}
where $\Gamma_\cusp[\alpha_s]$ is the universal cusp anomalous dimension~\cite{Korchemsky:1987wg} given in \eq{Gammacusp}, and the noncusp terms are given in \eq{gammaJi}.

\subsection{Soft Functions}
\label{subsec:SoftFuncs}

The usoft and csoft functions describe the contributions to the observable from particles softer than the jet energies.  Unlike collinear modes which contribute to only a single jet, the soft modes can contribute to all jets.  This means that these modes are sensitive to the invariant masses $s_{ij}$ between jets.  The csoft modes, while having smaller energy than the collinear modes, have collinear scaling and are needed to describe the soft interactions between the nearby pair of jets, because the usoft modes have too small energy to resolve these two jets. The results for the soft functions can be written for general $\kappa$, i.e., without having to specify a particular channel, since the dependence on the parton species solely arises through the SU(3) color representations of the Wilson lines.

\subsubsection{The Ultra-Soft Function $S_2$}

The operator definition of the usoft function $S_2$ is given in \eq{softdefs}, with the measurement function given in \eq{softmeas}. At one loop, the relevant integral we have to\pagebreak\ compute is
\begin{align}
&- 2\,(\T_1 + \T_2) \cdot \T_3 \Bigl(\frac{e^{\gamma_E}\mu^2}{4\pi}\Bigr)^\epsilon g^2\int
\frac{\df^d p}{(2\pi)^d}\, 2\pi\delta(p^2)\theta(p^0)
\nn\\* & \quad\times
\frac{n_3 \cdot n_t}{(n_3\cdot p)(n_t\cdot p)}\, \mathcal{M}_3^{\usoft}(k_1, k_2, k_3)
\,.\end{align}
For the color factor we have used that with respect to the external $3$-parton color space, the total color charge carried by the $\nt$ Wilson line is $\T_1 + \T_2$, i.e.\ the combined color of partons $1$ and $2$.

The usoft region for jets $1$ and $2$ is determined by \eq{Thetasoft}, where the boundary between jet $3$ and jets $1$ and $2$ depends only on $\hat{q}_t$ and $\hat{q}_3$. The division of the combined $\nt$ region between jets $1$ and $2$ is given by the additional $\theta[\pm \cos \phi_t(p)]$, whose only effect is to divide the azimuthal integral in half. In $d=4-2\e$ dimensions one gets
\begin{align}
\int_0^{\pi/2}\! \df \phi_t\, \sin^{-2\e} \phi_t
&= \int_{\pi/2}^\pi\! \df \phi_t\, \sin^{-2\e} \phi_t
\nn\\
&= \frac{1}{2} \int_0^\pi\! \df \phi_t\, \sin^{-2\e} \phi_t
\,.\end{align}
Hence, the $\nt$ hemisphere contribution is split in half between jets $1$ and $2$.

The final result for the renormalized usoft function at NLO is
\begin{widetext}
\begin{align}\label{eq:S2NLO}
S_2(k_1, k_2, k_3, \mu)
&= \delta(k_1)\, \delta(k_2)\, \delta(k_3)
+ \frac{\alpha_s(\mu)}{4\pi}\, (\T_1+\T_2)\cdot \T_3
\biggl\{ \frac{8}{\sqrt{\hs_Q}\mu} \biggl[
\frac{1}{2} \cL_1\Bigl(\frac{k_1}{\sqrt{\hs_Q}\mu}\Bigr) \delta(k_2)\, \delta(k_3)
\nn \\ & \quad
+ \frac{1}{2}\cL_1\Bigl(\frac{k_2}{\sqrt{\hs_Q}\mu}\Bigr) \delta(k_1)\, \delta(k_3)
+ \cL_1\Bigl(\frac{k_3}{\sqrt{\hs_Q}\mu}\Bigr) \delta(k_1)\, \delta(k_2) \biggr]
- \frac{\pi^2}{3}\delta(k_1)\, \delta(k_2)\, \delta(k_3) \biggr\}
\,,\end{align}
which is simply the sum of two hemisphere contributions. We can see explicitly that $S_2$ depends only on the scale $k_i/\sqrt{\hs_Q}$. It satisfies the RGE
\begin{equation} \label{eq:S2RGE}
\mu\frac{\df}{\df\mu}\, S_2(k_1, k_2, k_3, \mu)
= \int\!\df k_1'\,\df k_2'\,\df k_3'\,\gamma_{S_2}(k_1 - k_1',k_2 - k_2', k_3 - k_3', \mu)
S_2(k_1', k_2', k_3', \mu)
\,,\end{equation}
where the anomalous dimension at one loop is given by
\begin{align}\label{eq:gammaS2NLO}
\gamma_{S_2} (k_1, k_2, k_3, \mu)
&= -\frac{\alpha_s(\mu)}{4\pi}\,(\T_1 + \T_2)\cdot\T_3
\frac{4}{\sqrt{\hs_Q}\mu} \biggl\{
\cL_0\Bigl(\frac{k_1}{\sqrt{\hs_Q}\mu}\Bigr) \delta(k_2)\delta(k_3)
+ \cL_0\Bigl(\frac{k_2}{\sqrt{\hs_Q}\mu}\Bigr) \delta(k_1)\delta(k_3)
\nn\\* & \quad
+ 2\cL_0\Bigl(\frac{k_3}{\sqrt{\hs_Q}\mu}\Bigr) \delta(k_1)\delta(k_2) \biggr\}
\,.\end{align}

\subsubsection{The Collinear-Soft Function $S_+$}

\begin{figure*}[t!]
\subfigure[]{\includegraphics[scale=0.63]{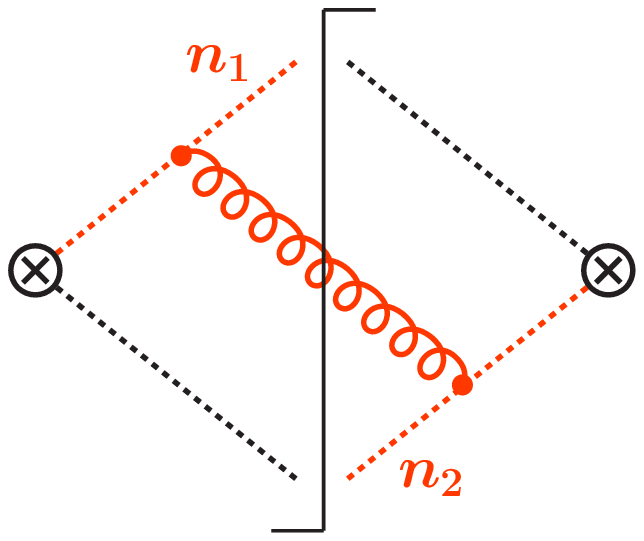}\label{fig:Sp12}}%
\hfill%
\subfigure[]{\includegraphics[scale=0.63]{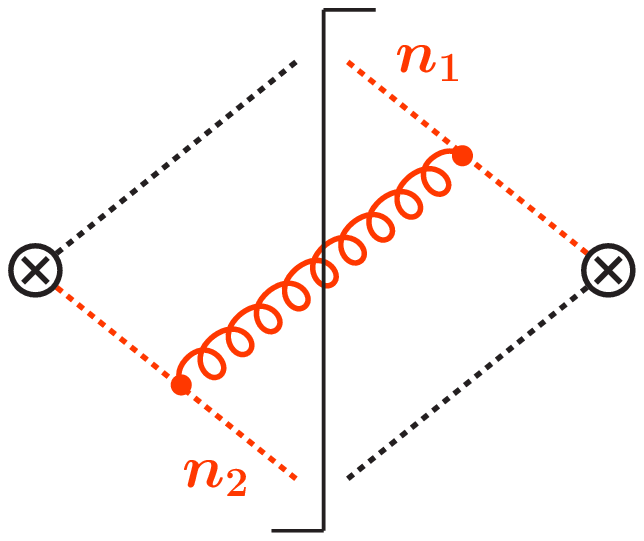}\label{fig:Sp21}}%
\hfill%
\subfigure[]{\includegraphics[scale=0.63]{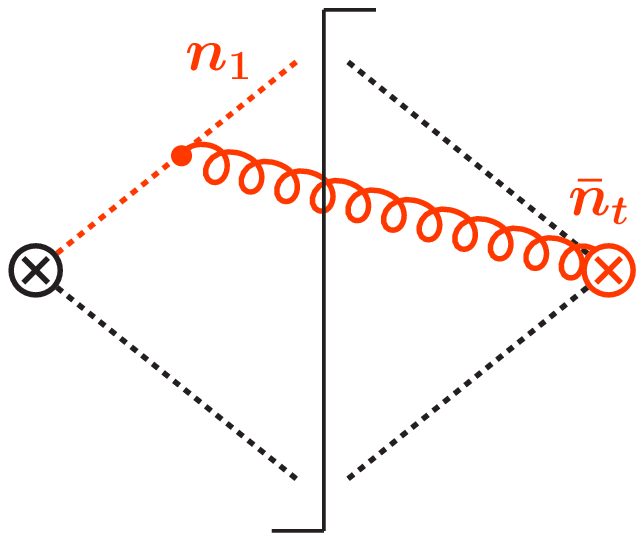}\label{fig:Sp1t}}%
\hfill%
\subfigure[]{\includegraphics[scale=0.63]{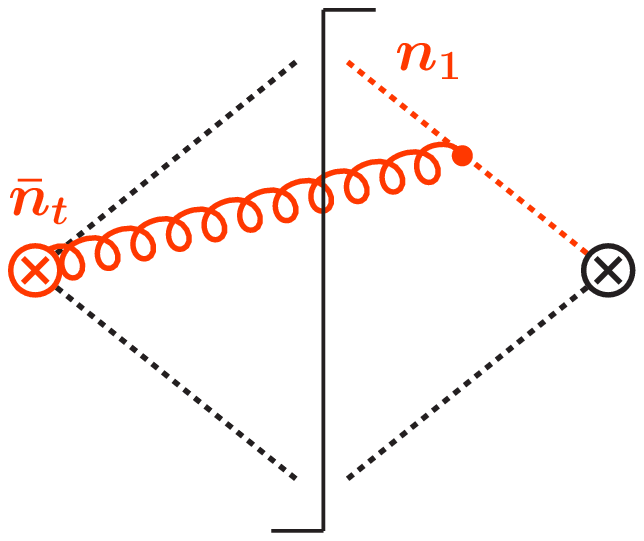}\label{fig:Spt1}}%
\vspace{-0.5ex}
\caption{One-loop diagrams for $S_+$. The vertical line denotes the final-state cut. There are also diagrams analogous to (c) and (d) with the gluon coupling to the $n_2$ Wilson lines. Virtual diagrams and diagrams with the gluon coupling to the same Wilson line are not shown.}
\label{fig:Splus}
\end{figure*}

The definition of the csoft function $S_+$ in terms of a matrix element of Wilson lines is given in \eq{softdefs}, with the measurement function given in \eq{softmeas}. The calculation for $S_+$ is more nontrivial due to additional csoft Wilson lines $V_\nt$, and we therefore provide some more details.

There are two basic types of diagrams at one loop, shown in \fig{Splus}. In the diagrams shown in \figs{Sp12}{Sp21}, a gluon is exchanged between the $X$ Wilson lines in the $n_1$ and $n_2$ directions, which corresponds to a csoft gluon exchanged between the nearby jets. These diagrams are the same as in a usual soft-function calculation. The analogous virtual diagrams vanish in pure dimensional regularization, and the diagrams with the gluon attaching to the same Wilson line vanish due to $n^2 = 0$. The diagrams do not require a zero-bin subtraction, and their contribution to the one-loop renormalized csoft function is the usual hemisphere contribution
\begin{equation}
S_{+,12}^{\kappa\one} (k_1,k_2,\mu)
= \frac{\alpha_s(\mu)}{4\pi}\, \T_1 \cdot \T_2
\biggr[\frac{8}{\sqrt{\hs_t}\mu}\cL_1\Bigl(\frac{k_1}{\sqrt{\hs_t}\mu}\Bigr)\delta(k_2)
+ \frac{8}{\sqrt{\hs_t}\mu}\cL_1\Bigl(\frac{k_2}{\sqrt{\hs_t}\mu}\Bigr) \delta(k_1)
- \frac{\pi^2}{3}\delta(k_1)\delta(k_2) \biggl]
\,.\end{equation}
Similarly, the contribution to the anomalous dimension from this diagram is the hemisphere contribution,
\begin{equation}
\gamma_{S_+,12}^{\kappa} (k_1,k_2,\mu)
= -\frac{\alpha_s(\mu)}{4\pi}\, \T_1 \cdot \T_2 \frac{8}{\sqrt{\hs_t}\mu}
\biggl[\cL_0\Bigl(\frac{k_1}{\sqrt{\hs_t}\mu}\Bigr)\delta(k_2) + \cL_0\Bigl(\frac{k_2}{\sqrt{\hs_t}\mu}\Bigr)\delta(k_1) \biggr]
\,.\end{equation}

The second type of diagram comes from exchange of a gluon between the $V$ and $\X$ Wilson lines, as shown in \figs{Sp1t}{Spt1}.  These diagrams have a nontrivial usoft limit, which means we must perform a zero-bin subtraction to remove double counting. As discussed in \sec{SCET+}, the zero-bin limit is obtained by expanding $n_1$ in terms of $\nt$, with the zero-bin measurement obtained from \eq{TauNcsoft0}. We focus on gluon exchange between $V_\nt$ and $\X_{n_1}$; the results for changing $n_1\to n_2$ are analogous.  Subtracting the zero-bin contribution from the naive part of the diagram yields
\begin{align}\label{eq:T1T3}
- 2\,\T_1 \cdot \T_3 \Bigl(\frac{e^{\gamma_E}\mu^2}{4\pi}\Bigr)^\epsilon g^2\int \frac{\df^d p}{(2\pi)^d}\, 2\pi\delta(p^2)\theta(p^0)
\biggl[\frac{\bnt\cdot n_1}{(\bnt\cdot p)(n_1\cdot p)}\, \cM^\csoft(k_1, k_2)
- \frac{\bnt\cdot n_t}{(\bnt\cdot p)(n_t\cdot p)}\, \cM^{\csoft\, 0}(k_1, k_2) \biggr]
\,.\end{align}
Here, $\T_1$ is the color charge carried by $\X_{n_1}$. Since $S_+$ is diagonal in color, the color charge carried by $V_\nt$ is $-\T_1 - \T_2 = \T_3$. The result for \eq{T1T3} can be extracted using the results of Ref.~\cite{Jouttenus:2011wh} in the limit $\hat{s}_t \ll \hat{s}_Q$. We split up the phase space for the naive and zero-bin csoft contribution into regions $ \hat{q}_3 \cdot p > \hat{q}_1 \cdot p$ and $\hat{q}_3 \cdot p < \hat{q}_1 \cdot p$, where in the latter region the naive and zero-bin contributions cancel. In the region  $ \hat{q}_3 \cdot p > \hat{q}_1 \cdot p$ the naive csoft contribution is given by the sum of the hemisphere and nonhemisphere contributions $S^{(1)}_{13,{\rm hemi}}+S^{(1)}_{13,2}$ of Ref.~\cite{Jouttenus:2011wh} expanded in the limit $\hat{s}_t \ll \hat{s}_Q$. It is straightforward to calculate the zero-bin contribution in \eq{T1T3} for $\hat{q}_3 \cdot p > \hat{q}_1 \cdot p$. Taking the difference between these terms gives the total contribution to the renormalized one-loop $S_+$ function from $V_\nt$ and $\X_{n_1}$ exchange
\begin{equation}
S_{+,13}^{\kappa\one} (k_1, k_2, \mu)
= \frac{\alpha_s(\mu)}{4\pi}\,\T_1 \cdot \T_3 \biggl[ \frac{4}{\sqrt{\hs_t}\mu}\cL_1\Bigl(\frac{k_1}{\sqrt{\hs_t}\mu}\Bigr)\delta(k_2)
- \frac{4}{\sqrt{\hs_t}\mu}\cL_1\Bigl(\frac{k_2}{\sqrt{\hs_t}\mu}\Bigr)\delta(k_1) - \frac{2\pi^2}{3} \delta(k_1)\delta(k_2) \biggr]
\,.\end{equation}
The corresponding contribution to the anomalous dimension is
\begin{equation}
\gamma_{S_+,13}^{\kappa} (k_1, k_2, \mu)
= -\frac{\alpha_s(\mu)}{4\pi}\, \T_1 \cdot \T_3\, \frac{4}{\xi}
\biggl[\cL_0\Bigl(\frac{k_1}{\xi}\Bigr)\delta(k_2)
- \cL_0\Bigl(\frac{k_2}{\xi}\Bigr)\delta(k_1) \biggr]
\,,\end{equation}
where $\xi$ is a dimension-one dummy variable which is only needed to make the argument of $\cL_0$ dimensionless, but cancels between the two terms. The analogous contribution with a gluon exchanged between $V_{n_t}$ and the $n_2$ Wilson line is the same with the replacement $1 \leftrightarrow 2$.

Combining everything, the final result for the one-loop renormalized $S_+$ function becomes
\begin{align} \label{eq:SCNLO}
S_+^\kappa(k_1,k_2,\mu)
&= \delta(k_1)\, \delta(k_2) + \frac{\alpha_s(\mu)}{4\pi}
\biggl\{ \T_1\cdot\T_2
\biggl[\frac{8}{\sqrt{\hs_t}\mu}\cL_1\Bigl(\frac{k_1}{\sqrt{\hs_t}\mu}\Bigr)\delta(k_2)
+ \frac{8}{\sqrt{\hs_t}\mu} \cL_1\Bigl(\frac{k_2}{\sqrt{\hs_t}\mu}\Bigr) \delta(k_1)
- \frac{\pi^2}{3}\delta(k_1)\, \delta(k_2)
\biggr]
\nn\\ &\quad
+ (\T_1 - \T_2) \cdot\T_3\, \frac{4}{\sqrt{\hs_t}\mu} \biggl[  \cL_1\Bigl(\frac{k_1}{\sqrt{\hs_t}\mu}\Bigr)\delta(k_2)
- \cL_1\Bigl(\frac{k_2}{\sqrt{\hs_t}\mu}\Bigr)\delta(k_1) \biggr]
- (\T_1 + \T_2)\cdot \T_3\,\frac{2\pi^2}{3} \delta(k_1)\, \delta(k_2)
\biggr] \biggr\}
\,.\end{align}
We can see explicitly that $S_+$ depends only on the scales $k_i/\sqrt{\hs_t}$. The RGE for $S_+$ has the form
\begin{equation} \label{eq:SCRGE}
\mu\frac{\df}{\df\mu} S_+^\kappa(k_1, k_2, \mu)
= \int\!\df k_1'\,\df k_2'\, \gamma_{S_+}^\kappa(k_1 - k_1',k_2 - k_2', \mu)\,
S_+^\kappa(k_1', k_2', \mu)
\,,\end{equation}
where the one-loop anomalous dimension is given by
\begin{align} \label{eq:gammaSCNLO}
\gamma_{S_+}^\kappa (k_1, k_2, \mu)
&= -\frac{\alpha_s(\mu)}{4\pi} \biggl\{\T_1\cdot\T_2\, \frac{8}{\sqrt{\hs_t}\mu} \biggl[\cL_0\Bigl(\frac{k_1}{\sqrt{\hs_t}\mu}\Bigr)\delta(k_2)
+ \cL_0\Bigl(\frac{k_2}{\sqrt{\hs_t}\mu}\Bigr)\delta(k_1) \biggr]
\\*\nn &\quad
+ (\T_1-\T_2)\cdot\T_3\, \frac{4}{\xi}\biggl[\cL_0\Bigl(\frac{k_1}{\xi}\Bigr)\delta(k_2)
- \cL_0\Bigl(\frac{k_2}{\xi}\Bigr) \delta(k_1) \biggr] \biggr\}
\,.\end{align}
\end{widetext}

\subsubsection{Soft Functions with Single Argument}

For our numerical analysis in \sec{Applications} we project the soft functions onto the sum of their arguments,
\begin{align} \label{eq:STau3}
S_2(k, \mu) &= \int\!\df k_1 \df k_2 \df k_3\, S_2(k_1, k_2, k_3, \mu)
\nn\\ & \qquad\times
\delta(k - k_1 - k_2 - k_3)
\,,\\\nn
S_+(k, \mu) &= \int\!\df k_1 \df k_2\, S_+^\kappa(k_1, k_2, \mu)\,\delta(k - k_1 - k_2)
\,.\end{align}
From \eqs{S2NLO}{SCNLO}, we obtain their NLO expressions,
\begin{align} \label{eq:STau3NLO}
S_2(k, \mu) &= \delta(k)
+ \frac{\alpha_s(\mu)C_F}{4\pi}
\nn\\ & \quad\times
\biggl[ -\frac{16}{\sqrt{\hs_Q} \mu} \cL_1\Bigl(\frac{k}{\sqrt{\hs_Q}\mu} \Bigr) + \frac{\pi^2}{3} \delta(k) \biggr]
\,, \nn \\
S_+(k, \mu)
&= \delta(k) + \frac{\alpha_s(\mu)}{4\pi} \biggl\{ -C_A \biggl[ \frac{8}{\sqrt{\hs_t}\mu} \cL_1\Bigl( \frac{k}{\sqrt{\hs_t}\mu} \Bigr)
\nn \\ & \quad
 - \frac{\pi^2}{6} \delta(k) \biggr] + C_F \frac{2\pi^2}{3} \delta(k) \biggr\}
\,.\end{align}
Note that this projection removes the dependence on $(\T_1 - \T_2)\cdot \T_3$, which makes $S_+(k, \mu)$ independent of $\kappa$. The single-argument soft functions satisfy the RGE
\begin{equation} \label{eq:S_RGE}
\mu\frac{\df}{\df\mu} S(k, \mu) = \int\! \df k'\, \gamma_S(k - k', \mu)\, S(k', \mu)
\,,\end{equation}
where the anomalous dimensions after projecting onto $k$ simplify to
\begin{align} \label{eq:S2tau3}
\gamma_{S_2}(k, \mu)
&= \frac{\alpha_s(\mu) C_F}{4\pi} \frac{16}{\sqrt{\hs_Q}\mu} \cL_0 \Bigl(\frac{k}{\sqrt{\hs_Q}\mu} \Bigr)
\,, \nn \\
\gamma_{S_+}(k, \mu)
&= \frac{\alpha_s(\mu) C_A}{4\pi} \frac{8}{\sqrt{\hs_t} \mu}\,
 \cL_0 \Bigl( \frac{k}{\sqrt{\hs_t}\mu}\Bigr)
\,.\end{align}

\subsection{All-Order Anomalous Dimensions}
\label{subsec:Consistency3jets}

In this section we discuss the consistency constraints on our factorized cross section in \eq{factSCETp}. This allows us to derive the general form of the anomalous dimensions for the \SCETp matching coefficient, $C_+$, and csoft function, $S_+$, which are the new ingredients in the factorization from \SCETp. In particular, we demonstrate that the convolution of the csoft and usoft functions at one loop reproduces the known result for the $3$-jettiness soft function in regular SCET in the limit $s_{12} \ll s_{13}\sim s_{23}$. This demonstrates that the csoft modes are necessary for \SCETp to reproduce the correct IR structure of QCD in this limit. We then show that the factorized cross section obeys exact renormalization group consistency.

\subsubsection{Hard-Function Consistency and Derivation of $\gamma_{C_+}$}

The factorized $3$-jettiness cross section in \SCETI is given by \cite{Jouttenus:2011wh}
\begin{align}\label{eq:factSCET1}
&\frac{\df \sigma}{\df\tN_1\,\df\tN_2\,\df\tN_3\, \df t \, \df z}
\nn\\ & \qquad
=  \frac{\sigma_0}{Q^2}
\sum_{\kappa} H_3^{\kappa}(s_{12}, s_{13}, s_{23}, \mu)
\prod_i \, \int \df s_i \,  J_{\kappa_i}(s_i, \mu)
\nn\\ & \qquad\quad\times
S_3^\kappa\Bigl(\tN_1 - \frac{s_1}{Q_1} , \tN_2 - \frac{s_2}{Q_2}, \tN_3 - \frac{s_3}{Q_3}, \mu\Bigr)
\,.\end{align}
Here, all dijet invariant masses are counted as $s_{ij} \sim Q^2$. This means that the hard function, $H_3^\kappa$, is evaluated at their exact values given in terms of $t$ and $z$,
\begin{equation} \label{eq:sijfull}
s_{12} = t
\,,\quad
s_{13} = z Q^2 - (1 - z)\,t
\,,\quad
s_{23} = (1-z)Q^2 - z\,t
\,,\end{equation}
which follow from momentum conservation for $e^+e^-\!\to 3$ massless jets. At tree level,
\begin{align} \label{eq:H3tree}
&H_3^{\{q,g,\bq\}}(s_{12}, s_{13}, s_{23}, \mu)
\nn\\ & \qquad
= \frac{\alpha_s(\mu)C_F}{2\pi}\,
\frac{(s_{13} + s_{23})^2 + (s_{12}+s_{13})^2}{s_{12} s_{23}}
\,.\end{align}

In SCET, all loop diagrams contributing to the bare matrix element of $\mae{qg\bq}{O_3}{0}$ vanish in pure dimensional regularization, and consequently the $3$-jet hard function in SCET, $H_3(\{s_{ij}\},\mu)$, is directly given by the IR finite terms of the full QCD amplitude $\abs{\cM_\mathrm{QCD}(0\to q\bar qg)}^2$. Comparing with \eq{qcd1loop}, it follows that the hard functions in \SCETI and \SCETp to all orders in perturbation theory have to satisfy
\begin{equation}
H_3^\kappa(\{s_{ij}\},\mu) \Big|_{s_{12}\ll s_{13}\sim s_{23}}
= H_2( Q^2,\mu)\, H_+^\kappa(t, x, \mu)
\,.\end{equation}
At tree level, this can be seen immediately: to expand \eq{H3tree} in the limit $s_{12}\ll s_{13}\sim s_{23}$, we set $s_{13} = x Q^2$, $s_{23} = (1-x)Q^2$ [see \eq{si3x}], and $t = s_{12}$ and drop any terms subleading in $t/Q^2$, which gives the tree-level result for $H_+^{\{q,g,\bq\}}(t, x, \mu)$ in \eq{H3Plus}.

The above argument also applies directly to the Wilson coefficients before squaring them, so
\begin{equation} \label{eq:hardconsistency}
C_3^\kappa(\{s_{ij}\},\mu) \Big|_{s_{12}\ll s_{13}\sim s_{23}}
= C_2( Q^2,\mu)\, C_+^\kappa(t, x, \mu)
\,.\end{equation}
Taking the derivative with respect to $\mu$, it follows that
\begin{equation} \label{eq:gammahardconsistency}
\gamma_{C_3}^\kappa(\{s_{ij}\}, \mu) \Big|_{s_{12}\ll s_{13}\sim s_{23}}
= \gamma_{C_2}(Q^2, \mu) + \gamma_{C_+}^\kappa(t, z, \mu)
\,.\end{equation}
The general all-order forms of the anomalous dimensions $\gamma_{C_2}$ and $\gamma_{C_3}^\kappa$ are~\cite{Manohar:2003vb, Chiu:2008vv, Becher:2009qa}
\begin{align} \label{eq:gammaC23}
\gamma_{C_2}(Q^2, \mu)
&= - \Gamma_\cusp[\alpha_s(\mu)]\,(\T_1\! +\! \T_2)\cdot \T_3\, \ln\!\frac{- Q^2 - \img 0}{\mu^2}
\nn\\* & \quad
+ \gamma_C^q[\alpha_s(\mu)] + \gamma_C^\bq[\alpha_s(\mu)]
\,,\nn\\
\gamma_{C_3}^\kappa(\{s_{ij}\}, \mu)
&= - \Gamma_\cusp[\alpha_s(\mu)] \sum_{i<j} \T_i\cdot \T_j\, \ln\frac{- s_{ij} - \img 0}{\mu^2}
\nn\\* & \quad
+ \gamma_C^q[\alpha_s(\mu)] + \gamma_C^\bq[\alpha_s(\mu)] + \gamma_C^g[\alpha_s(\mu)]
\,,\end{align}
where the individual quark and gluon contributions in the noncusp terms are given in \eq{gammaCi}.
Compared to \eq{gammaC2} we have identified the color structure in $\gamma_{C_2}$ as
\begin{equation}\label{eq:3jetTt}
\T_t \cdot \T_3\Big|_{q\bq} = (\T_1 + \T_2)\cdot \T_3\Big|_{q\bq g} = -C_F
\,.\end{equation}
Here $\T_t$ denotes the combined color charge of the quark or antiquark that splits into partons $1$ and $2$, and $\T_t\cdot \T_3$ is evaluated in the corresponding $2$-parton $q\bq$ color space, i.e., $\T_q\cdot \T_\bq|_{q\bq} = - C_F$. In the second step, we wrote the same total color charge using the individual color charges of the daughter partons $1$ and $2$, which are now evaluated with respect to the $3$-parton $q\bq g$ color space. Explicitly, using \eq{3color} we have $(\T_q + \T_g)\cdot \T_\bq|_{q\bq g} = (C_A/2 - C_F) - C_A/2 = -C_F$, and with the same result for $q\leftrightarrow \bq$.

Using \eqs{gammahardconsistency}{gammaC23} and expanding $\gamma_{C_3}^\kappa$, we obtain the general form of $\gamma_{C_+}^\kappa$, valid to all orders in perturbation theory,
\begin{align}\label{eq:gammaC3Plus}
\gamma_{C_+}^\kappa(t, x, \mu)
&=- \Gamma_\cusp[\alpha_s(\mu)]  \,  \T_1 \cdot \T_2 \ln\frac{-t-\img 0}{\mu^2}
\nn\\ & \quad
+ \gamma_{C_+}^\kappa[\alpha_s(\mu), x]
\,,\nn\\
\gamma_{C_+}^\kappa[\alpha_s, x]
&= - \Gamma_\cusp[\alpha_s] \Bigl[
\T_1 \mcdot \T_3 \ln x
+ \T_2 \mcdot \T_3  \ln (1-x) \Bigr]
\nn\\ & \quad
+ \gamma_C^g[\alpha_s]
\,.\end{align}
Note that this provides a nontrivial example of a hard anomalous dimension, where the nonlogarithmic term, $\gamma_{C_+}[\alpha_s, x]$, depends on a kinematic variable, whose overall coefficient however is still determined by $\Gamma_\cusp$. At one loop, \eq{gammaC3Plus} reproduces \eq{gammaH3Plus} exactly using that $\gamma_{H_+}^\kappa(t, x, \mu) = 2\mathrm{Re}[\gamma_{C_+}^\kappa(t, x, \mu)]$.

\subsubsection{Soft-Function Consistency and Derivation of $\gamma_{S_+}$}

In \secs{PhysPicture}{SCET+} we have seen that \SCETp arises from expanding \SCETI in the limit $t\ll Q$. It follows that the \SCETp $3$-jet cross section in \eq{factSCETp} has to reproduce the $3$-jet cross section \eq{factSCET1} computed in \SCETI when the latter is expanded in the limit $s_{12}\ll s_{13}\sim s_{23}$,
\begin{equation} \label{eq:sigmaconsistency}
\df\sigma^\text{\SCETI} \Big|_{s_{12}\ll s_{13}\sim s_{23}} = \df\sigma^\text{\SCETp}
\,.\end{equation}
(This is exactly analogous to the statement that the \SCETI cross section must reproduce the QCD cross section expanded in the limit $m\ll Q$.) As we have seen above, the product of hard functions in \SCETp reproduces the full \SCETI hard function, and the jet functions are the same in both cases. Hence, for the cross sections to satisfy \eq{sigmaconsistency}, the soft functions have to satisfy
\begin{align} \label{eq:softconsistency}
&S_3^\kappa(k_1,k_2,k_3, \mu) \Big|_{\hs_t \ll \hs_{13} = \hs_{23}}
\\\nn &\quad
= \int\! \df k_{1}'\, \df k_{2}' \, S_2(k_1 - k_1',k_2 - k_2', k_3, \mu)\, S_+^\kappa(k_1',k_2', \mu)
\,.\end{align}
For the soft functions the limit $s_{12} \ll s_{13}\sim s_{23}$ is taken using \eq{hqtscaling} by setting $\hs_t = \hs_{12}$, $\hs_{13} = \hs_{23} = \hs_Q$ and expanding in $\hs_t \ll \hs_Q$.

The fact that the hard and soft functions separately factorize in the limit $t\ll Q$ as in \eqs{hardconsistency}{softconsistency} is a direct consequence of factorization in \SCETI and \SCETp. Since the soft sectors in both theories are decoupled from the collinear sectors, the soft sector of \SCETp has to reproduce the soft sector of \SCETI expanded in $t\ll Q$. Since the factorization applies also in the kinematic region where the soft functions become nonperturbative, the relation in \eq{softconsistency} between the soft functions in the two theories holds both at the perturbative and also the nonperturbative level.

We can check explicitly that \eq{softconsistency} is satisfied by our one-loop results. Since \SCETI correctly reproduces the IR structure of QCD, this also provides an explicit demonstration at the one-loop level that the csoft modes are necessary to reproduce the IR structure of QCD in the limit $m\ll t \ll Q$, and thus that \SCETp is the appropriate effective theory of QCD in this limit.

The full $N$-jettiness soft function at NLO has been calculated explicitly in Ref.~\cite{Jouttenus:2011wh}, where the final result is given in terms of a single integral, which can be evaluated numerically. In Ref.~\cite{Bauer:2011hj} a general algorithm was developed to calculate a wide class of soft functions with an arbitrary number of collinear directions numerically. In the limit $\hs_t \ll \hs_{Q}$, the required integrals for $S_3^\kappa$ in Ref.~\cite{Jouttenus:2011wh} can be obtained analytically, and we find
\begin{widetext}
\begin{align}\label{eq:S3NLO}
&S_3^\kappa(k_1, k_2, _3, \mu) \Big|_{\hs_t \ll \hs_{13} = \hs_{23}}
\nn\\ & \quad
= \delta(k_1)\, \delta(k_2)\, \delta(k_3)
+ \frac{\alpha_s(\mu)}{4\pi}
\biggl\{\T_1\cdot\T_2 \biggl[
\frac{8}{\sqrt{\hs_t}\mu} \cL_1\Bigl(\frac{k_1}{\sqrt{\hs_t}\mu}\Bigr)\delta(k_2)
+ \frac{8}{\sqrt{\hs_t}\mu}\cL_1\Bigl(\frac{k_2}{\sqrt{\hs_t}\mu}\Bigr)\delta(k_1)
- \frac{\pi^2}{3}\delta(k_1)\, \delta(k_2) \biggr] \delta(k_3)
\nn \\ &\qquad
+ \T_1\cdot\T_3 \biggl[
\frac{8}{\sqrt{\hs_Q}\mu} \cL_1\Bigl(\frac{k_1}{\sqrt{\hs_Q}\mu}\Bigr)\delta(k_3)
+ \frac{8}{\sqrt{\hs_Q}\mu} \cL_1\Bigl(\frac{k_3}{\sqrt{\hs_Q}\mu}\Bigr)\delta(k_1)
- \frac{\pi^2}{3}\delta(k_1)\, \delta(k_3) \biggr] \delta(k_2)
\nn \\ &\qquad
+ \T_2\cdot\T_3 \biggl[
\frac{8}{\sqrt{\hs_Q}\mu} \cL_1\Bigl(\frac{k_2}{\sqrt{\hs_Q}\mu}\Bigr)\delta(k_3)
+ \frac{8}{\sqrt{\hs_Q}\mu} \cL_1\Bigl(\frac{k_3}{\sqrt{\hs_Q}\mu}\Bigr)\delta(k_2)
- \frac{\pi^2}{3}\delta(k_2)\, \delta(k_3) \biggr] \delta(k_1)
\nn \\ &\qquad
+ (\T_1 - \T_2) \cdot\T_3\, 2\ln\Bigl(\frac{\hs_Q}{\hs_t}\Bigr) \frac{1}{\xi} \biggl[\cL_0\Bigl(\frac{k_1}{\xi}\Bigr)\delta(k_2)
- \cL_0\Bigl(\frac{k_2}{\xi}\Bigr)\delta(k_1)\biggr]
- (\T_1 + \T_2) \cdot \T_3\, \frac{2\pi^2}{3}\delta(k_1)\, \delta(k_2)\,\delta(k_3)
\biggr\}
\,.\end{align}
Here, the first three terms proportional to $\T_i\cdot\T_j$ are the hemisphere contributions, which contain the explicit $\mu$ dependence. The last two terms come from the nonhemisphere contributions, where the dummy variable $\xi$ again cancels between the two terms and is only needed to make the argument of $\cL_0$ dimensionless. As a cross check, we have compared this result with the numerical result obtained using Ref.~\cite{Bauer:2011hj}, and the two agree in the limit $\hs_t \ll \hs_{13} = \hs_{23} = \hs_Q$.  As expected, this soft function depends on both $\hs_t$ and $\hs_Q$, and there is no choice of renormalization scale for which the logarithms of $\hs_t/\hs_Q$ are absent.

Since the soft functions at tree level are all just $\delta$ functions, \eq{softconsistency} simplifies at one loop to
\begin{align}
\label{eq:softconsistency1loop}
&S_3^{\kappa\one}(k_1,k_2,k_3, \mu)\Big|_{\hs_t \ll \hs_{13} = \hs_{23}}
= S_2^\one(k_1, k_2, k_3, \mu)
+ S_+^{\kappa\one}(k_1,k_2, \mu)\,\delta(k_3)
\,.\end{align}
Subtracting the $\alpha_s$ corrections in \eqs{SCNLO}{S2NLO} from those in \eq{S3NLO}, the terms proportional to $\T_1\cdot \T_2$ and those involving $\cL_1(k_3)$ or only $\delta$ functions immediately cancel. For the remaining terms, we obtain
\begin{align}
\frac{\alpha_s(\mu)}{4\pi} (\T_1 - \T_2) \cdot\T_3
&\biggl\{\frac{4}{\sqrt{\hs_Q}\mu} \biggl[
\cL_1\Bigl(\frac{k_1}{\sqrt{\hs_Q}\mu}\Bigr)\delta(k_2)
- \cL_1\Bigl(\frac{k_2}{\sqrt{\hs_Q}\mu}\Bigr) \delta(k_1) \biggr]
+ 2\ln\Bigl(\frac{\hs_Q}{\hs_t}\Bigr) \frac{1}{\xi} \biggl[\cL_0\Bigl(\frac{k_1}{\xi}\Bigr)\delta(k_2)
- \cL_0\Bigl(\frac{k_2}{\xi}\Bigr)\delta(k_1)\biggr]
\nn\\ & \quad
- \frac{4}{\sqrt{\hs_t}\mu} \biggl[ \cL_1\Bigl(\frac{k_1}{\sqrt{\hs_t}\mu}\Bigr)\delta(k_2)
- \cL_1\Bigl(\frac{k_2}{\sqrt{\hs_t}\mu}\Bigr)\delta(k_1)
\biggr] \biggr\}
= 0
\,,\end{align}
where we used the rescaling identity
\begin{equation}
\lambda \cL_1(\lambda x) = \cL_1(x) + \ln\lambda\, \cL_0(x) + \frac{1}{2}\ln^2\!\lambda \, \delta(x)
\,.\end{equation}
Thus, \eq{softconsistency} is satisfied by our one-loop results.

We can also use \eq{softconsistency} to derive the all-order structure of the anomalous dimension of $S_+$. Taking the derivative of \eq{softconsistency} with respect to $\mu$, we get
\begin{align} \label{eq:gammasoftconsistency}
&\gamma_{S_3}^\kappa(k_1,k_2,k_3,\mu) \Big|_{\hs_t \ll \hs_{13} = \hs_{23}}
= \gamma_{S_2}(k_1, k_2, k_3,\mu) + \gamma_{S_+}^\kappa(k_1, k_2, \mu) \delta(k_3)
\,.\end{align}
The all-order structure of the anomalous dimension of the $N$-jettiness soft function was derived in Ref.~\cite{Jouttenus:2011wh}. For $\gamma_{S_3}^\kappa$ in this limit we have
\begin{align} \label{eq:gammaS3}
\gamma_{S_3}^\kappa(k_1, k_2, k_3, \mu) \Big|_{\hs_t \ll \hs_{13} = \hs_{23}}
&= -2 \Gamma_\cusp [\alpha_s(\mu)] \biggl\{
\T_1\cdot\T_2 \frac{1}{\sqrt{\hs_t}\mu} \biggl[
\cL_0\Bigl(\frac{k_1}{\sqrt{\hs_t}\mu}\Bigr)\delta(k_2)
+ \cL_0\Bigl(\frac{k_2}{\sqrt{\hs_t}\mu}\Bigr)\delta(k_1) \biggr] \delta(k_3)
\nn \\ &\quad
+ \T_1\cdot\T_3 \frac{1}{\sqrt{\hs_Q}\mu} \biggl[
\cL_0\Bigl(\frac{k_1}{\sqrt{\hs_Q}\mu}\Bigr)\delta(k_3)
+ \cL_0\Bigl(\frac{k_3}{\sqrt{\hs_Q}\mu}\Bigr)\delta(k_1)\biggr]\delta(k_2)
\nn \\ &\quad
+ \T_2\cdot\T_3 \frac{1}{\sqrt{\hs_Q}\mu} \biggl[
\cL_0\Bigl(\frac{k_2}{\sqrt{\hs_Q}\mu}\Bigr)\delta(k_3)
+ \cL_0\Bigl(\frac{k_3}{\sqrt{\hs_Q}\mu}\Bigr)\delta(k_2)\biggr]\delta(k_1) \biggr\}
\nn \\* &\quad
+ \gamma_{S_3}^\kappa[\alpha_s(\mu)]\,\delta(k_1)\, \delta(k_2)\, \delta(k_3)
\,.\end{align}

To deduce the all-order structure for $\gamma_{S_2}$, we first note that upon projecting $S_2(k_1, k_2, k_3)$ onto $k_t = k_1 + k_2$,
\begin{equation}
S_2(k_t, k_3, \mu) = \int\!\df k_1\,\df k_2\, S_2(k_1, k_2, k_3, \mu)\, \delta(k_t - k_1 - k_2)
\,,\end{equation}
it reduces to the normal $2$-jettiness soft function. Therefore, we know that to all orders
\begin{align}\label{eq:gammaS2t}
&\int\!\df k_1\,\df k_2\, \gamma_{S_2}(k_1, k_2, k_3, \mu)\, \delta(k_t - k_1 - k_2)
\nn \\ & \quad
= -2\Gamma_\cusp[\alpha_s(\mu)]\,(\T_1 + \T_2)\cdot\T_3\, \frac{1}{\sqrt{\hs_Q}\mu} \biggl[
\cL_0\Bigl(\frac{k_t}{\sqrt{\hs_Q}\mu}\Bigr)\delta(k_3)
+ \cL_0\Bigl(\frac{k_3}{\sqrt{\hs_Q}\mu}\Bigr) \delta(k_t) \biggr]
+ \gamma_{S_2}[\alpha_s(\mu)]\,\delta(k_t)\,\delta(k_3)
\,.\end{align}
From its definition, we know that the full $S_2(k_1, k_2, k_3)$ is symmetric in $k_1$ and $k_2$, and since
the distinction between $k_1$ and $k_2$ only comes from the measurement function, this symmetry cannot be changed by the renormalization and the anomalous dimension must therefore also be symmetric in $k_1$ and $k_2$. Furthermore, from \eq{gammasoftconsistency} we know that the dependence on $k_3$ must exactly cancel between $\gamma_{S_2}$ and $\gamma_{S_3}^\kappa$. The most general form of $\gamma_{S_2}$ that satisfies these requirements, \eq{gammaS2t}, and is only single-logarithmic in $\mu$ is
\begin{align}\label{eq:gammaS2}
\gamma_{S_2} (k_1, k_2, k_3, \mu)
&= -2\Gamma_\cusp[\alpha_s(\mu)]\,(\T_1 + \T_2)\cdot\T_3\, \frac{1}{\sqrt{\hs_Q}\mu} \biggl\{
\frac{1}{2}\cL_0\Bigl(\frac{k_1}{\sqrt{\hs_Q}\mu}\Bigr) \delta(k_2)\,\delta(k_3)
+ \frac{1}{2}\cL_0\Bigl(\frac{k_2}{\sqrt{\hs_Q}\mu}\Bigr) \delta(k_1)\,\delta(k_3)
\nn \\ & \quad
+ \cL_0\Bigl(\frac{k_3}{\sqrt{\hs_Q}\mu}\Bigr) \delta(k_1)\,\delta(k_2) \biggr\}
+ \gamma_{S_2}[\alpha_s(\mu)]\,\delta(k_1)\,\delta(k_2)\,\delta(k_3)
\,.\end{align}
The noncusp terms in \eqs{gammaS3}{gammaS2} are
\begin{align}
\gamma_{S_3}^\kappa[\alpha_s]
&= -\!\! \sum_{i = \{q,\bq,g\}}\!\! \bigl( \gamma_J^i[\alpha_s] + 2\gamma_C^i[\alpha_s] \bigr)
= 0 + \ord{\alpha_s^2}
\,,\qquad
\gamma_{S_2}[\alpha_s]
= -\!\! \sum_{i = \{q,\bq \}}\!\! \bigl(\gamma_J^i[\alpha_s] + 2\gamma_C^i[\alpha_s] \bigr)
= 0 + \ord{\alpha_s^2}
\,,\end{align}
where $\gamma_J^i[\alpha_s]$ and $\gamma_C^i[\alpha_s]$ are the noncusp terms in the jet and hard anomalous dimensions \eqs{gammaJ}{gammaC23}, and are given in \eqs{gammaJi}{gammaCi}. This form of \eq{gammaS2} agrees with our one-loop result in \eq{gammaS2NLO}. Taking the difference between \eqs{gammaS3}{gammaS2}, we obtain the general form of $\gamma_{S_+}^\kappa$,
\begin{align}\label{eq:gammaSC}
\gamma_{S_+}^\kappa (k_1, k_2, \mu)
&= -2 \Gamma_\cusp[\alpha_s(\mu)]\, \T_1\cdot\T_2 \frac{1}{\sqrt{\hs_t}\mu} \biggl[\cL_0\Bigl(\frac{k_1}{\sqrt{\hs_t}\mu}\Bigr)\delta(k_2)
+ \cL_0\Bigl(\frac{k_2}{\sqrt{\hs_t}\mu}\Bigr)\delta(k_1) \biggr]
+ \gamma_{S_+}^\kappa[\alpha_s(\mu), k_1, k_2]
\,,\\\nn
\gamma_{S_+}^\kappa[\alpha_s, k_1, k_2]
&= -\Gamma_\cusp[\alpha_s]\, (\T_1-\T_2)\cdot\T_3\, \frac{1}{\xi} \biggl[\cL_0\Bigl(\frac{k_1}{\xi}\Bigr)\delta(k_2)
- \cL_0\Bigl(\frac{k_2}{\xi}\Bigr) \delta(k_1) \biggr]
- \bigl(\gamma_J^g[\alpha_s] + 2\gamma_C^g[\alpha_s] \bigr)\,\delta(k_1)\,\delta(k_2)
\,,\end{align}
\end{widetext}
which again agrees with our explicit one-loop result in \eq{gammaSCNLO}. The part of the anomalous dimension of $S_+$ which does not explicitly depend on $\mu$ has a more complicated structure than for $S_2$ and $S_3$. It has a nontrivial color structure and dependence on the kinematic variables $k_1$ and $k_2$, which effectively behaves as $\ln(k_1/k_2)$. The coefficient of that dependence is however still determined by $\Gamma_\cusp$. This is the soft analog of what we saw for the anomalous dimension of $C_+$ in \eq{gammaC3Plus}, which contains terms like $\ln(s_{13}/Q^2) = \ln(x)$ and $\ln(s_{23}/Q^2) = \ln(1-x)$.

\subsubsection{Combined Consistency of Factorized Cross Section}

As we have seen above, the sum of the hard and soft anomalous dimensions in \SCETp each reproduce the hard and soft anomalous dimension $\gamma_{C_2}$ and $\gamma_{S_3}$ in \SCETI in the limit $s_{12} \ll s_{13}\sim s_{23}$. The full cross section in \eq{factSCETp} is a physical observable and cannot depend on the arbitrary renormalization scale $\mu$. This implies that the anomalous dimensions must satisfy the consistency relation
\begin{align} \label{eq:3jetconsistency}
0 &= 2\mathrm{Re} \bigl[\gamma_{C_2} (Q^2, \mu) + \gamma_{C_+}^\kappa(t, z, \mu) \bigr]\prod_i \delta(k_i)
\nn\\ & \quad
+ \sum_i  Q_i \gamma^i_J (Q_i k_i, \mu) \prod_{j\neq i} \delta(k_j)
\nn\\ & \quad
+ \gamma_{S_2} (k_1,k_2,k_3,\mu) + \gamma_{S_+}^\kappa (k_1, k_2, \mu)\,\delta(k_3)
\,,\end{align}
which is derived by taking the derivative of \eq{factSCETp} with respect to $\mu$, and following the same steps as in Ref.~\cite{Jouttenus:2011wh} to derive the analogous relation for the cross section in SCET given in \eq{factSCET1}.

Since the SCET cross section satisfies the RGE consistency, we already know that \eq{3jetconsistency} must be satisfied as well. Nevertheless it is an instructive and straightforward exercise to show that \eq{3jetconsistency} is indeed satisfied to all orders by the results for the anomalous dimensions given in Eqs.~\eqref{eq:gammaJ}, \eqref{eq:gammaC23}, \eqref{eq:gammaC3Plus}, \eqref{eq:gammaS2}, and \eqref{eq:gammaSC}. The cancellation of the different logarithmic dependence in the hard, jet, and soft functions for the three color structures $\T_1\cdot \T_2$, $\T_1\cdot\T_3$, and $\T_2\cdot\T_3$, now happens as follows,
\begin{align} \label{eq:logcons}
0 &= - \ln\frac{t}{\mu^2} + \ln\frac{Q_1\xi}{\mu^2} + \ln\frac{Q_2\xi}{\mu^2}
- \ln\frac{\xi^2}{\mu^2 \hs_t}
\,,\\\nn
0 &= - \ln\frac{Q^2}{\mu^2} - \ln x
+ \ln\frac{Q_1\xi}{\mu^2} + \ln\frac{Q_3\xi}{\mu^2}
- \ln\frac{\xi^2}{\mu^2 \hs_Q}
\,,\\\nn
0 &= - \ln\frac{Q^2}{\mu^2} - \ln(1-x)
+ \ln\frac{Q_2\xi}{\mu^2} + \ln\frac{Q_3\xi}{\mu^2}
- \ln\frac{\xi^2}{\mu^2\hs_Q}
\,.\end{align}
The $\ln (Q_i\xi/\mu^2)$ terms are supplied by the jet functions. Note that this cancellation crucially relies on a consistent power expansion in $\lambda_t$, as in \eqs{hqtscaling}{si3}, which implies $s_{13} \equiv x Q^2 = Q_1 Q_3 \hs_Q$ and $s_{23} \equiv (1-x) Q^2 = Q_2 Q_3 \hs_Q$, so \eq{logcons} is satisfied exactly without requiring any further expansion.

\section{\boldmath Generalization to $pp\to N$ Jets}
\label{sec:njets}

In \secs{3jetfact}{3jetNLO}, we have applied our new effective theory \SCETp to the simple case of $e^+ e^-\! \to$ 3 jets. This allowed us to discuss in detail how \SCETp is applied to derive the factorized cross section, and to obtain all its ingredients at NLO. In this section, we extend our discussion to the general case of $pp\to N$ jets plus additional leptons relevant for the LHC. In particular this requires adding hadrons to the initial state, as well as generalizing to more final state jets and the resulting more complicated color structure.

The key ingredients needed to derive the factorization theorem are the same here as in our $3$-jet example in \sec{3jetfact}: We have to define a consistent power counting, determine the relevant operators, and show the factorization of the measurement function. Many aspects in this discussion are completely analogous to the $3$-jet case, so we will focus on those where the extension to more jets is nontrivial, which are the kinematic dependence and the color structure.

\subsection{Kinematics and Power Counting}

For our observable we again use $N$-jettiness defined in \eq{TauN_def}. The two beams are included using two reference momenta $q_a$ and $q_b$, which correspond to the momenta of the incoming partons, and the corresponding dimensionless reference vectors $\hq_{a,b} = q_{a,b}/Q_{a,b}$, which determined the separation between the beam and jet regions (see Refs.~\cite{Stewart:2010tn, Jouttenus:2011wh} for more details). We will consider the cross section differential in the $N$-jettiness contributions $\tN_a, \tN_b, \tN_1,\ldots, \tN_{N}$, where $\tN_{a,b}$ measure the contribution from the beam regions. In addition we measure the small dijet invariant mass $t$ and the energy ratio $z = E_1/(E_1 + E_2)$ for the two nearby jets.

As before, we label jets 1 and 2 as the two nearby jets, and consider the limit in \eq{tscaling}, with all jet energies parametrically of the same size, such that we have
\begin{align}\label{eq:Ns12}
t \equiv s_{12} \ll s_{ij} \sim Q^2
\,,\qquad
\hs_t \equiv \hs_{12} \ll \hs_{ij} \sim 1
\,.\end{align}
(To have a manageable notation, we specify $1$ and $2$ to be two final-state jets. The case where jet $1$ is close to a beam, such that $s_{1a} = 2q_1\cdot q_a \ll s_{ij}$ is completely analogous and does not involve additional complications.)

The power expansion in $\lambda_t^2 = t/Q^2$ is again defined by choosing a common reference vector $\hq_t$ for jets $1$ and $2$, as in \eq{hqtscaling}. This gives
\begin{align} \label{eq:NhatSij}
\hat{s}_{Qi} &\equiv 2 \hat{q}_t \cdot \hat{q}_i = \hat{s}_{1i}[1+\ord{\lambda_t}] = \hat{s}_{2i}[1+\ord{\lambda_t}]
\end{align}
for each jet $i \neq 1,2$, which generalizes $\hs_Q$ from the $3$-jet case. The corresponding dijet invariant masses in the SCET above $\sqrt{t}$ are then given by
\begin{align} \label{eq:Nsij}
s_{1i} &= Q_1 Q_i\, \hs_{Qi}
\,, \qquad s_{2i} = Q_2 Q_i\, \hs_{Qi}
\,, \nn \\
s_{ti} &= (s_{1i} + s_{2i}) = [(q_1 + q_2) +q_i]^2 [1+\ord{\lambda_t}]
\,,\end{align}
which we can also express as
\begin{align} \label{eq:Nsijz}
s_{1i} = x\, s_{ti}
\,,\quad
s_{2i} = (1-x)\, s_{ti}
\,,\quad
x = \frac{Q_1}{Q_1 + Q_2}
\,.\end{align}
For the geometric measure, $Q_i = E_i$, we have $x = z$ as before.

The factorization of the measurement function follows the same logic as in the $3$-jet case.  By taking the collinear, csoft, and usoft limits of the full $N$-jettiness measurement function, we obtain the generalizations of the measurement functions in the $3$-jet case. The collinear and csoft measurement functions are not affected by the presence of additional jets and so are unchanged from the 3-jet case. The generalization of the usoft measurement function for the $N$-jet case is given by
\begin{align}
\label{eq:usoftmeas}
&\mathcal{M}_N^\usoft(\tN_1^\usoft, \tN_2^\usoft, \tN_a^\usoft, \ldots, \tN_{N}^\usoft)
\\\nn & \quad
= \prod_{i=1,2} \delta\bigl(\tN_i^\usoft - 2\hq_t\cdot \hP_i^\usoft\bigr)
\prod_{i\neq 1,2} \delta\bigl(\tN_i^\usoft - 2\hq_i\cdot \hP_i^\usoft\bigr)
\,,\end{align}
where the momentum operator $\hP_i^\usoft$ is defined in \eq{hPidef}, and the $\Theta_i^\usoft(p)$ are defined by the obvious generalization of \eq{Thetasoft},
\begin{align} \label{eq:ThetasoftN}
\Theta_1^\usoft(p) &=  \prod_{i\neq 1,2} \theta(\hq_i\cdot p - \hq_t\cdot p)\,\theta[\cos\phi_{t}(p)]
\,, \nn \\
\Theta_2^\usoft(p) &= \prod_{i\neq 1,2} \theta(\hq_i\cdot p - \hq_t\cdot p)\,\theta[-\cos\phi_{t}(p)]
\,, \\\nn
\Theta_{i\neq 1,2}^\usoft(p) &=  \theta(\hq_t\cdot p - \hq_i\cdot p) \prod_{j\neq 1,2,i}  \theta(\hq_j\cdot p - \hq_i\cdot p)
\,.\end{align}
The reference vectors $\hq_i$ for $i\neq 1,2$ can now have a nonzero component in the transverse plane, and can therefore have $\phi_t$ dependence.  This implies that the jet regions for jets 1 and 2 are in general not symmetric (unlike the 3-jet case, where they were symmetric up to power corrections in $t/Q^2$).

\subsection{Factorization}\label{subsec:njetFact}

The hard factorization in \SCETp proceeds through the same basic steps as for the 3-jet case. We first match from QCD to \SCETI at the hard scale $Q$,
\begin{equation} \label{eq:ON-1match}
\cM_\mathrm{QCD}(2\to N-1) = \Mae{N-1}{\vO_{N-1}^\dagger(\mu)}{2} \vC_{N-1}(\{s_{ij}\}, \mu)
\,,\end{equation}
where $\cM_\mathrm{QCD}(2\to N-1)$ is the $2\to N-1$ QCD amplitude for the process we are interested in, and $\vO_{N-1}$ is the corresponding $2\to N-1$-jet operator in \SCETI, discussed below. We will again use $\kappa$ to denote the dependence on a specific partonic channel when needed, but to simplify the notation, we mostly suppress the label $\kappa$ in what follows. As before, the bare loop diagrams of $\vO_{N-1}$ vanish in pure dimensional regularization, so including counterterms the renormalized matrix element of $\vO_{N-1}$ equals the tree-level result plus pure $1/\epsilon$ IR divergences which precisely cancel against the IR divergences in the QCD amplitude. Therefore, to all orders in perturbation theory, $\vC_{N-1}$ is given by the finite parts of $\cM_\mathrm{QCD}(2\to N-1)$.

The operator $\vO_{N-1}$ in the matching in \eq{ON-1match} has the form
\begin{align} \label{eq:ON-1}
\vO_{N-1}^\dagger
&= \Gamma \, \bigl[\collF_{\nt} \bigr] \bigl[\collF_{n_a} \bigr] \bigl[\collF_{n_b} \bigr] \bigl[\collF_{n_3} \bigr] \dotsb \bigl[\collF_{n_{N}}\bigr]
\nn\\ & \quad\times
\bigl[Y_{\nt} Y_{n_a} Y_{n_b} Y_{n_3} \dotsb Y_{n_N} \bigr]
\,.\end{align}
We let $\collF_{n_i}$ denote a (usoft-decoupled) gauge-invariant collinear field in the $n_i$ direction, which can be a collinear quark, antiquark, or gluon, and $\Gamma$ represents the spin structure connecting the different fields together. In general there are many such structures possible, so \eq{ON-1} really represents a set of operators. As before, jets $1$ and $2$ are described by a single collinear field $\collF_\nt$ in the $\nt$ direction, $\collF_{n_{a}}$ and $\collF_{n_{b}}$ are the fields for the incoming partons, and $\collF_{n_3}$ to $\collF_{n_{N}}$ are the fields for the outgoing partons that initiate the remaining final-state jets for $i \geq 3$. The usoft Wilson lines are written generically as $Y_{n_i}$ without any reference to their particular color representation.

The operator $\vO_{N-1}$ and Wilson coefficient $\vC_{N-1}$ in \eqs{ON-1match}{ON-1} are now vectors in the color space spanned by the $2 + N-1$ external partons, as indicated by the vector symbols. That is,
\begin{equation}
\vO_{N-1}^\dagger \vC_{N-1}
\equiv O_{N-1}^{\dagger\,\alpha_t \dotsb\alpha_N}
C_{N-1}^{\alpha_t \dotsb\alpha_N}
\,,\end{equation}
where $\alpha_i$ is the color index of the $i$th external particle. The product of all usoft Wilson lines in \eq{ON-1} is a matrix in the same color space,
\begin{equation}
\hY \equiv \bigl[ Y_{n_t}\dotsb Y_{n_N} \bigr]^{\beta_t \dotsb \beta_N \vert \alpha_t \dotsb\alpha_N }
\,.\end{equation}
The vertical bar separates the column indices (on the left) and row indices (on the right) of the matrix. The color charges $\T_k^A$ act in the external color space as
\begin{align} \label{eq:Tdef}
(\T^A_k\,\vC)^{\dotsb i_k\dotsb} &= T^A_{i_k j_k}\, C^{\dotsb j_k \dotsb}
\,,\nn\\
(\T^A_k\,\vC)^{\dotsb i_k \dotsb} &= - T^A_{j_k i_k}\, C^{\dotsb j_k \dotsb}
\,,\nn\\
(\T^A_k\,\vC)^{\dotsb A_k\dotsb} &= \img f^{A_k A B_k}\, C^{\dotsb B_k \dotsb}
\,,\end{align}
where the three lines are for the $k$th particle being an outgoing quark or incoming antiquark, an incoming quark or outgoing antiquark, or a gluon, respectively. The products $\T_i\cdot \T_j = \sum_A T_i^A T_j^A$ are matrices in color space. From \eq{Tdef} it is clear that $\T_i$ for different $i$ commute.

In the next step we match from \SCETI to \SCETp at the scale $\sqrt{t}$. From the construction of the effective theory in~\sec{SCET+}, it should be clear that the relevant operator in \SCETp is constructed out of $N+2$ collinear fields, for the two incoming and $N$ outgoing partons in the hard interaction, csoft fields that interact with the collinear fields in directions 1 and 2, and usoft fields that interact with all collinear degrees of freedom. The $2\to N$-jet operator in \SCETp, $\vO_N^+$, is obtained from \eq{ON-1} by the analogous replacement as in \eq{chint_matching},
\begin{align} \label{eq:Cnt_matching}
\collF_\nt^{\alpha_t}
&\to \collF_{n_1}^{\beta_1}\, \collF_{n_2}^{\beta_2}\, \T_{t}^{\beta_1\beta_2 \vert \beta_t}\, V_\nt^{\beta_t \vert \alpha_t}
\nn\\
&\to \bigl[\collF_{n_1}\bigr]^{\beta_1}\, \bigl[\collF_{n_2}\bigr]^{\beta_2} \bigl[X_{n_1} X_{n_2} \T_{t}  V_\nt\bigr]^{\beta_1\beta_2 | \alpha_t}
\,.\end{align}
In the second line we performed the csoft decoupling \eq{csoftBPS}, which produces the csoft Wilson lines $X_{n_1}$ and $X_{n_2}$ (dropping the superscripts that distinguish the fields before and after the field redefinition). The color generator $\T_{t}$ is contracted with the color indices of the daughter fields as shown in the first line of \eq{Cnt_matching}. From the product of csoft Wilson lines we define
\begin{equation} \label{eq:Xopcolor}
\hX \equiv \bigl[X_{n_1} X_{n_2} \T_t V_\nt \bigr]^{\beta_1\beta_2 \vert \alpha_t}
\id^{\beta_a\dotsb \beta_N \vert \alpha_a \dotsb \alpha_N}
\,,\end{equation}
which is a color space matrix that takes us from $(N+1)$-parton to $(N+2)$-parton color space, and $\id^{\beta_a\dotsb \beta_N \vert \alpha_a \dotsb \alpha_N } = \delta^{\beta_a\alpha_a }\dotsb \delta^{\beta_N\alpha_N}$ denotes the identity in the remaining $N$-parton color space for partons $a,b,3,\ldots,N$. The operator $\vO_N^+$ then has the form
\begin{align} \label{eq:ON+def}
\vO_N^{+\dagger}
&= \Gamma \, \bigl[\collF_{n_1}\bigr] \bigl[\collF_{n_2}\bigr] \bigl[\collF_{n_a} \bigr] \dotsb \bigl[\collF_{n_{N}}\bigr] \bigl[\hX\bigr] \bigl[\hY\bigr]
\,,\end{align}
where the product of all collinear fields is a row vector in $(N+2)$-parton color space. We have grouped the different factors in square brackets belonging to different sectors, which do not interact with one another through the leading-order \SCETp Lagrangian in \sec{SCET+}.

The matching from \SCETI to \SCETp takes the form
\begin{equation} \label{eq:ON+match}
\Mae{N}{\vO^\dagger_{N-1}(\mu)}{2} = \Mae{N}{\vO^{+\dagger}_{N}(\mu)}{2}\, C_+(t, x, \mu)
\,.\end{equation}
Since $\vO_{N-1}$ and $\vO_N^+$ are both vectors in color space, in principle $C_+$ could be a matrix in color space. However, since the different sectors in both \SCETI and \SCETp are explicitly decoupled, the matching coefficient $C_+$ is actually determined by the $1\to 2$ matching in \eq{Cnt_matching}%
\begin{equation}
\collF_\nt = C_+(t, x)\,\collF_{n_1}\collF_{n_2} X_{n_1} X_{n_2} \T_{t} V_\nt
\,.\end{equation}
In other words, $C_+ \equiv C_+^\kappa$ is universal and only depends on the specific $1\to 2$ splitting channel $q\to qg$, $g\to gg$, or $g\to q \bq$. In \app{RPIMatchingHard}, we use reparameterization invariance to show that $C_+$ depends only on $t$, $x$, and the azimuthal angle of the splitting.

Using the same arguments as in \subsec{HardFuncs}, we can relate $C_+$ to the collinear limit of the $2\to N$ QCD amplitude. Since the matching onto $\vO_{N-1}$ in \eq{ON-1match} is independent of the external state, we have
\begin{align}\label{eq:NjetHardmatching}
&\cM_\mathrm{QCD}(2\to N) \Big\vert_{t \ll s_{ij}}
\nn\\ & \quad
= \Mae{N}{\vO_{N-1}^\dagger}{2}\, \vC_{N-1}(\{s_{ij}\})
\nn\\ & \quad
= \Mae{N}{\vO_N^{+\dagger}}{2}\, \vC_{N-1}(\{s_{ij}\})\, C_+(t, x)
\,,\end{align}
where $\cM_\mathrm{QCD}(2\to N)\vert_{t\ll s_{ij}}$ is the $2\to N$ QCD amplitude expanded in the collinear limit of partons $1$ and $2$ becoming close, and in the second step we used \eq{ON+match}.\pagebreak\
The loop diagrams of $\vO_N^+$ again vanish in pure dimensional regularization, so the product $\vC_{N-1} C_+$ is given by the IR-finite part of $\cM_\mathrm{QCD}(2\to N)\vert_{t\ll s_{ij}}$. On the other hand, as we saw above, $\vC_{N-1}$ contains the IR-finite parts of $\cM_\mathrm{QCD}(2\to N-1)$. Therefore,
\begin{equation} \label{eq:QCDC+}
\cM^\mathrm{IR-fin}_\mathrm{QCD}(2\to N) \Big\vert_{t \ll s_{ij}}
= C_+(t, x)\, \cM^\mathrm{IR-fin}_\mathrm{QCD}(2\to N-1 )
\,.\end{equation}
It is well-known that the $N$-point QCD amplitudes in the collinear limit $t\ll s_{ij}$ factorize into $(N-1)$-point amplitudes times universal splitting amplitudes~\cite{Berends:1988zn, Mangano:1990by, Bern:1994zx, Kosower:1999xi, Kosower:1999rx, Bern:1999ry}. Just like the IR-finite parts of the full amplitude determine the hard matching coefficient, it follows from \eq{QCDC+} that the same is true for the splitting amplitudes: The IR-finite parts of the splitting amplitudes directly determine the matching coefficients $C_+$ for the different partonic channels. Taking the square of the one-loop results for the $q\to q g$ splitting amplitudes from Ref.~\cite{Bern:1994zx} and summing over helicities reproduces the expression for $H_+$ in \eq{H3Plus}. In the same way, the one-loop results for $C_+$ for the other splitting channels can be obtained.

The cross section in \SCETp is obtained from the forward matrix element of $\vO_N^+$ in \eq{ON+def} with the measurement function inserted,
\begin{align}
\df\sigma &\sim \abs{C_+}^2 \vec{C}_{N-1}^{\dagger}
\Mae{2}{ \widehat{Y}^{\dagger} \hX^\dagger \prod_i \collF_{n_i}^{\dagger}
\nn \\ & \quad \times
\cM_N(\{\tN_k\}) \, \prod_j \collF_{n_j} \hX \, \hY }{2} \vec{C}_{N-1}
\,.\end{align}
Using the factorization of the measurement together with that of the operator, the matrix element factorizes into independent collinear, csoft, and usoft matrix elements. The collinear matrix elements produce $N$ jet functions and two beam functions, which are all diagonal in color and contribute a factor of $\id = \prod_i \delta^{\alpha_i\beta_i}$. The remaining soft matrix element is given by
\begin{widetext}
\begin{align} \label{eq:softcolor}
\vec{C}_{N-1}^{\dagger\,\beta_t \dotsb \beta_N}
\Mae{0}{
\hY^{\dagger\, \beta_t \dotsb \beta_N \vert \beta_t' \dotsb \beta_N'}
\hX^{\dagger\, \beta_t' \dotsb \beta_N' \vert \alpha_1'' \alpha_2'' \dotsb \alpha_N'' }
\cM_N^\usoft \cM^\csoft
\hX^{\alpha_1'' \alpha_2'' \dotsb \alpha_N'' \vert \alpha_t' \dotsb \alpha_N'}
\hY^{\alpha_t' \dotsb \alpha_N' \vert \alpha_t \dotsb \alpha_N} }{0}
\, \vec{C}_{N-1}^{\, \alpha_t \dotsb \alpha_N}
,\end{align}
where we explicitly wrote out the color indices in the product of csoft and usoft Wilson lines. From \eq{Xopcolor} we know that $\hX$ is diagonal in color except for the $1,2,t$ subspace, so the product $\hX^{\dagger} \hX$ has only two nontrivial color indices $\beta_t'\vert\alpha_t'$. The only object we can form from these is $\delta^{\beta_t' \alpha_t'}$, which implies that the csoft matrix element is entirely color diagonal,
\begin{equation} \label{eq:Spluscolor}
\Mae{0}{ \bigl[\hX^\dagger \cM^\csoft(k_1, k_2)\, \hX \bigr]^{\beta_t \dotsb \beta_N \vert \alpha_t \dotsb \alpha_N} }{0}
= S_+(k_1, k_2)\, \id^{\beta_t \ldots \beta_N \vert \alpha_t \ldots \alpha_N}
\,.\end{equation}
The csoft function $S_+$ is the same as in the $3$-jet case,
\begin{equation} \label{eq:Splusgeneral}
S_+^\kappa(k_1, k_2,\mu)
= \frac{1}{c^\kappa}\, \tr \Mae{0}{\bar{T} \bigl[V_{\nt}^{\dagger} \T_t^\dagger\, \X_{n_2}^\dagger \X_{n_1}^\dagger]\, \cM^\csoft(k_1, k_2)\, T\bigl[ \X_{n_1} \X_{n_2} \T_t V_\nt\bigr] }{0}
\,,\end{equation}
where we restored the proper time-ordering, the trace is over color indices, and the color normalization constant, $c^\kappa$, is such that at tree level $S_+^\kappa(k_1, k_2) = \delta(k_1)\,\delta(k_2)$. Like $C_+^\kappa$, the csoft function $S_+^\kappa$ is universal and only depends on the color representations of the partons $1,2,t$ involved in the $1\to 2$ splitting. The explicit form for $q_t\to q_1 g_2$ was given in \eq{softdefs}. Using \eq{Spluscolor} in \eq{softcolor}, the remaining usoft matrix element yields the usoft function
\begin{equation} \label{eq:softN}
\hS^\kappa_{N-1}(\{k_i\}, \mu)
= \frac{1}{c^\kappa_{N-1}}
\Mae{0}{\bar{T} \bigl[ \hY^\dagger \bigr] \cM_N^\usoft(\{k_i\})\, T \bigl[\hY \bigr] }{0}
\,,\end{equation}
which is a matrix in $N+1$-parton color space, and the color normalization factor, $c^\kappa_{N-1}$, is such that at tree level $\hS^\kappa_{N-1}(\{k_i\}) = \id \prod_i\delta(k_i)$.

Having discussed the color structure, assembling the full factorization theorem for the $N$-jet case now follows the usual steps. For the cross section differential in the $\tN_i$, the small dijet invariant mass $t$, and the energy fraction $z$, we find
\begin{align}
\label{eq:factSCETpN}
&\frac{\df \sigma}{\df\tN_a\,\df\tN_b\, \df \tN_1 \dotsb \df \tN_{N}\, \df t \, \df z}
\nn\\* &\quad
= \int \! \df^4 q \, \df\Phi_L(q) \int \! \df \Phi_N(\{q_i\})\, M_N(\Phi_N,\Phi_L) \,
(2\pi)^4 \delta^4\Bigl(q_a + q_b - \sum_{i} q_i - q \Bigr)\, \delta \bigl(t- s_{12} \bigr)\, \delta\Bigl(z-\frac{E_1}{E_1+E_2} \Bigr)
\nn\\* &\qquad \times
\sum_{\kappa} \int \! \df x_a \df x_b \int\! \df s_a \df s_b\,  B_{\kappa_a}(s_a ,x_a, \mu)\,  B_{\kappa_b}(s_b ,x_b,\mu) \prod_i \int\! \df s_i\,  J_{\kappa_i}(s_i, \mu)\,
\abs{C_+^\kappa (t,z,\mu)}^2 \int\!\df k_1\,\df k_2\, S^{\kappa}_+ (k_1, k_2, \mu)
\nn \\* &\qquad \times
\vC_{N-1}^{\kappa \dagger} (\Phi_N, \Phi_L, \mu)\,
\widehat{S}^\kappa_{N-1}\Bigl(\tN_1 - \frac{s_1}{Q_1} - k_1, \tN_2 - \frac{s_2}{Q_2} - k_2, \tN_a - \frac{s_a}{Q_a}, \ldots, \tN_N - \frac{s_N}{Q_N}, \mu \Bigr) \vC_{N-1}^{\kappa} (\Phi_N, \Phi_L, \mu)
\,.\end{align}
\end{widetext}
Here we have included the possibility for the hadronic system to recoil against a color-singlet final state with total momentum $q$ and internal phase space $\Phi_L(q)$. This allows us, for example, to describe $W/Z$ + jets at the LHC. The massless $N$-jet phase space for the $N$ final-state jets is denoted as $\Phi_N(\{ q_i \})$. The incoming momenta $q_{a,b}$ are given by $q_{a,b} = x_{a,b} E_\mathrm{cm}(1,\pm \hat z)/2$, where $E_\mathrm{cm}$ is the total hadronic center-of-mass energy, $\hat z$ points along the beam axis, and $x_{a,b}$ are the light-cone momentum fractions of the colliding hard partons. The restriction $M_N$ enforces any phase space cuts on the final state, such as requiring that the jets be energetic and only one dijet invariant mass be small. The sum over $\kappa$ again runs over all relevant partonic channels.  The jet functions, $J_{\kappa_i}$, and beam functions, $B_{\kappa_{a,b}}$, arise from the collinear matrix elements as described above. The beam functions depend on the momentum fractions $x_{a,b}$ and describe the collinear initial-state radiation of the incoming hard partons~\cite{Stewart:2009yx, Stewart:2010qs}, whereas the jet functions describe collinear final-state radiation of the outgoing hard partons and are the same as in the $3$-jet case.

Since each of the ingredients in \eq{factSCETpN} only depends on a single scale, the standard RG procedure can now be followed to resum the large logarithms in the cross section, by evaluating each function at its natural scale and then evolving all functions to a common scale $\mu$. The solutions to the RG equations are straightforward generalizations of the explicit results presented for the case of $e^+e^-\to 3$ jets in \sec{Applications}.

\subsection{Consistency}
\label{subsec:njetsConsistency}

Having derived the factorization theorem for $pp\to N$ jets, we now discuss its renormalization group consistency. We already know the structure of the anomalous dimensions of the new objects $C_+$ and $S_+$ from \subsec{Consistency3jets}. Nevertheless, it is instructive to see how they fit together with the anomalous dimensions of $\vC_{N-1}$ and $\hS_{N-1}$ including the more complicated color structure now.

We will follow the same logic in our discussion here as in \subsec{Consistency3jets}. Our starting point is the factorized $pp\to N$-jet cross section for $N$-jettiness in \SCETI given in Ref.~\cite{Jouttenus:2011wh}, which has the form
\begin{equation} \label{eq:sigmaNjet}
\sigma_N \sim \sum_{\kappa} \biggl[ B_{\kappa_a} B_{\kappa_b} \prod_{i=1}^N J^\kappa_i \biggr]
\otimes \bigl[ \vC_{N}^{\kappa \dagger} \hS^\kappa_N \vC_{N}^{\kappa} \bigr]
\,.\end{equation}
We then argue that the combined anomalous dimensions of $C_+ \vC_{N-1}$ and $S_+ \hS_{N-1}$ must reproduce those of $\vC_N$ and $\hS_N$ in the limit $t\ll s_{ij}$. Using this, we rederive the general form of the anomalous dimensions for $C_+$ and $S_+$, reproducing those found in \subsec{Consistency3jets}.

\subsubsection{Consistency of the Hard Functions to Determine $\gamma_{C_+}$}

The Wilson coefficient $\vC_N$ in \eq{sigmaNjet} is obtained from
\begin{equation}\label{eq:scet1match}
\cM_\mathrm{QCD}(2 \to N)= \mae{N}{\vec{O}^\dagger_{N}(\mu)}{2} \vC_N (\{ s_{ij} \}, \mu)
\,,\end{equation}
and given by the IR-finite terms of $\cM_\mathrm{QCD}(2\to N)$, since as before the loop corrections to the bare matrix element of $\vO_N$ vanish in pure dimensional regularization. Thus, from \eq{NjetHardmatching} and the discussion below it, it follows that
\begin{equation}
\vC_N(\{s_{ij} \},\mu) \Big\vert_{t \ll s_{ij}}
 = \T_t \,\vC_{N-1}(\{ s_{ij}\}, \mu)\,  C_+(t,x,\mu)
\,.\end{equation}
Note that $\vec{C}_N$ and $\vec{C}_{N-1}$ are vectors in the color space of $2+N$ and $2+N-1$ external partons, respectively. We can write this with explicit color indices as
\begin{equation}
\vC_{N}^{\alpha_1\alpha_2 \cdots \alpha_N}
= \T_t^{ \alpha_1 \alpha_2 \vert \alpha_t}  \vC_{N-1}^{\alpha_t \dotsb \alpha_N} \,C_+ (t,x,\mu)
\,. \end{equation}
Taking the derivative with respect $\mu$ and using
\begin{equation}
\mu \frac{\df}{\df \mu} \vec{C}_N   =  \widehat{\gamma}_{C_N} \vec{C}_N
\,,\end{equation}
we get
\begin{align} \label{eq:NgammaCrelation}
\mu \frac{\df}{\df \mu} C_+ (t,x,\mu)
&= \gamma_{C_+} (t,x,\mu) \, C_+ (t,x,\mu)
\,,\nn\\
\gamma_{C_+} (t,x,\mu)\, \id
&= \frac{1}{C_t} \T_t^\dagger \,   \widehat{\gamma}_{C_N} (\{ s_{ij}\} , \mu)\, \T_t \Big\vert_{t \ll s_{ij}}
\nn \\* &\quad
- \widehat{\gamma}_{C_{N-1}} (\{ s_{ij}\} , \mu)
\,,\end{align}
where $\T_t^\dagger \T_t = \id\, C_t $, with $C_t$ determined by the parent parton splitting to $1$ and $2$, $C_q=C_{\bar{q}} =C_F$ and $C_g=C_A$. Since $C_+$ is a scalar function and does not know about the color space, the difference between the anomalous dimensions in the second equation must be proportional to the color identity.

The $\mu$-dependence of $\widehat{\gamma}_{C_N}(\mu)$ has the all-order structure~\cite{Chiu:2008vv, Becher:2009cu, Gardi:2009qi, Becher:2009qa, Dixon:2009ur}
\begin{align}
\widehat{\gamma}_{C_{N}}(\mu)
&= -\Gamma_\cusp [\alpha_s(\mu)]  \sum_{i\neq j} \frac{\T_i \cdot\T_j}{2}
\nn \\* &\quad \times
\ln \frac{ (-1)^{\Delta_{ij} } s_{ij} - \img 0}{\mu^2}
+ \widehat{\gamma}_{C_{N} } [\alpha_s(\mu)]
\,,\end{align}
where $\Delta_{ij}=1$ if both $i$ and $j$ are incoming or outgoing and $\Delta_{ij} = 0$ otherwise; $s_{ij}$ is always positive. We do not make additional assumptions about the all-order structure of the noncusp term $\widehat{\gamma}_{C_{N}}[\alpha_s]$, which can in general be a matrix in color space. Beyond two loops and for four or more external directions it can also depend on the $s_{ij}$ through conformal cross ratios of the form $s_{ij} s_{kl}/ s_{ik}s_{jl}$~\cite{Gardi:2009qi, Dixon:2009ur}. Hence, we can write $\widehat{\gamma}_{C_{N-1}}$ as
\begin{widetext}
\begin{align}\label{eq:gammaCN-1}
\widehat{\gamma}_{C_{N-1}}(\mu)
&= -\Gamma_\cusp[\alpha_s(\mu)]  \biggl\{ \sum_{i\neq1,2} \T_t \cdot \T_i \ln\frac{(-1)^{\Delta_{ti}} s_{ti} - \img 0}{\mu^2}
+ \sum_{i \neq j \neq1,2} \frac{\T_i\cdot \T_j}{2} \, \ln\frac{(-1)^{\Delta_{ij}} s_{ij} - \img 0}{\mu^2} \biggr\} +\widehat{\gamma}_{C_{N-1} } [\alpha_s(\mu)]
\,.\end{align}
Using \eqs{Nsij}{Nsijz} to expand in the limit $t \ll s_{ij}$, we find
\begin{align} \label{eq:gammaCN}
\frac{1}{C_t} \T_t^\dagger \, \widehat{\gamma}_{C_{N}}(\mu)  \T_t \Big\vert_{t \ll s_{ij}}
&= -\Gamma_\cusp[\alpha_s(\mu)]  \biggl\{
\sum_{i\neq 1,2} \T_t \cdot \T_i \ln \frac{(-1)^{\Delta_{t i}} s_{ti} - \img 0}{\mu^2}+
\sum_{i\neq j\neq 1,2} \frac{\T_i\cdot \T_j}{2}\, \ln\frac{(-1)^{\Delta_{ij}} s_{ij} - \img 0}{\mu^2}  \nn   \\ & \quad
+ \frac{1}{2} \Bigl( \T_t^2 -\T_1^2-\T_2^2 \Bigr) \ln \frac{- t - \img 0}{\mu^2}
- \frac{1}{2} \Bigl( \T_t^2 +\T_1^2-\T_2^2 \Bigr) \ln x
- \frac{1}{2} \Bigl( \T_t^2 -\T_1^2+\T_2^2 \Bigr)  \ln(1-x)
\biggr\}
\nn   \\ & \quad
+ \frac{1}{C_t} \T_t^\dagger \, \widehat{\gamma}_{C_{N} } [\alpha_s(\mu)]\, \T_t
\,,\end{align}
where we have used the identity $\sum_{i\neq1,2}\T_i = -(\T_1+\T_2)$ and the following relations
\begin{align}\label{eq:TtRelations}
\frac{1}{C_t}  \T_t^\dagger (\T_1+\T_2) \cdot \T_i\, \T_t = \T_t \cdot \T_i
\,,  \qquad
\frac{1}{C_t}  \T_t^\dagger\, \T_1 \cdot \T_2\, \T_t
= \frac{1}{2} \bigl( \T_t^2-\T_1^2-\T_2^2 \bigr)
\,,\end{align}
to write the contribution of $\widehat{\gamma}_{C_{N}}$ entirely in $2+N-1$ parton color space. Equation~\eqref{eq:TtRelations} generalizes \eq{3jetTt} to the $N$-jet case. Taking the difference of \eqs{gammaCN}{gammaCN-1}, we obtain the result for $\gamma_{C_+}$ valid to all orders in perturbation theory,
\begin{align} \label{eq:gammaC+allOrd}
\gamma_{C_+}(t,x,\mu)\,\id
&= -\Gamma_\cusp[\alpha_s(\mu)]
\frac{1}{2} \Bigl( \T_t^2 -\T_1^2-\T_2^2 \Bigr) \ln \frac{- t - \img 0}{\mu^2}
+ \gamma_{C_+ } [\alpha_s(\mu)]\,\id
\,,\nn\\
\gamma_{C_+ } [\alpha_s]\,\id
&= -\Gamma_\cusp[\alpha_s] \biggl\{
-\frac{1}{2} \Bigl( \T_t^2 +\T_1^2 -\T_2^2 \Bigr) \ln x
-\frac{1}{2} \Bigl( \T_t^2 -\T_1^2+\T_2^2 \Bigr)  \ln(1-x)
\biggr\}
\nn \\ &\quad
+ \bigl( \gamma_C^1[\alpha_s]+\gamma_C^2[\alpha_s] -\gamma_C^{t}[\alpha_s] \bigr)\,\id
\,.\end{align}
Here we have used that the difference of the noncusp terms in the collinear limit must be diagonal in color to all orders since as above in \eq{NgammaCrelation} $\gamma_{C_+}[\alpha_s]$ must be color diagonal,
\begin{align} \label{eq:gammaCNcond}
\frac{1}{C_t} \T_t^\dagger \, \widehat{\gamma}_{C_{N} } [\alpha_s] \T_t
- \widehat{\gamma}_{C_{N-1} } [\alpha_s] \Big\vert_{t \ll s_{ij}}
=  \bigl( \gamma_C^1[\alpha_s]+\gamma_C^2[\alpha_s] -\gamma_C^{t}[\alpha_s]  \bigr)\,\id
\,.\end{align}
This condition provides a nontrivial constraint on the collinear limit of $\widehat\gamma_{C_N}$ and is equivalent to what was used in Refs.~\cite{Becher:2009cu, Becher:2009qa} in deriving constraints on the form of $\widehat\gamma_{C_N}$ beyond two loops. (It cannot be spoiled by conformal cross ratios appearing in $\widehat\gamma_{C_N}$, which would require at least four distinct momenta and thus cannot be reduced to functions of $t$ and $x$ only.)

To see that \eq{gammaC+allOrd} agrees with the result for $\gamma_{C_+}$ in \eq{gammaC3Plus}, we note that in the three-parton color space used in \eq{gammaC3Plus} we have $\T_3 = -(\T_1 + \T_2)$. Equivalently, we can write the color factors in \eq{gammaC+allOrd} in the external $N+2$ color space using the inverse of \eq{TtRelations}, which amounts to the replacements
\begin{equation}
\frac{1}{2} \Bigl( \T_t^2 -\T_1^2-\T_2^2 \Bigr) \to \T_1\cdot \T_2
\,,\quad
\frac{1}{2} \Bigl( \T_t^2 + \T_1^2-\T_2^2 \Bigr) \to \T_1\cdot(\T_1 + \T_2)
\,,\quad
\frac{1}{2} \Bigl( \T_t^2 -\T_1^2+\T_2^2 \Bigr) \to  \T_2\cdot(\T_1 + \T_2)
\,.\end{equation}
With these replacements, \eq{gammaC+allOrd} agrees with the anomalous dimensions of the splitting amplitudes given in Ref.~\cite{Becher:2009qa}. Note that $\gamma_{C_+}$ in $N+2$ parton color space is not color diagonal but essentially proportional to $\T_t \T_t^\dagger$. Working in the $N+1$ parton color space has the advantage that $\gamma_{C_+}$ is manifestly color diagonal as in \eq{gammaC+allOrd}.

\subsubsection{Consistency of the Soft Functions to Determine $\gamma_{S_+}$}

The cross section for $pp\to N$ jets in \SCETp must reproduce the corresponding cross section in \SCETI expanded in the limit $t \ll s_{ij}$. Since the product of hard matching coefficients in \SCETp, $C_+ \vC_{N-1}$ reproduces $\vec{C}_N$ in \SCETI, as we saw in the previous section, and the beam and jet functions are the same in both theories, the soft functions must satisfy
\begin{align}
\hS_{N}(\{ k_i \},\mu) \Big\vert_{\hs_t \ll \hs_{ij} }
= \frac{1}{C_t^2} \int \df k'_1 \df k'_2 \, S_+(k'_1,k'_2,\mu) \, \T_t \,  \hS_{N-1} ( k_1 -k'_1,k_2 -k'_2,k_a,\ldots,k_N,\mu) \,  \T_t^\dagger
\,, \end{align}
which is the $N$-jet generalization of \eq{softconsistency}. Taking the derivative with respect to $\mu$ and using the known evolution equations for $\hS_{N}$ in \SCETI from Ref.~\cite{Jouttenus:2011wh}, we find
\begin{align} \label{eq:gammSNrelation}
\mu \frac{\df}{\df\mu} S_+ (k_1,k_2,\mu)
&=  \int\! \df k'_1 \, \df k'_2\, \gamma_{S_+}(k_1-k'_1, k_2-k'_2,\mu)\, S_+(k'_1,k'_2,\mu)
\,,\\\nn
\gamma_{S_+}(k_1,k_2,\mu)\,\id \prod_{i\neq 1,2}\delta(k_i)
&= \frac{1}{C_t} \T_t^\dagger \, \frac{1}{2}\bigl[\widehat{\gamma}_{S_N}(\{k_i \},\mu) + \widehat{\gamma}_{S_N}^\dagger(\{k_i \},\mu)\bigr] \,\T_t \Big\vert_{\hat{s}_t \ll \hat{s}_{ij}}
 - \frac{1}{2}\bigl[\widehat{\gamma}_{S_{N-1}} (\{k_i \},\mu) + \widehat{\gamma}^\dagger_{S_{N-1}} (\{k_i \},\mu)\bigr]
\,. \end{align}
The $\T_t^\dagger \widehat{\gamma}_{S_N} \T_t$ projects the anomalous dimension of $\widehat{S}_N$ on the $2+N-1$ parton color space. Similar to what we saw for $C_+$, the difference on the right-hand side of the second equation must be color diagonal, since $S_+$ knows nothing about the larger color space.

We can now use the all orders form of the $N$-jettiness soft anomalous dimension derived in Ref.~\cite{Jouttenus:2011wh},
\begin{align}
\widehat{\gamma}_{S_N}(\{ k_i\},\mu) = -2 \Gamma_\cusp[ \alpha_s(\mu)] \sum_{i \neq j} \T_i \cdot  \T_j
\biggl[ \frac{1}{\sqrt{\hs_{ij} }\mu}\cL_0 \Bigl( \frac{k_i}{\sqrt{\hs_{ij}}\mu} \Bigr)  -\frac{\img \pi}{2}\Delta_{ij} \, \delta(k_i) \biggr]\prod_{m\neq i} \delta(k_m)
+ \widehat{\gamma}_{S_N} [\alpha_s(\mu)] \prod_j \delta(k_j)
\, , \end{align}
to determine $\widehat{\gamma}_{S_N}$ in the limit $\hs_t\ll \hs_{ij}$, which corresponds to setting $\hs_{1i} = \hs_{2i} = \hs_{Qi}$ [see \eq{NhatSij}]. It is then straightforward to extract the anomalous dimension of $S_+$ using \eq{gammSNrelation}. We obtain
\begin{align} \label{eq:gammaS+allOrd}
\gamma_{S_+} (k_1,k_2,\mu) \, \id
&= -2 \Gamma_\cusp[ \alpha_s(\mu)]
\frac{1}{2} (  \T_t^2 -\T_1^2-\T_2^2) \frac{1}{\sqrt{\hs_t}\mu}
\biggl[ \cL_0 \Bigl( \frac{k_1}{\sqrt{\hs_t}\mu} \Bigr) \delta(k_2)
+  \cL_0 \Bigl( \frac{k_2}{\sqrt{\hs_t}\mu} \Bigr) \delta(k_1) \biggr]
+ \gamma_{S_+} [\alpha_s(\mu), k_1,k_2] \,\id
\,,\nn\\
\gamma_{S_+}[\alpha_s, k_1,k_2] \, \id
&= - \Gamma_\cusp[\alpha_s] (\T_2^2 -\T_1^2) \frac{1}{\xi} \biggl[ \cL_0\Bigl( \frac{k_1}{\xi} \Bigr) \delta(k_2) - \cL_0\Bigl( \frac{k_2}{\xi}  \Bigr)\delta(k_1) \biggr]
\nn \\ & \quad
- \bigl(\gamma_J^1[\alpha_s] + 2\gamma_C^1[\alpha_s] + \gamma_J^2[\alpha_s] + 2\gamma_C^2[\alpha_s] - \gamma_J^t[\alpha_s] - 2\gamma_C^t[\alpha_s]
\bigr)\,\id\,\delta(k_1)\,\delta(k_2)
\,.\end{align}
This result agrees with \eq{gammaSC} for $e^+ e^- \to 3$ jets as it must since the csoft function is universal for a given splitting channel.

To arrive at \eq{gammaS+allOrd}, we have used that the difference of the noncusp terms in $\hga_{S_N}$ and $\hga_{S_{N-1}}$ in the collinear limit must be color diagonal because $\gamma_{S_+}$ is, i.e.,
\begin{align}
\frac{1}{C_t} \T_t^\dagger \, \widehat{\gamma}_{S_N} [\alpha_s] \,\T_t
- \widehat{\gamma}_{S_{N-1}}[\alpha_s] \Big\vert_{\hs_t \ll \hs_{ij}}
= - \bigl(\gamma_J^1[\alpha_s] + 2\gamma_C^1[\alpha_s] + \gamma_J^2[\alpha_s] + 2\gamma_C^2[\alpha_s] - \gamma_J^t[\alpha_s] - 2\gamma_C^t[\alpha_s]
\bigr) \, \id
\,.\end{align}
This condition is equivalent to \eq{gammaCNcond}, since $\hga_{S_N}$ and $\hga_{C_N}$ are related by consistency. We have also used the relations in \eq{TtRelations} to rewrite the different color structures appearing in $\hga_{S_N}$. The terms proportional to $\T_1\cdot \T_2$ in $\hga_{S_N}$ directly give the terms proportional to $(\T_t^2 - \T_1^2 - \T_2^2)/2$ in the first line of \eq{gammaS+allOrd}. The terms in $\hga_{S_N}$ involving neither $\T_1$ nor $\T_2$ immediately cancel with the corresponding ones in $\hga_{S_{N-1}}$. The nontrivial terms in $\hga_{S_N}$ are those proportional to $\T_1\cdot \T_i$ and $\T_2\cdot \T_i$ with $i\neq 1,2$, which we can rewrite as
\begin{align} \label{eq:gammaSNT1Ti}
&\qquad
\T_1\cdot\T_i\, \frac{1}{\sqrt{\hs_{Qi}}\, \mu}
\biggr[ \cL_0\Bigl(\frac{k_1}{\sqrt{\hs_{Qi}}\,\mu}\Bigr) \delta(k_i) + \cL_0\Bigl(\frac{k_i}{\sqrt{\hs_{Qi}}\,\mu}\Bigr) \delta(k_1)
\biggr] \prod_{j\neq 1,i} \delta(k_j)
\nn\\ & \qquad
+ \T_2\cdot\T_i \frac{1}{\sqrt{\hs_{Qi}}\, \mu}
\biggr[ \cL_0\Bigl(\frac{k_2}{\sqrt{\hs_{Qi}}\,\mu}\Bigr) \delta(k_i) + \cL_0\Bigl(\frac{k_i}{\sqrt{\hs_{Qi}}\,\mu}\Bigr) \delta(k_2)
\biggr] \prod_{j\neq 2,i} \delta(k_j)
\nn\\ & \quad
= (\T_1 + \T_2)\cdot \T_i\, \frac{1}{\sqrt{\hs_{Qi}}\, \mu}\biggr[
\frac{1}{2}\cL_0\Bigl(\frac{k_1}{\sqrt{\hs_{Qi}}\,\mu}\Bigr) \delta(k_2)
+ \frac{1}{2}\cL_0\Bigl(\frac{k_2}{\sqrt{\hs_{Qi}}\,\mu}\Bigr) \delta(k_1)
\biggr] \prod_{j \neq 1,2}\delta(k_j)
\nn\\ & \qquad
+ (\T_1 + \T_2)\cdot \T_i\, \frac{1}{\sqrt{\hs_{Qi}}\, \mu}
\cL_0\Bigl(\frac{k_i}{\sqrt{\hs_{Qi}}\,\mu}\Bigr) \prod_{j\neq i}\delta(k_j)
\nn\\ & \qquad
+ \frac{1}{2}(\T_1 - \T_2)\cdot \T_i\, \frac{1}{\xi}
\biggr[
\cL_0\Bigl(\frac{k_1}{\xi}\Bigr) \delta(k_2) - \cL_0\Bigl(\frac{k_2}{\xi}\Bigr) \delta(k_1)
\biggr] \prod_{j \neq 1,2}\delta(k_j)
\,.\end{align}
\end{widetext}
After projecting $(\T_1+\T_2)\cdot \T_i \to \T_t\cdot \T_i$, the first two lines on the right-hand side cancel against $\hga_{S_{N-1}}$. To see this first note that as in the $3$-jet case, our $\hS_{N-1}(k_1, k_2, \ldots)$ is related to the usual $\hS_{N-1}(k_t, \ldots)$ from Ref.~\cite{Jouttenus:2011wh} by projecting onto $k_t = k_1 + k_2$. To determine how the $k_t$ measurement is split between jets $1$ and $2$ notice that for general $N$, $\hS_{N-1}(k_1, k_2, \ldots)$ is no longer symmetric under $k_1\leftrightarrow k_2$, because the boundary to the $i$th jet region can be different for jet regions $1$ and $2$. However, the fact that the anomalous dimension of $S_+$ must be color diagonal requires that the first line on the right-hand side cancels against $\hga_{S_{N-1}}$. This tells us that the contribution to the anomalous dimension of $\hS_{N-1}(k_t, \ldots)$ involving $\T_t\cdot\T_i\,\cL_0(k_t)$ must again be split up symmetrically into $\T_t\cdot\T_i[\cL_0(k_1)\delta(k_2) + \cL_0(k_2)\delta(k_1)]/2$ in  $\hga_{S_{N-1}}$ in our case, just as for $\gamma_{S_2}$ in \eq{gammaS2}.%
\footnote{To see that $\gamma_{S_{N-1}}$ must still be symmetric under $k_1\leftrightarrow k_2$, we can split up the phase space region associated to jets $1$ and $2$ into a $\phi_t(p)$ independent region enclosing $\hat{q}_1$ and $\hat{q}_2$ and the UV-finite difference to the boundaries of the $i$th jet region. In Ref.~\cite{Jouttenus:2011wh} it was demonstrated at one loop that the UV divergences associated with jet regions $i$ and $j$ only arise from $i$,$j$ hemispheres, whereas the contributions depending on the boundary with the remaining jet directions are UV finite and do not affect the anomalous dimension.}
Finally, since the last line on the right-hand side of \eq{gammaSNT1Ti} has no dependence on $i$ other than $\T_i$ itself, we can sum over $i$ using $\sum_{i\neq 1,2}\T_i = -(\T_1 + \T_2)$. This yields the term proportional to $\T_2^2 - \T_1^2$ in \eq{gammaS+allOrd}.

\section{Numerical Results}
\label{sec:Applications}

In this section we present some numerical results for our example of $e^+e^-\!\to 3$ jets. For simplicity, we project onto the total $3$-jettiness of the event [see \eq{TauNsum}],
\begin{equation} \label{eq:Tau3def}
\tN \equiv \tN_{N=3} = \tN_1 + \tN_2 + \tN_3
\,.\end{equation}
Just as thrust characterizes how 2-jet-like an event is, the total $N$-jettiness of an event characterizes how $N$-jet-like an event is, and can be used as a veto against additional jets. The factorized cross section for $\tN$ is obtained from \eq{factSCETp} by projecting onto $\tN$ using \eq{Tau3def},
\begin{align} \label{eq:sigmaTau3}
&\frac{\df\sigma}{\df\tN \df t \, \df z}
\\ & \quad
= \frac{\sigma_0}{Q^2}
\sum_{\kappa} H_2(Q^2, \mu) H_+^{\kappa} (t,z,\mu) \prod_i \int\!\df s_i\, J_{\kappa_i}(s_i, \mu)
\nn\\ & \qquad\times
\int\!\df k\, S_+(k, \mu)\, S_2\Bigl(\Tau - \frac{s_1}{Q_1}  - \frac{s_2}{Q_2}  - \frac{s_3}{Q_3} - k, \mu\Bigr)
\nn \,. \end{align}
The different hard, jet, and soft functions were discussed in detail in \sec{3jetNLO}. They are renormalized objects, and in the cross section must all be evaluated at the same scale $\mu$.
To resum the large logarithms in \eq{sigmaTau3} we have to evaluate each function at a scale where its perturbative series does not involve large logarithms and then RG evolve all functions to the common scale $\mu$,
\begin{align} \label{eq:sigmaTau3resum}
&\frac{\df\sigma}{\df\tN \df t \, \df z }
\nn\\ & \quad
= \frac{\sigma_0}{Q^2}
\sum_{\kappa}\, H_2(Q^2, \mu_{H_2})\, U_{H_2} (Q^2, \mu_{H_2}, \mu)
\nn\\ & \qquad\times
H_+^{\kappa} (t,z, \mu_{H_+})\, U_{H_+}^\kappa(t,z, \mu_{H_+}, \mu)\,
\nn\\ & \qquad\times
\prod_i \int\!\df s_i \df s_i'\, J_{\kappa_i}(s_i - s_i', \mu_J)\, U_{J_{\kappa_i}}(s_i', \mu_J, \mu)
\nn\\ & \qquad\times
\int\!\df k\, \df k'\, S_+(k - k', \mu_{S_+})\, U_{S_+}(k', \mu_{S_+}, \mu)
\nn\\ & \qquad\times
\int\!\df \ell\, S_2\Bigl(\Tau - \frac{s_1}{Q_1}  - \frac{s_2}{Q_2}  - \frac{s_3}{Q_3} - k - \ell, \mu_{S_2}\Bigr)
\nn\\ & \qquad\times
U_{S_2}(\ell, \mu_{S_2}, \mu)
\,.\end{align}
The evolution kernels $U_X$ for each function in \eq{sigmaTau3resum} are given in \app{resum}.  After RG evolving the hard, jet, and soft functions, we obtain a distribution with all large logarithms of $\tN$ and $m_{jj}$ resummed.

We are interested in the distribution in $m_{jj}$, for which we integrate the cross section over $\tN$ and $z$,
\begin{align} \label{eq:mjjdist}
\frac{\df \sigma}{\df m_{jj}}
= 2m_{jj} \int_{z_\cut}^{1 - z_\cut}\! \df z \int_0^{\tN_\cut} \! \df\tN\, \frac{\df\sigma}{\df \tN\, \df t\, \df z}
\,,\end{align}
where the factor of $2m_{jj}$ is the Jacobian from changing variables from $t = m_{jj}^2$ to $m_{jj}$, and we suppress the dependence of $\df\sigma/\df m_{jj}$ on $\Tau_\cut$ and $z_\cut$. For our numerical analysis, we choose
\begin{equation} \label{eq:kincuts}
Q = 500 \GeV
\,, \quad
\tN_\cut = 10 \GeV
\,, \quad
z_\cut = \frac13
\,.\end{equation}
The $Q_i$ are given in terms of $z$ and $t$ in \eq{Qi}. For this $z_\cut$, the ratio of nearby jet energies ranges between $0.5$ and $2$. For $\tN_\cut = 10\GeV$ the two nearby jets have a typical jet mass of $35\GeV$. In terms of $m_{jj}$ and $\Tau_\cut$, the scales for the different functions in \eq{sigmaTau3resum} have the scaling
\begin{align} \label{eq:canonicalscale}
\mu_{H_2} &\simeq Q
\,,\qquad
\mu_{H_+} \simeq m_{jj}
\,,\qquad
\mu_{J_i} \simeq \sqrt{Q_i \Tau_\cut}
\,,\nn\\
\mu_{S_+} &\simeq \frac{\Tau_\cut}{\sqrt{\hs_t}}
\,,\qquad
\mu_{S_2} \simeq \frac{\Tau_\cut}{\sqrt{\hs_Q}}
\,.\end{align}

The simultaneous resummation of logarithms of $\Tau_\cut$ and $m_{jj}$ is challenging because the hierarchy between the scales depends on several kinematic variables. At large $m_{jj}$ we are in the kinematic region of 3 equally separated jets. In this $3$-jet limit, power corrections in $m_{jj}^2 / Q^2$ become important and the resummation in $m_{jj}$ must be turned off, which requires $\mu_{H_+} \simeq \mu_{H_2}$ and $\mu_{S_+} \simeq \mu_{S_2}$. For any $m_{jj}$ we also have the requirement that $\mu_{H_+} \mu_{S_+}\sim \mu_{J_1} \mu_{J_2}$. To satisfy these requirements we choose the scales as follows:
\begin{align} \label{eq:profilescales}
\mu_H &= Q
\,,\quad&
\mu_{H_+}(m_{jj}) &= \mu_\mathrm{run}(\mu_H, m_{jj})
\,,\nn\\*
\mu_{J_i} &= \sqrt{Q_i \Tau_\cut}
\,,\nn\\*
\mu_{S_2} &= \frac{\Tau_\cut}{\sqrt{\hs_Q}}
\,,\quad&
\mu_{S_+}(m_{jj}) &= \frac{Q\, \mu_{S_2}}{\mu_\mathrm{run}(Q, m_{jj})}
\,,\end{align}
where $\hs_Q = Q^2/[Q_3(Q_1 + Q_2)]$. The scales $\mu_H$, $\mu_{J_i}$, and $\mu_{S_2}$ do not depend on $m_{jj}$ so we simply use their canonical scales in \eq{canonicalscale}. The $m_{jj}$-dependent scales $\mu_{H_+}(m_{jj})$ and $\mu_{S_+}(m_{jj})$ are given in terms of the profile function
\begin{align} \label{eq:murun}
\mu_\mathrm{run}(\mu, m_{jj}) &=
\begin{cases}
2a\, m_{jj} + \mu_{0} & m_{jj} < m_1\,, \\
\mu - a \dfrac{(m_{jj}-m_2)^2}{m_2 - m_1} &  m_1 < m_{jj} < m_2\,, \\
\mu & \quad m_{jj} > m_2
\,,\end{cases}
\nn\\
a &= \frac{\mu_{H_2} - \mu_{0}}{m_1 + m_2}
\,,\end{align}
such that they have a quadratic approach to $\mu_H$ and $\mu_{S_2}$ over the region $m_1< m_{jj} < m_2$. Analogous profile functions have been used previously in different contexts~\cite{Ligeti:2008ac, Abbate:2010xh, Berger:2010xi}. We choose the parameters of $\mu_\mathrm{run}$ as
\begin{equation}
m_1 = 200 \GeV
\,, \quad
m_2 = 300 \GeV
\,, \quad
\mu_{0} = 0 \textrm{ GeV}
\,.\end{equation}

To get a measure of how important the resummation in $m_{jj}$ is, we will compare to results with the resummation in $m_{jj}$ turned off. This is achieved by setting
\begin{equation} \label{eq:3jscales}
\mu_{H_+} = \mu_{H_2}\,, \quad \mu_{S_+} = \mu_{S_2}
\,,\end{equation}
with $\mu_{H_2}$, $\mu_{J_i}$, $\mu_{S_2}$ as in \eq{profilescales}. Note that the distribution without resummation in $m_{jj}$ is what one would obtain from directly using \SCETI with $3$ jet directions (up to the power corrections in $t/Q^2$ which we have not included).

The approach to the $2$-jet region where the two nearby jets merge into a single large jet is more complicated. This happens for $m_{jj} \lesssim m_j \sim 35\GeV$, i.e.\ when the dijet invariant mass becomes smaller than the individual jet masses. Here, $\mu_{H_+}$ and $\mu_{S_+}$ become equal to $\mu_{J_i}$ and eventually $\mu_{S_+}$ becomes larger than $\mu_{H_+}$. The $3$-jet observables $m_{jj}$ and $3$-jettiness are not meaningful anymore once the two close jets merge into each other. In addition, the proper theory would now be \SCETI with two collinear sectors. We leave a more detailed investigation of this transition to future work. For illustrative purposes, we will plot our results all the way down to $m_{jj} = 5\GeV$, where $\mu_{H_+}$ becomes equal to $\mu_{S_2}$.

A detailed numerical study is beyond the scope of this work. Below, we will present results at next-to-leading logarithmic (NLL) order, which uses the one-loop noncusp and two-loop cusp anomalous dimensions in the running and tree-level matching, and at NLL$'$ order, which combines the NLL running with the one-loop matching corrections. It is straightforward to extend our results to next-to-next-to-leading logarithmic (NNLL) order. To obtain a rough estimate of the perturbative uncertainties we vary $\mu_{H_2}$, $\mu_{J_i}$, $\mu_{S_2}$ by factors of two. By varying $\mu_{H_2}$ and $\mu_{S_2}$ we automatically let $\mu_{H_+}$ and $\mu_{S_+}$ vary accordingly. This tends to give the largest effect in the variation of $\mu_{H_+}$ and $\mu_{S_+}$, whereas varying the parameters in $\mu_\mathrm{run}$ within reasonable ranges has smaller effects. We then take the envelope from varying each of $\mu_{H_2}$, $\mu_{J_1}$, $\mu_{J_2}$, $\mu_{J_3}$, $\mu_{S_2}$ individually while keeping the other scales fixed and in addition from varying all of them up and down at the same time. The largest variation mostly comes from the individual variations of $\mu_{S_2}$ and the gluon jet scale.

\begin{figure}[t!]
\includegraphics[width=\columnwidth]{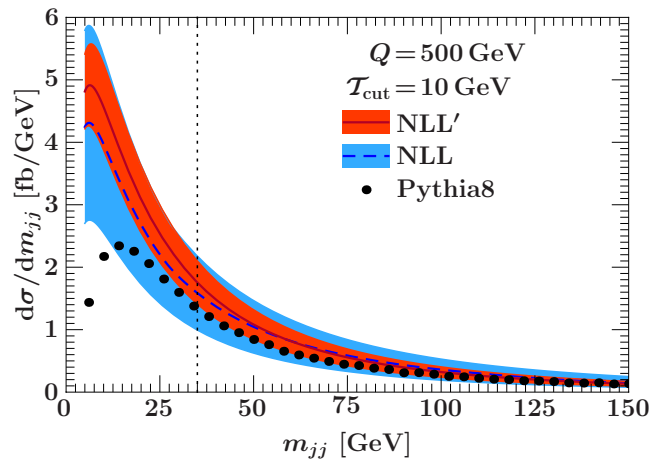}
\vspace{-0.5ex}
\caption{The distribution in $m_{jj}$ for $Q = 500\GeV$, $\Tau_\cut = 10\GeV$, and $z_\cut = 1/3$ as in \eq{kincuts}.  The bands show the perturbative scale uncertainties at NLL (light blue) and NLL$'$ (dark orange) as explained in the text, with the central values given by the center of the bands. The dots show the result obtained from Pythia. The dotted line indicates the value $m_{jj} = 35\GeV$ below which we enter the $2$-jet region where our expansion breaks down.}
\label{fig:dsdmjj}
\end{figure}

\begin{figure*}[t!]
\includegraphics[width=0.99\columnwidth]{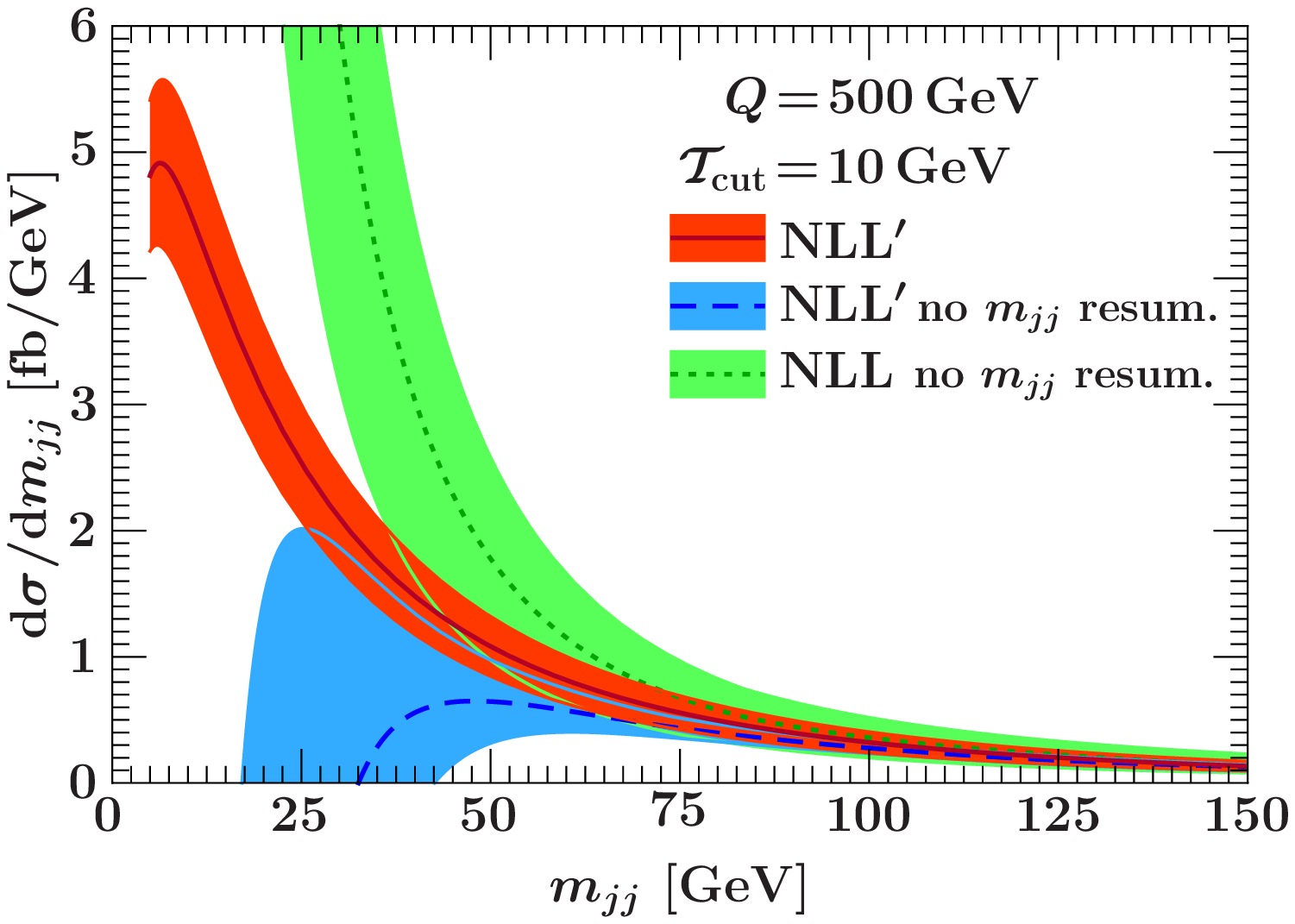}%
\hfill\includegraphics[width=1.05\columnwidth]{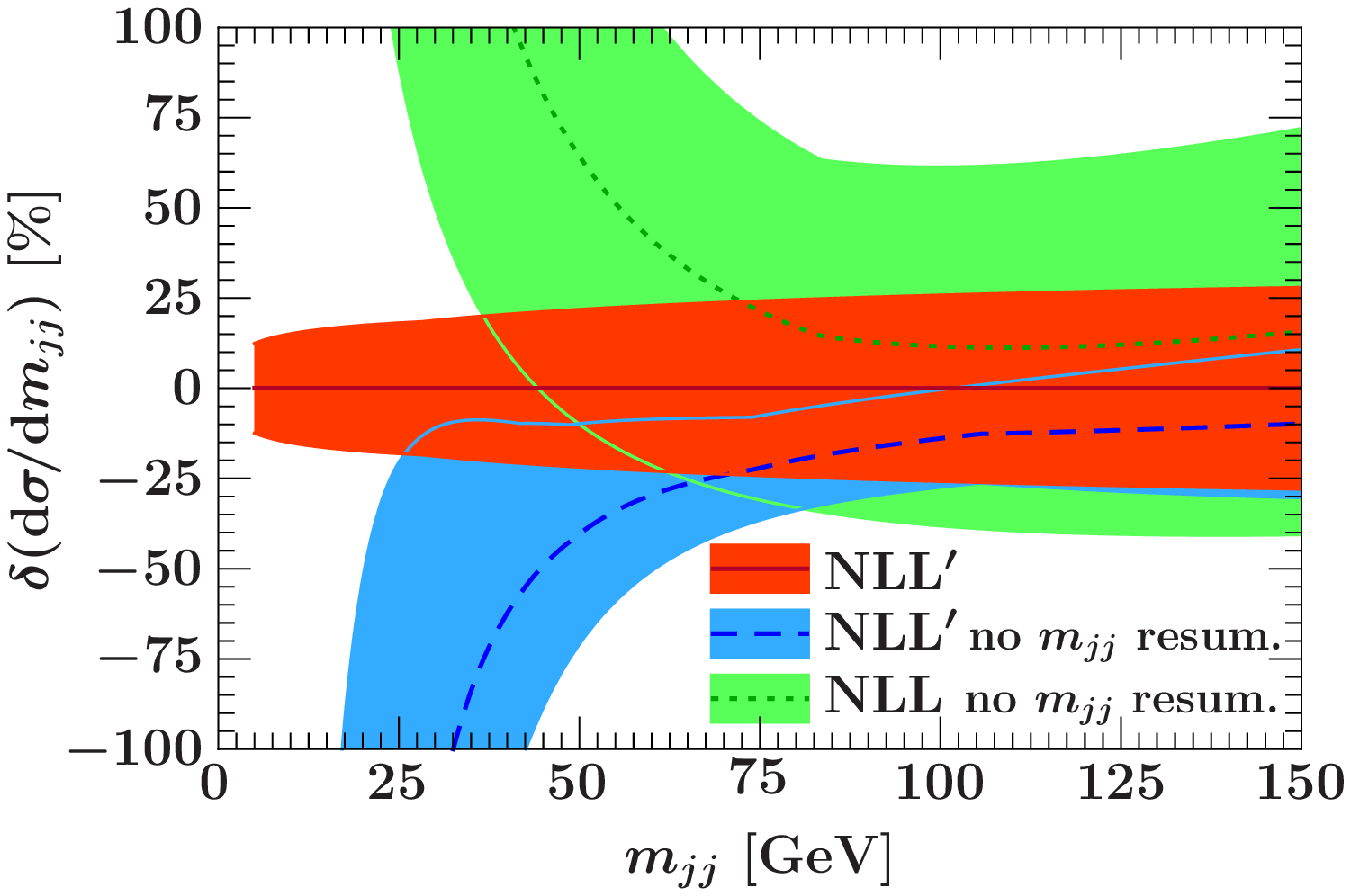}
\vspace{-0.5ex}
\caption{Left panel: The distribution in $m_{jj}$ for $Q = 500\GeV$, $\Tau_\cut = 10\GeV$, and $z_\cut = 1/3$ as in \eq{kincuts}.  Right panel: Same as the left plot but shown as the percent difference relative to the NLL$'$ central value. The bands show the perturbative scale uncertainties as explained in the text. The dark orange band shows our result including the resummation (resum.) in $m_{jj}$ at NLL$'$. The medium blue and light green bands show the results at NLL and NLL$'$ with the resummation only in $\Tau_\cut$ but not in $m_{jj}$.}
\label{fig:dsdmjjnomjj}
\end{figure*}

In \fig{dsdmjj}, we plot the NLL and NLL$'$ distributions from \eq{mjjdist} for our default scale choices.
The distribution is stable in going from NLL to NLL$'$, with the expected reduction in the scale uncertainties. The ninja region, which has the hierarchy of scales shown in \fig{SCET+_3jets}, corresponds to the region to the right of the dotted line in the plot. For smaller $m_{jj}$ we enter the $2$-jet region where our expansion breaks down. For $m_{jj} \gtrsim 150\GeV$ we enter the $3$-jet region where power corrections of order $m_{jj}/Q$ become important.

In \fig{dsdmjj} we also compare our resummed predictions to the distribution obtained from Pythia8~\cite{Sjostrand:2007gs}. To make the distribution in Pythia, we simulated $e^+e^-\! \to q\bar{q}$ events and determined the reference momenta $q_i$ from the first hard emission. For simplicity we use this as an approximate alternative instead of determining the $\hq_i$ by fully minimizing the $3$-jettiness for each event. (We found that using a jet algorithm to determine the reference momenta introduces a significant bias as the nearby jets get close to each other and the events become more $2$-jet-like. In this limit, jet algorithms tend to merge nearby clusters of high energy into a single jet~\cite{Ellis:2009me}, and hence become unsuitable to determine the $\hq_i$.) Pythia resums the kinematic logarithms of $m_{jj}$ in the parton shower, which is formally correct at leading-logarithmic order, but also includes various other effects that contribute at NLL.  We see that the Pythia distribution agrees well with both the NLL and NLL$'$ distributions even down to $m_{jj} \sim 25$ GeV. Note that Pythia is producing exclusive samples of events and does not report any systematic uncertainties. This makes it challenging to interpret the theory uncertainties in the distributions generated by the Monte Carlo.  Even though the NLL and NLL$'$ distributions are more accurate than the parton shower, since Pythia contains many NLL effects it is reasonable to take the uncertainty band of the NLL curve as a proxy for the systematic uncertainties in the Pythia distribution in the range of $m_{jj}$ where we trust our calculation.

In the left panel of \fig{dsdmjjnomjj}, we compare the NLL$'$ distribution including the resummation in $m_{jj}$ to the NLL$'$ and NLL distributions without $m_{jj}$ resummation.  In the right panel we show the same results in terms of the percent difference from the central NLL$'$ curve with $m_{jj}$ resummation. We see that for $m_{jj} \lesssim 75\GeV$ the resummation in $m_{jj}$ becomes important, where the results without $m_{jj}$ resummation become unstable due to the presence of unresummed $\ln(m_{jj}/Q)$.

\section{Conclusions}
\label{sec:Conclusions}

In this paper we have constructed a new effective field theory, \SCETp, which can describe exclusive multijet events at the LHC where the kinematic configuration of final-state jets give rise to a hierarchy of scales. We focused on the case where the dijet invariant mass $m_{jj}$ of a pair of jets is much smaller than all other dijet invariant masses, which are of order the total center-of-mass energy $Q$. This results in the presence of three separated scales $m \ll m_{jj} \ll Q$, where $m$ is the typical jet mass. We have shown that using \SCETp we can simultaneously and systematically resum the logarithms of both the individual jet mass, $\ln(m/Q)$, as well as the dijet mass, $\ln(m_{jj}/Q)$, to higher orders in perturbation theory, which has a wide range of applications.

Separating the additional kinematic scales that arise requires a successive hard matching from QCD onto \SCETI with $N-1$ jet directions and then onto \SCETp with $N$ jet directions. The matching from \SCETI onto \SCETp introduces a new hard matching coefficient, $C_+$, which encodes the splitting to two nearby jets, is universal for a given splitting channel, and depends only on the scale $m_{jj}$ to all orders in perturbation theory. We showed that it is given by the finite parts of the known universal splitting amplitudes of QCD in the collinear limit.

\SCETp is an extension of \SCETI with an additional csoft mode that has virtuality $m^4/m_{jj}^2$ and is necessary to describe the soft radiation between the nearby jets. The usoft modes of \SCETI have virtuality $m^4/Q^2$ and therefore cannot resolve the two nearby jets. Thus, \SCETp is required to properly separate the physics between these two scales. We constructed the Lagrangian of \SCETp and showed how to decouple the collinear, csoft, and usoft interactions. We discussed the structure of gauge invariance in \SCETp and used this to construct gauge invariant $N$-jet operators, which require new csoft Wilson lines. The factorization of the csoft and usoft interactions leads to the factorization of the regular soft function into a new universal csoft function, $S_+$, and a usoft function.

As an example of our new effective theory, we derived the factorization for the dijet invariant mass spectrum of $e^+ e^-\!\to 3$ jets, using $N$-jettiness to define the jets. We determined all contributions to the factorized cross section at NLO and derived the form of the anomalous dimensions of $C_+$ and $S_+$ to all orders in perturbation theory. We also extended our factorization theorem to $pp \to N$ jets plus leptons, discussing in detail the kinematic scales that appear and the nontrivial color structure for $N$ jets. We demonstrated that the new ingredients, $C_+$ and $S_+$, are color diagonal and independent of the number of additional jets in the process.

We gave explicit numerical results for the example of $e^+ e^-\!\to 3$ jets using $3$-jettiness, $\Tau$, resumming both the jet-mass logarithms of $\Tau/Q$ and the kinematic logarithms of $m_{jj}/Q$ to NLL'. We found that the resummation of $m_{jj}$ is important over a wide range. Our results show good agreement with Pythia, which resums the kinematic logarithms of $m_{jj}$ at leading order.

\SCETp has a wide range of applications, and expands our ability to make accurate theoretical predictions relevant for jet-based new-physics searches at the LHC. This includes cases where dijet invariant masses lead to a natural scale hierarchy, ranging from the study of jet substructure to improving parton-shower programs, which are currently the only tool available to resum logarithms of kinematic scales in exclusive jet cross sections.

\begin{acknowledgments}
This work was supported
by the Director, Office of Science, Offices of High Energy and Nuclear
Physics of the U.S.  Department of Energy under the Contract No.
DE-AC02-05CH11231 and by the Office of Nuclear Physics of the U.S.\ Department of
Energy under the Grant No. DE-FG02-94ER40818. J.W. was supported by
the National Science Foundation Grant No. PHY-0705682.
C.W.B. and S.Z. received support from the Early Career Award No. DE-AC02-05CH11231.
\end{acknowledgments}

\appendix

\section{All-Order Scale Sensitivity of the Hard Function}
\label{app:RPIMatchingHard}

We have argued that the matching coefficient $C_+(t, z, \mu)$, which arises from matching \SCETI to \SCETp, can only depend on the scale $t$, which is the scale where the matching is performed. In \subsec{HardFuncs} we have seen that at NLO $C_+$ contains only on a single dimensionful scale $t$. In this appendix, we prove that this holds to all orders in perturbation theory.  In \sec{njets} we argued that for a given parton channel ($q\to qg$, $g\to gg$, or $g\to q\bar{q}$) the matching coefficient $C_+$ is a universal function, and our arguments here apply to all splitting channels. The main tool we use is RPI~\cite{Manohar:2002fd}.  RPI restores the symmetry that is broken by choosing a particular basis of light-cone coordinates to label the collinear fields. For example, by choosing specific light-cone vectors $n$ and $\bn$ to define the $2$-jet operator $O_2$, we have picked out a specific direction in an equivalence class of directions close to the jet directions that would all give the same results. RPI transformations correspond to reparameterizing to a different light-cone basis within the equivalence class, and invariance under such reparametrizations effectively restores the broken symmetry. In Ref.~\cite{Baumgart:2010qf}, RPI has been used in a similar way to what we follow here to understand the matching between operators with different multiplicities in SCET.

There are three types of RPI transformations for each pair of basis vectors $\{n,\bn\}$.  For our argument in this section we are only concerned with what is known as RPI-III, under which each direction $n$ and its conjugate direction $\bn$ transform as
\begin{equation}
n_i^{\mu} \to e^{\alpha_i} n_i^{\mu}
\,, \qquad \bn_i^{\mu} \to e^{-\alpha_i} \bn_i^{\mu}
\,.\end{equation}
Invariance under this transformation implies that any occurrence of $n_i^\mu$ must be accompanied by an occurrence of $\bn_i^\mu$. Each jet momentum $q_i^{\mu}$ is RPI-III invariant, as it can be written as
\begin{equation}
q_i^{\mu} = \frac12 (\bn_i \cdot q_i) n_i^{\mu}
\,.\end{equation}
As discussed in Ref.~\cite{Marcantonini:2008qn}, the hard matching onto an $N$-jet operator in SCET is RPI invariant,
\begin{equation}
\cM(2\to N) = \int\!\Bigl[\prod_i \df \w_i\Bigr] \, C_N (\{\w_i\}) \, \mae{N}{O_N(\{\w_i\})}{2}
\,,\end{equation}
where $\w_i = \bn_i \cdot q_i$, and we ignore the color structure as it is irrelevant for this discussion. Since the operator $O_N$ is built out of gauge-invariant quark and gluon jet fields with given label momenta,
\begin{align}
\chi_{n,\w}
&= \bigl[ \delta(\w - \bn \cdot \Pop_n) \chi_n\bigr]
\,, \nn \\
\cB_{n, \w\perp}^\mu
&= \bigl[ \delta(\w - \bn \cdot \Pop_n) \cB_{n \perp}^\mu\bigr]
\,,\end{align}
it transforms under RPI-III as $\langle O_N(\{\w_i\}) \rangle \to \exp(\alpha_1 + \cdots +\alpha_N) \langle O_N(\{\w_i\}) \rangle$. It follows that both the matching coefficient, $C_N(\{\w_i\})$, and the operator matrix element together with the measure, $\langle O_N(\{\w_i\}) \rangle\prod_i\df \w_i$, are separately RPI-III invariant. In what follows we suppress the integral over the $\w_i$ labels.

Consider the matching of $O_{N-1}$ in \SCETI onto $O^+_N$ in \SCETp, evaluated by calculating the $2\to N$ matrix elements of both operators with external momenta $q_i$ such that $s_{12} / s_{ij} \sim \lambda_t^2$. In the \SCETI above the matching scale $t = s_{12}$, the final states with momenta $q_1$ and $q_2$ are described by the same $n_t$ collinear sector with scaling $Q (\lambda_t^2, 1, \lambda_t)$. Below this scale in \SCETp, the final states with $q_1$ and $q_2$ are described by two separate collinear sectors. The matching then takes the schematic form [see \eq{ON+match}]
\begin{equation}
\mae{N}{O_{N-1}}{2} = C_{+} \mae{N}{O^+_N}{N}
\,.\end{equation}
where in pure dimensional regularization, $C_+$ is given by the IR-finite terms in the matrix element $\mae{N}{O_{N-1}}{2}$.

After performing the BPS field redefinition, the matrix element can be factorized as
\begin{align}
\mae{N}{O_{N-1}}{2}
&= \mae{2}{\collF_{\nt}}{0} \mae{0}{\collF_{n_a}}{1} \dotsb \mae{1}{\collF_{n_N}}{0}
\nn\\ & \quad\times
\Mae{0}{ Y_{N-1}}{0}
\,,\end{align}
where $Y_{N-1}$ stands for the product of $N-1$ usoft Wilson lines. Since a separate RPI exists for each collinear sector, each collinear matrix element $\mae{1}{\collF_{n_i}}{0}$ must separately be RPI-III invariant.  However, the only objects that each collinear matrix element can depend on are
\begin{equation}
\bigl\{n_i^{\mu}, \bn_i^{\mu}, q_i^{\mu} \bigr\}
\,.\end{equation}
The only RPI-III invariant scale that we can define from these objects is $q_i^2 = 0$.  Therefore, the loop corrections to the collinear matrix elements are scaleless and hence zero in pure dim reg. A similar argument goes through for the usoft matrix element $\mae{0}{Y_{N-1}}{0}$. This is of course the familiar result that the virtual loop corrections to jet and soft functions vanish in pure dimensional regularization.

If we apply a similar argument to the $\nt$ collinear matrix element $\mae{2}{\collF_{\nt}}{0}$, we do not find that it vanishes in pure dimensional regularization. Since there are two particles in the final state, the set of objects that the matrix element can depend on is
\begin{equation} \label{eq:RPIparams}
\{ n_t^{\mu}, \,\bn_t^{\mu},\, n_1^{\mu},\, \bn_1^{\mu},\, n_2^{\mu},\, \bn_2^{\mu},\, q_1^{\mu},\, q_2^{\mu}\} \,.
\end{equation}
What RPI-III invariant, dimensionful parameters can be constructed out of these? While there are many choices, they fall into three categories:
\begin{equation} \label{eq:RPIscales}
\{ q_{1,2}^2 = 0,\, q_1\cdot q_2 , \, \big(n_i \cdot q_{1,2}\big)\big(\bn_i \cdot q_{1,2}\big) \}\,, \quad i = 1,2,t
\,.\end{equation}
We now argue that all of the nonzero scales are the same order as $t = 2q_1\cdot q_2$.  To understand this scaling, note that all objects in \eq{RPIparams} can be constructed from the components of $q_1^{\mu}$, $q_2^{\mu}$, and $q_t^{\mu}$.  Enforcing $q_t^\mu = q_1^\mu + q_2^\mu$ and $q_1^2 = q_2^2 = 0$ leaves only 6 independent variables in the components of $q_1$, $q_2$, and $q_t$.  We can choose these 6 variables to be $t$, the energy $E_t = E_1 + E_2$, the energy fraction $z = E_1/E_t$, where we are working in the regime where $z$ is not near $0$ or $1$, i.e., parametrically $E_1/E_2 \sim 1$. The last three variables are three angles. One is the azimuthal angle of the $1\to 2$ splitting, while the other two are trivial and merely specify the coordinate system. The dependence on the azimuthal angle is determined by the spin structure and therefore only appears as an overall dependence.

In principle, there are two scales, $t$ and $E_t$, which the objects in \eq{RPIscales} can depend on, and $t \ll E_t^2$.  However, using
\begin{align}
t = 2q_1\cdot q_2 = (\bn_1\cdot q_1)(n_1\cdot q_2) = z E_t (n_1\cdot q_2)
\end{align}
it follows that $\bn_1\cdot q_1 \sim E_t$ and $n_1\cdot q_2 \sim t/E_t$, and it is easy to see that all other nonzero $\bn_i \cdot q_{1,2}$ and $n_i\cdot q_{1,2}$ have the same parametric scaling, $n_i\cdot q_{1,2} \sim t/E_t$ and $\bn_i\cdot q_{1,2}\sim E_t$. Therefore all possible RPI-III invariant combinations in \eq{RPIscales} are of order $t$, while the $E_t$ dependence can only enter through higher-order power corrections of $t/E_t^2$.

This scaling dependence then implies that at leading order in the power counting the only scale that all loop corrections can depend on is $t$, and after integrating over the azimuthal angle, the matching coefficient will only depend on $t$ and $z$, as we have seen in the NLO result for $C_+$ in \eq{H3Plus}.  This is the desired result, valid at all orders in perturbation theory.

\section{All-Order Scale Sensitivity of the Soft Functions}
\label{sec:RPIMatchingSoft}

In \sec{njets} we derived the factorization theorem for the $N$-jettiness cross section in the limit $t \ll s_{ij}$. In this appendix we use RPI-III to derive the scale dependence of the csoft and usoft functions and show that they each depend on only a single parametric scale to all orders. The csoft function $S_+(k_1,k_2,\mu)$ only depends on the scale $\mu_{S_+} \sim k_i/\sqrt{\hs_t}$, and the usoft function $S_{N-1} (k_1, k_2, \ldots, k_N,\mu)$ depends on the scale $\mu_{S_{N-1}} \sim k_i/\sqrt{\hs_Q}$.

We start with the csoft function, $S_+ (k_1,k_2,\mu)$, defined in \eq{Splusgeneral}, and determine how it transforms under RPI-III.  The structure of $S_+$ is independent of the partonic splitting channel, which only determines the color representation of the Wilson lines. The structure of the measurement function depends on the observable. Recall that our discussion of the measurement in \subsec{fact} applies to the geometric measure.

In general, $S_+$ depends on $\{k_1,k_2\}$ through its measurement function in \eq{softmeas} and on the reference vectors  $\{ \hq_1,\hq_2, \hq_t, \hq_{\bar t} \}$.  The dependence on $\hq_1$ and $\hq_2$ comes from the $X$ Wilson lines and the measurement function, while the dependence on $\hq_{\bar t}$ comes from the $V$ Wilson line and the dependence on $\hq_t$ comes from the zero-bin subtraction. The directions $\hq_{1,2,t} $ are in the same csoft RPI equivalence class and therefore transform under RPI-III as
\begin{equation} \label{eq:RPItransform}
\hq_{1,2,t} \to e^\alpha \hq_{1,2,t}
\,,\qquad \hq_{\bar t} \to e^{-\alpha} \hq_{\bar t}
\,.\end{equation}
To determine the transformation properties of $S_+$ under RPI-III, we only need to consider the transformation of the measurement function since the Wilson lines are RPI-III invariant. The csoft measurement function transforms under the RPI-III transformation in \eq{RPItransform} as
\begin{align}
\cM^\csoft(k_1,k_2)
&\to e^{-2 \alpha} \prod_{i=1,2} \delta\bigl(k_i - 2 \hq_i \cdot \hP_i^\csoft \bigr)
\end{align}
where we have rescaled the $k_i$ convolution variables to $e^\alpha k_i$, which determines their transformation properties under RPI-III. The analogous result also holds for the zero-bin measurement function. The csoft function therefore transforms as $S_+ \to  e^{-2\alpha} S_+$. We can see this scaling in the RPI transformation of the tree level result
\begin{equation}
S_+^{\rm tree} (k_1, k_2, \mu) = \delta(k_1)\, \delta(k_2) \to e^{-2\alpha} \delta(k_1)\,\delta(k_2)
\,,\end{equation}
which is simply the statement that the $S_+$ has mass dimension $-2$ and therefore has an overall scaling factor of $1/(k_1 k_2)$ at all orders in perturbation theory multiplying a RPI-III invariant function with nontrivial $k_{1,2}$ and $\mu$ dependence. The convolution measure $\df k_1 \df k_2$ has a compensating $e^{2 \alpha}$ dependence and therefore the combination $\df k_1\, \df k_2 \, S_+ (k_1,k_2,\mu) $ is RPI-III invariant. It is useful to rescale the measure to factor out this $1/(k_1 k_2)$ scaling from the csoft function, which effectively changes the measurement function to an RPI-III invariant measurement,
\begin{align}
\widetilde\cM^{\csoft} (k_1,k_2,\mu)
&= \prod_{i=1,2} \delta\biggl(1 - \frac{2 \hq_i\cdot \hP_i^\csoft }{k_i} \biggr)
\end{align}
and similarly for the zero-bin measurement function. This allows us to write the csoft function in an RPI-III invariant form, while the combination $\df k_1\, \df k_2/(k_1 k_2) \widetilde S_+$ is unchanged. We can now consider all RPI-III invariant scales that arise in the rescaled csoft function. This method of changing the convolution variable to write the delta function in an RPI-III invariant form is analogous to the method used in Ref.~\cite{Marcantonini:2008qn} to analyze the RPI transformation properties in the hard matching onto \SCETI.
The only RPI-III invariant dimension-one combinations that we can construct in the rescaled csoft function are
\begin{equation}
\biggl\{  \frac{k_1}{\sqrt{\hs_{ij}}},\frac{k_2}{\sqrt{\hs_{ij}}} \biggr\}
\,,\end{equation}
where $i,j \in \{1,2,t\}$, and recall that $\hs_{ij} = 2\hq_i\cdot \hq_j$.  Additionally, we can form the dimensionless RPI-III invariants $\hq_{\bar t}\cdot \hq_i$ and $\hs_{ij}/\hs_{ik}$, but these are order 1, and so we can ignore them.  Since all of the $\hs_{ij}$ are of order $\hs_t = 2\hq_1\cdot \hq_2$, this shows that the only parametric scales that the csoft function depends on are $k_{1,2} / \sqrt{\hs_t}$, as advertised.

The scale dependence of the usoft function $\widehat{S}_{N-1}$ can be constrained using analogous arguments.  The usoft function depends on the $N+2$ measurements $k_a, k_b, k_1, \ldots, k_N$ and the $N+1$ directions $\hq_a, \hq_b, \hq_t, \hq_3, \ldots \hq_N$.  Note that all of these directions are well separated, so that any $\hs_{ij} = 2\hq_i \cdot \hq_j \sim 1$.  Additionally, in the $k_1$ and $k_2$ measurement functions the jet boundaries depend on the orientation of the nearby jets relative to the rest of the event, which is parameterized by an angle $\phi_t$.  Since this angle is parametrically order 1, we need not consider its dependence.

The RPI-III transformation properties are more intricate now, as there are $N+1$ separate collinear sectors contributing usoft Wilson, each of which can in principle have a separate RPI-III transformation parameter $\alpha_i$.  The jet boundaries are not invariant under arbitrary RPI-III transformations, since the measurement contains comparisons of the form
\begin{equation}
\theta(\hq_i \cdot p_k < \hq_j \cdot p_k) \,,
\end{equation}
which do not have simple transformation properties under $\hq_i \to e^{\alpha_i} \hq_i$ and $\hq_j \to e^{\alpha_j} \hq_j$.

However, the usoft function has simple transformation properties under a restricted set of RPI-III transformations with all $\alpha_i$ set equal, so that all $\hq_i$ transform as
\begin{equation} \label{eq:globalRPI}
\hq_i \to e^{\alpha} \hq_i
\,.\end{equation}
We call this a global RPI-III transformation, since it universally affects all collinear sectors.  Note that in $\hS_{N-1}$ there are no $\bn$ directions to consider, since they do not enter in the Wilson lines or the measurement function.  Under a global RPI-III transformation, the jet boundaries are invariant, and the measurement function has a simple transformation property:
\begin{equation}
\mathcal{M}_N(k_1, \ldots , k_N) \to e^{-(N+2)\alpha} \mathcal{M}_N(k_1, \ldots , k_N)
\,,\end{equation}
where $k_i \to e^{\alpha} k_i$ under global RPI-III transformations, just as for $S_+$. The usoft function therefore transforms as $\widehat{S}_{N-1} \to e^{-(N+2)\alpha} \widehat{S}_{N-1}$. This is compensated for by the transformation of the measure of the convolution variables under global RPI-III $\prod_i \df k_i \to e^{(N+2)\alpha} \prod_i \df k_i$, such that the combination $\prod_i \df k_i \, \widehat{S}_{N-1}$  is invariant. As before, it is useful to rescale the measure to factor out the $\prod_i 1/k_i$ scaling dependence of $\hS_{N-1}$. The usoft measurement function then becomes
\begin{equation}
\widetilde\cM_N^\usoft(\{k_i\})
= \!\prod_{i = 1,2}\! \delta\biggl( 1 - \frac{2\hq_t \cdot \hP_i^\usoft}{k_i}\biggr)
\!\prod_{i\neq 1,2}\! \delta\biggl( 1 - \frac{2\hq_i \cdot \hP_i^\usoft}{k_i}\biggr)
.\end{equation}

The rescaled usoft function is now invariant under global RPI-III. To determine its kinematic dependence, we construct all global RPI-III invariant dimension-one combinations, which are
\begin{equation} \label{eq:softRPIscales}
\biggl\{ \frac{k_i}{\sqrt{\hs_{jk}}} \biggr\}
\end{equation}
for all $i \in a,b,1,2,\ldots, N$ and $j,k \in a,b,t,\ldots, N$ with $j\neq k$.  Unlike the $S_+$ function, there is no $\bn_i$ dependence in $\hS_{N-1}$. The only dimensionless RPI-III invariant combinations of the $\hq_i$ that we can construct are the ratios $\hs_{ij}/\hs_{kl}$, which are all order 1 and can again be ignored.

We can further restrict the set in \eq{softRPIscales} using the fact that in the rescaled function, each $k_i$ only appears in the measurement function and always comes with $\hq_i$ in the combination $ \hq_i^{\mu}/k_i$.  (Since the $\hq_i$ can also arise from the Wilson lines, they can also appear without corresponding factors of $k_i$.)  This requires either $j = i$ or $k = i$ in \eq{softRPIscales}, reducing the set of possible combinations $\hS_{N-1}$ can depend on to
\begin{equation} \label{eq:softRPIscalesfull}
\biggl\{ \frac{k_i}{\sqrt{\hs_{ij}}} \biggr\}
\,.\end{equation}
Therefore, the only scales that the soft function can depend on to all orders in perturbation theory are those in \eq{softRPIscalesfull}.  Since all of the $\hs_{ij} \sim \hs_Q$, these scales are all of the same parametric order.

\section{Resummation Formulas}
\label{app:resum}

\begin{table}
\begin{tabular}{c|cc|c}
\hline\hline
      & $C_X$  & $j_X$ & $\gamma_X^0$ \\\hline
$C_2$ & $C_F$   & $2$ & $-6C_F$ \\
$C_+^{\{\cdot, g,\cdot\}}$ & $ C_A/2$   & $2$ & $2(2C_F - C_A) \ln x + 2C_A \ln(1-x) -\beta_0 $ \\
$C_+^{\{g,\cdot,\cdot\}}$ & $ C_A/2$   & $2$ & $2(2C_F - C_A) \ln(1-x) + 2C_A \ln x -\beta_0 $ \\
$J_q$ & $-2C_F$ & $2$ & $6C_F$ \\
$J_g$ & $-2C_A$ & $2$ & $2\beta_0$ \\
$S_+$ & $2C_A$  & $1$ & 0 \\
$S_2$ & $4C_F$  & $1$ & 0 \\\hline\hline
\end{tabular}
\caption{Anomalous dimension coefficients for the various functions in the factorization theorem.}
\label{table:anom}
\end{table}

In this appendix we collect all inputs needed for the resummation at NLL in \sec{Applications}. The inputs required for the resummation at NNLL can be found for example in Appendix D.2 of Ref.~\cite{Stewart:2010qs} and Appendix B.3 of Ref.~\cite{Berger:2010xi}.

The evolution kernels $U_X$ for the different functions in \eq{sigmaTau3resum} which solve their evolution equations are written in terms of the evolution functions
\begin{align} \label{eq:Ketadef}
K_X(\mu_0,\mu) &= - j_X C_X K_\Gamma(\mu_0,\mu) + K_{\gamma_X}(\mu_0,\mu)
\,, \nn \\
\eta_X(\mu_0,\mu) &= C_X \eta_\Gamma(\mu_0,\mu)
\,,\end{align}
where $j_X$ is the dimension of the evolution variable and $C_X$ the coefficient of $\Gamma_\cusp(\alpha_s)$ in the anomalous dimension for each function, which are summarized in Table~\ref{table:anom}. The evolution functions $K_\Gamma(\mu_0, \mu)$, $\eta_\Gamma(\mu_0, \mu)$, and $K_{\gamma_X}(\mu_0, \mu)$ are given further below.

The RG evolution of the hard function $H_2$ following from \eqs{C2RGE}{gammaC2} is
\begin{align} \label{eq:Hrun2}
H_2(Q^2, \mu) &= H_2(Q^2, \mu_0)\, U_{H_2}(Q^2, \mu_0, \mu)
\,,\\\nn
U_{H_2}(Q^2, \mu_0, \mu)
&= \biggl\lvert e^{K_{C_2}(\mu_0, \mu)} \Bigl(\frac{- Q^2 - \img 0}{\mu_0^2}\Bigr)^{\eta_{C_2}(\mu_0, \mu)} \biggr\rvert^2
\,.\end{align}
The RG evolution of the hard function $H_+^\kappa$ following from \eq{RGEgammaH3Plus} is
\begin{align} \label{eq:Hrun3p}
H_+^\kappa(t,z, \mu) &= H_+^\kappa(t,z, \mu_0) \, U_{H_+}^\kappa(t,z, \mu_0, \mu)
\,,\nn\\
U_{H_+}^\kappa(t,z, \mu_0, \mu)
&= \biggl\lvert e^{K_{C_+}^\kappa(\mu_0, \mu, z)} \Bigl(\frac{- t - \img 0}{\mu_0^2}\Bigr)^{\eta_{C_+}(\mu_0, \mu)} \biggr\rvert^2
\,.\end{align}
Here, $K_{C_+}^\kappa(\mu_0, \mu, x)$ explicitly depends on $\kappa$ and $x$ via $\gamma_{C_+}^\kappa[\alpha_s, x]$.

The solution for the soft function RGE in \eq{S_RGE} is given by~\cite{Balzereit:1998yf, Neubert:2004dd, Fleming:2007xt, Ligeti:2008ac}
\begin{align} \label{eq:Srun}
S(k,\mu) & =  \int\! \df k'\, S(k - k',\mu_0)\, U_S(k',\mu_0,\mu)
\,, \\
U_S(k, \mu_0, \mu) & = \frac{e^{K_S -\gamma_E\, \eta_S}}{\Gamma(1 + \eta_S)}\,
\biggl[\frac{\eta_S}{\sqrt{\hs} \mu_0} \cL^{\eta_S} \Bigl( \frac{k}{\sqrt{\hs} \mu_0} \Bigr) + \delta(k) \biggr]
\,,\nn \end{align}
where $K_S \equiv K_S(\mu_0, \mu)$, $\eta_S \equiv \eta_S(\mu_0, \mu)$, and $\hs=\hs_t$ or $\hs = \hs_Q$ for $S_+$ or $S_2$, respectively. The plus distribution $\cL^\eta$ is defined as
\begin{equation}
\cL^{\eta}(x)  = \biggl[ \frac{\theta(x)}{x^{1-\eta}}\biggr]_+
= \lim_{\beta\to 0} \biggl[ \frac{\theta(x-\beta)}{x^{1-\eta}} + \delta(x-\beta) \frac{x^\eta-1}{\eta} \biggr]
\,.\end{equation}
The RGE for the jet functions is given in \eq{J_RGE}. The solution has exactly the same structure as \eq{Srun},
\begin{align} \label{eq:Jrun}
J(s,\mu) & =  \int\! \df s'\, J(s - s',\mu_0)\, U_J(s',\mu_0,\mu)
\,, \nn \\
U_J(s, \mu_0,\mu) & = \frac{e^{K_J -\gamma_E\, \eta_J}}{\Gamma(1 + \eta_J)}\,
\biggl[\frac{\eta_J}{\mu_0^2} \cL^{\eta_J} \Bigl( \frac{s}{\mu_0^2} \Bigr) + \delta(s) \biggr]
\,,\end{align}
where $K_J \equiv K_J(\mu_0, \mu)$ and $\eta_J \equiv \eta_J(\mu_0, \mu)$ as in \eq{Ketadef}. The convolutions of the plus distributions in the final cross section are easily evaluated analytically and can be found in Appendix B of Ref.~\cite{Ligeti:2008ac}.

The RGE functions are defined as
\begin{align} \label{eq:Keta_def}
K_\Gamma(\mu_0, \mu)
& = \int_{\alpha_s(\mu_0)}^{\alpha_s(\mu)}\!\frac{\df\alpha_s}{\beta(\alpha_s)}\,
\Gamma_\cusp(\alpha_s) \int_{\alpha_s(\mu_0)}^{\alpha_s} \frac{\df \alpha_s'}{\beta(\alpha_s')}
\,,\nn\\
\eta_\Gamma(\mu_0, \mu)
&= \int_{\alpha_s(\mu_0)}^{\alpha_s(\mu)}\!\frac{\df\alpha_s}{\beta(\alpha_s)}\, \Gamma_\cusp(\alpha_s)
\,,\nn \\
K_{\gamma_X}(\mu_0, \mu)
& = \int_{\alpha_s(\mu_0)}^{\alpha_s(\mu)}\!\frac{\df\alpha_s}{\beta(\alpha_s)}\, \gamma_X(\alpha_s)
\,.\end{align}
Expanding the beta function and anomalous dimensions in powers of $\alpha_s$,
\begin{align} \label{eq:Gammacusp}
\beta(\alpha_s) &=
- 2 \alpha_s \sum_{n=0}^\infty \beta_n\Bigl(\frac{\alpha_s}{4\pi}\Bigr)^{n+1}
\,, \nn\\
\Gamma_\cusp(\alpha_s) &= \sum_{n=0}^\infty \Gamma_n \Bigl(\frac{\alpha_s}{4\pi}\Bigr)^{n+1}
\,, \nn\\
\gamma(\alpha_s) &= \sum_{n=0}^\infty \gamma_n \Bigl(\frac{\alpha_s}{4\pi}\Bigr)^{n+1}
\,,\end{align}
the explicit expressions of the evolution functions at NLL are
\begin{align} \label{eq:Keta}
K_\Gamma(\mu_0, \mu) &= -\frac{\Gamma_0}{4\beta_0^2}\,
\biggl[ \frac{4\pi}{\alpha_s(\mu_0)}\, \Bigl(1 - \frac{1}{r} - \ln r\Bigr)
\nn\\ & \quad
   + \biggl(\frac{\Gamma_1 }{\Gamma_0 } - \frac{\beta_1}{\beta_0}\biggr) (1-r+\ln r)
   + \frac{\beta_1}{2\beta_0} \ln^2 r \biggr]
\,, \nn\\
\eta_\Gamma(\mu_0, \mu) &=
 - \frac{\Gamma_0}{2\beta_0}\, \biggl[ \ln r
 + \frac{\alpha_s(\mu_0)}{4\pi}\, \biggl(\frac{\Gamma_1 }{\Gamma_0 }
 - \frac{\beta_1}{\beta_0}\biggr)(r-1)    \biggr]
\,, \nn\\
K_\gamma(\mu_0, \mu) &=
 - \frac{\gamma_0}{2\beta_0}\, \ln r
\,.\end{align}
Here, $r = \alpha_s(\mu)/\alpha_s(\mu_0)$ and the running coupling is given by the three-loop expression
\begin{align} \label{eq:alphas}
\frac{1}{\alpha_s(\mu)} &= \frac{X}{\alpha_s(\mu_0)}
  +\frac{\beta_1}{4\pi\beta_0}  \ln X
\\\nn & \quad
  + \frac{\alpha_s(\mu_0)}{16\pi^2} \biggr[
  \frac{\beta_2}{\beta_0} \Bigl(1-\frac{1}{X}\Bigr)
  + \frac{\beta_1^2}{\beta_0^2} \Bigl( \frac{\ln X}{X} +\frac{1}{X} -1\Bigr) \biggl]
\,,\end{align}
where $X\equiv 1+\alpha_s(\mu_0)\beta_0 \ln(\mu/\mu_0)/(2\pi)$.

The coefficients of the beta function~\cite{Tarasov:1980au, Larin:1993tp} and cusp anomalous dimension~\cite{Korchemsky:1987wg} in $\overline{\mathrm{MS}}$ are
\begin{align} \label{eq:Gacuspexp}
\beta_0 &= \frac{11}{3}\,C_A -\frac{4}{3}\,T_F\,n_f
\,,\nn\\
\beta_1 &= \frac{34}{3}\,C_A^2  - \Bigl(\frac{20}{3}\,C_A\, + 4 C_F\Bigr)\, T_F\,n_f
\,, \nn\\
\beta_2 &=
\frac{2857}{54}\,C_A^3 + \Bigl(C_F^2 - \frac{205}{18}\,C_F C_A
 - \frac{1415}{54}\,C_A^2 \Bigr)\, 2T_F\,n_f
\nn\\ & \quad
 + \Bigl(\frac{11}{9}\, C_F + \frac{79}{54}\, C_A \Bigr)\, 4T_F^2\,n_f^2
\\[2ex]
\Gamma_0 &= 4
\,,\nn\\
\Gamma_1 &= 4 \Bigl[\Bigl( \frac{67}{9} -\frac{\pi^2}{3} \Bigr)\,C_A  -
   \frac{20}{9}\,T_F\, n_f \Bigr]
\,.\end{align}
The individual quark and gluon contributions to the noncusp parts of the hard and jet anomalous dimensions in \eqs{gammaJ}{gammaC23} at one loop are
\begin{align}
\label{eq:gammaJi}
\gamma_J^q[\alpha_s] = \gamma_J^{\bar q}[\alpha_s]
&= \frac{\alpha_s}{4\pi}\, 6 C_F + \ord{\alpha_s^2}
\,,\nn\\
\gamma_J^g[\alpha_s]
&= \frac{\alpha_s}{4\pi}\, 2\beta_0 + \ord{\alpha_s^2}
\,,\\[1ex]
\label{eq:gammaCi}
\gamma_C^q[\alpha_s] = \gamma_C^{\bq}[\alpha_s]
&= \frac{\alpha_s}{4\pi}\, (-3 C_F) + \ord{\alpha_s^2}
\,,\nn\\
\gamma_C^g[\alpha_s]
&= \frac{\alpha_s}{4\pi}\, (-\beta_0) + \ord{\alpha_s^2}
\,.\end{align}

\bibliographystyle{../physrev4}
\bibliography{../scetp}

\end{document}